\documentclass[10pt,showpacs,aps,prb,reprint,superscriptaddress,floatfix]{revtex4-1}
\usepackage{graphicx}
\usepackage{multirow}
\usepackage{epstopdf}
\usepackage{amssymb}
\usepackage{dcolumn}
\newcolumntype{d}[1]{D{.}{.}{#1}}
\usepackage{color}
\usepackage{amsmath}

\begin{document}

% number in the upper right-hand corner of the title page in preprint mode.

\title{ Vibrational and dielectric properties of the bulk transition metal dichalcogenides}

\author{Nicholas A. Pike}
\email[]{Nicholas.pike@ulg.ac.be}
%\thanks{}
\affiliation{Centre for Materials Science and Nanotechnology, University of Oslo, NO-0349 Oslo, Norway }
\affiliation{Nanomat/Q-Mat/CESAM, Universit\'{e} de Li\`{e}ge \& European Theoretical Spectroscopy Facility, B-4000 Li\`{e}ge, Belgium}
\author{Antoine Dewandre}
\affiliation{Nanomat/Q-Mat/CESAM, Universit\'{e} de Li\`{e}ge \& European Theoretical Spectroscopy Facility, B-4000 Li\`{e}ge, Belgium}
\author{Benoit Van Troeye} 
\affiliation{Universit\'{e} Catholique de Louvain, Institute of Condensed Matter and Nanosciences 
(IMCN) \& European Theoretical Spectroscopy Facility, B-1348 Louvain-la-Neuve, Belgium}
\author{Xavier Gonze}
\affiliation{Universit\'{e} Catholique de Louvain, Institute of Condensed Matter and Nanosciences 
(IMCN) \& European Theoretical Spectroscopy Facility, B-1348 Louvain-la-Neuve, Belgium}
\author{Matthieu J. Verstraete}
\affiliation{Nanomat/Q-Mat/CESAM, Universit\'{e} de Li\`{e}ge \& European Theoretical Spectroscopy Facility, B-4000 Li\`{e}ge, Belgium}

\date{\today}

\begin{abstract}
Interest in the bulk transition metal dichalcogenides for their electronic, photovoltaic, and optical properties has grown and led to their use in many technological applications.  We present a systematic investigation of their interlinked vibrational and dielectric properties, using density functional theory and density functional perturbation theory, studying the effects of the spin-orbit interaction and of the long-range e$^-$- e$^-$ correlation as part of our investigation.  This study confirms that the spin-orbit interaction plays a small role in these physical properties, while the direct contribution of dispersion corrections is of crucial importance in the description of the interatomic force constants. Here, our analysis of the structural and vibrational properties, including the Raman spectra, compare well to experimental measurement. Three materials with different point groups are showcased and data trends on the full set of fifteen existing hexagonal, trigonal, and triclinic materials are demonstrated. This overall picture will enable the modeling of devices composed of these materials for novel applications.
\end{abstract}

\pacs{Density Functional Theory +GGA \\  Van der Waals Interaction \\ Transition metal Dichalcogenides}
%\keywords{}

\maketitle

\section{Introduction}

Interest in the Transition Metal Dichalcogenides (TMDs) is widespread in materials science since they display a number of useful properties for industrial and engineering applications~\cite{Chhowalla2013, Wang2012}.  Indeed, the bulk electronic properties of these materials vary greatly depending on their chemical makeup and underlying crystal symmetry~\cite{Jariwala2014, Rasmussen2015}.  For electronic and photovoltaic applications these materials display promising features, including an indirect to direct band gap transition that can be controlled via the number of layers or by the application of strain~\cite{Gupta2015, Bhimanapati2015, Wang2012, Yun2012, Ataca2012}. They are known to have high electron mobilities~\cite{Kuc2015, Radisavjevic2011, Ganatra2014} and have excellent lubricative properties due to their layered natures. Some of these materials exhibit charge-density waves (CDW)~\cite{VanBakel1992,Johannes2006,Calandra2009,Zhu2015,Leroux2012,McMullan1985,Wang1974,Chen2015,Guster2018,Dolui2016}. Several recent high-throughput studies have shown the power of density functional theory to identify materials which can be exfoliated and have useful physical properties~\cite{2017_arxiv_mounet_2D, 2017_choudhary_scirep,Sofer2017}.

Several optoelectronic devices~\cite{Li2015, Wei2015, Liu2016} have been proposed which contain heterostructures of TMD materials in the hopes of blending their unique properties. The understanding of these heterostructures, either as truly two-dimensional layers of materials or semi-bulk layered materials, requires detailed knowledge of the mechanical, electronic, optical, thermal, elastic and acoustic properties of the individual components.  These properties are intimately linked and a full picture is essential to understand heterostructured systems composed of TMDs and other two-dimensional materials.

To understand the properties of these materials, and find trends amongst them, we undertake an investigation using the same functionals and approximations for a wide range of commonly used TMD materials. Our examination uses Density Functional Theory (DFT)~\cite{Martin2004} and Density Functional Perturbation Theory (DFPT)~\cite{Baroni2001} where the particular effects of different commonly used approximations are examined.  We find that many DFPT properties are strongly dependent on the lattice parameters and on the inclusion of long-range e$^-$-e$^-$ correlation, approximated via a van der Waals (vdWs) dispersion scheme from Grimme known as  DFT-D~\cite{Grimme2006, Grimme2010, Grimme2011}.

\begin{figure*}[!t]
\includegraphics[width=0.8\textwidth]{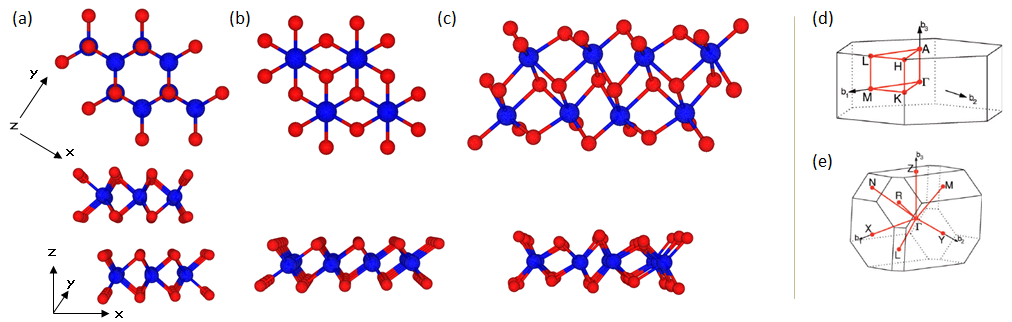} 
\caption{\label{fig:unitcell}(Color online) The top and side view of a (a) h-TMD, (b) t-TMD, and (c) a tc-TMD. Here the transition metal atoms are shown in blue and the chalcogen atoms are in red.  In each case, a coordinate system is used such that the x and y-axes lie in the plane of atoms and the z-axis lies perpendicular to the plane. Additionally, the corresponding high-symmetry paths~\cite{Setyawan2010} in momentum space used in our band structure calculations for (d) the h- and t-TMDs and (e) the tc-TMDs are given. }
\end{figure*}

In section~\ref{methods}, a brief outline of the methods used and a brief description of the crystal structures of the compounds investigated here are followed by a discussion of the effects of the different approximations used to calculate the phonon band structures and interatomic force constants where the most important contributions are highlighted. In sections~\ref{properties} - V, we present the mechanical, electrical, optical, thermal and acoustic properties of three prototypical TMD systems, show their corresponding electronic and phonon band structures, phonon density of states, and Raman spectra, and display global trends and outliers in the properties of the full set of TMDs. The numerical data for all systems is given in the Supplemental Material (SM)~\cite{supmat}.  Finally, we summarize and conclude in section~\ref{conclusions}.

\section{Calculation Methods}\label{methods} 

We employ DFT~\cite{Martin2004} and DFPT~\cite{Gonze1997b,Baroni2001,Gonze1997a} as implemented in the~\textit{ABINIT} software package~\cite{Gonze2005,Gonze2009,Gonze2016}, with a plane wave basis set and norm-conserving pseudopotentials. For a review of DFT and its role in electronic structure calculations see Ref.~\citenum{Jones2015} and for a summary of DFPT and its role in the calculation of the vibrational properties see Refs.~\citenum{Baroni2001},~\citenum{Gonze2005a} and~\citenum{verstraete_2014_dfpt}. More details of our DFT and DFPT calculation parameters for these materials, including energy cut-offs and grid sampling~\cite{Monkhorst1976}, are found in the SM~\cite{supmat}.  For our calculations of the Raman spectra, we assume a fixed Lorentz broadening of the spectral lines of $1.0\times10^{-5}$s and a laser wavelength of 532 nm representing a frequency typically used in Raman experiments~\cite{Zeng2013, Wolverson2016}.  The intensity of each Raman spectra should only be used to demonstrate the existence of Raman peaks near the calculated frequencies as the calculations presented here use LDA exchange-correlation during the calculation of the third derivative of the total energy with wave functions calculated using a GGA exchange-correlation functional and a vdWs dispersion correction.  Our plotted Raman intensities are also rescaled by its corresponding maximum value and thus only relative differences in intensity can be compared directly to experiment.

\begin{figure}[!b]
\centering
\includegraphics[width=0.5\textwidth]{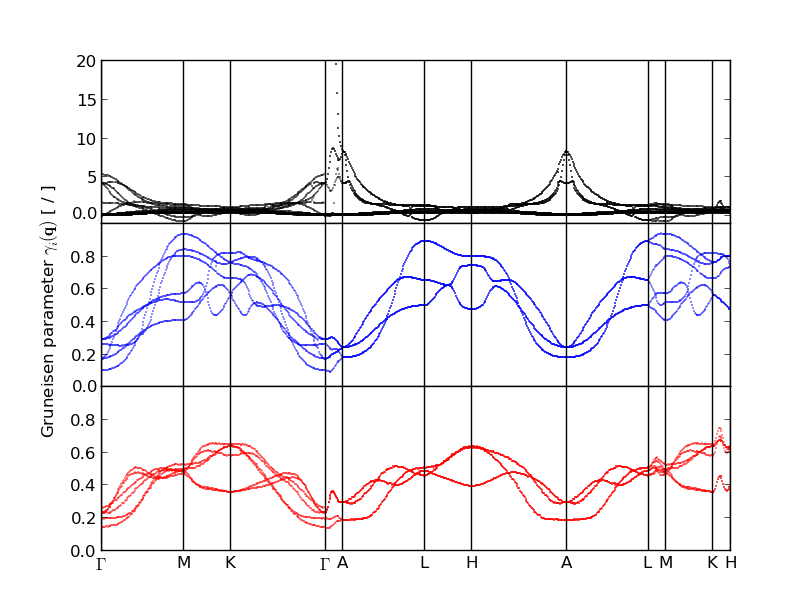}
\caption{\label{Gruneisen}(Color Online) Mode-dependent and momentum dependent Gr\"{u}neisen parameters of MoTe$_2$. Top: acoustic and lattice Gr\"{u}neisen parameters; Middle and bottom: intralayers Gr\"{u}neisen parameters.  The high-symmetry path is from Ref.~\citenum{Setyawan2010}.  }
\end{figure}

There are three main families of TMDs: hexagonal (h-TMDs) with space group P6$_3$/mmc, trigonal (t-TMDs) with space group P$\overline{3}$m1, and triclinic (tc-TMDs) with space group symmetry P$\overline{1}$. In Fig.~\ref{fig:unitcell}, we show example structures for each of the three crystal symmetries.  Namely, we consider the h-TMDs:  MoS$_2$, MoSe$_2$, MoTe$_2$. WS$_2$, WSe$_2$, NbS$_2$, and NbSe$_2$, the t-TMDs: TiS$_2$, TiSe$_2$, TiTe$_2$, and ZrS$_2$, ZrSe$_2$, and finally the tc-TMDs: ReS$_2$, ReSe$_2$, and TcS$_2$. The variety of crystal symmetries found in the TMDs gives rise to unique physical properties with each material being composed of layers of metal atoms packed between two layers of chalcogen atoms.  In the case of the bulk h-TMDs, the unit cell consists of two trilayers. As shown in Fig.~\ref{fig:unitcell}(a) this leads to a well-known graphene-like hexagonal configuration in the xy plane. The AB stacking between the layers can be clearly identified in the yz plane. For the t-TMDs, shown in Fig.~\ref{fig:unitcell}(b) the chalcogenide layers are trigonally coordinated. Finally, the tc-TMDs contain a single-layer of metal atoms sandwiched between the chalcogen atoms which are buckled along the x and y directions as shown in Fig.~\ref{fig:unitcell}(c).  The buckling along both of the in-plane directions has physical implications for the tc-TMDs, as many of their tensorial properties are asymmetric.  

Before determining the physical properties of these materials we first examine the influence of the choice of the pseudopotentials, the treatment of relativistic effects (spin-orbit coupling), and of the approximation of the long-range e$^-$-e$^-$ correlation on the proper description of the TMDs.

\begin{figure*}[!t]
\centering
\includegraphics[width=0.90\textwidth]{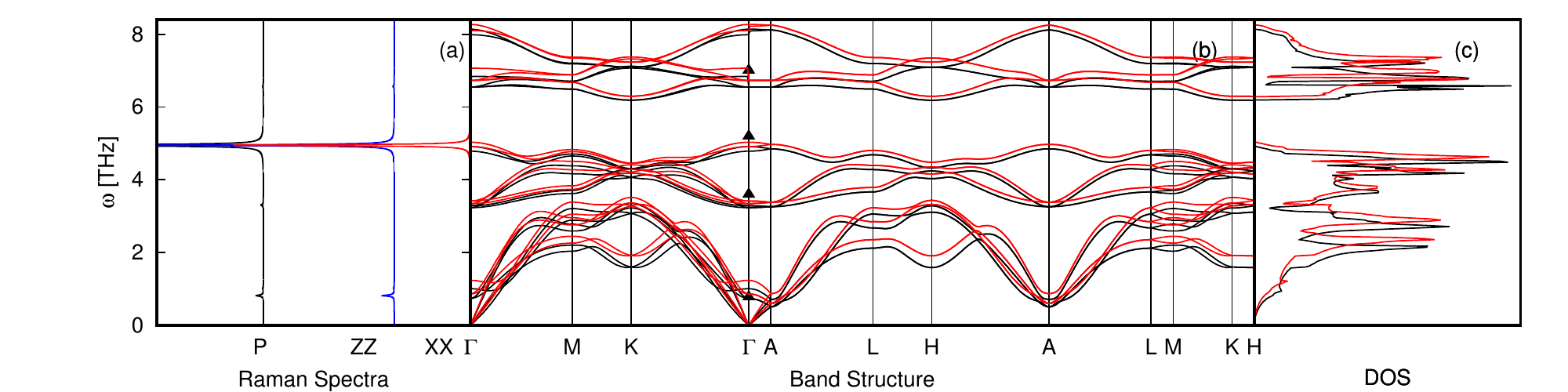} 
\caption{\label{MoTe2_lattice_phon}(Color online)(a) Raman spectra for MoTe$_2$ with three different polarizations corresponding to a DFT-D3/DFT-D3 calculation in which the Raman signals relate to our calculation (in black) of the $\Gamma$ point phonons in (b).  (b)  Calculated phonon band structure of MoTe$_2$ along a path of high symmetry points in $q$ space. The red lines correspond to a DFT-D3/DFT calculation, and the black lines correspond to a DFT-D3/DFT-D3 calculation in which the DFT-D3 scheme is used during both the DFT and DFPT calculation. (c) Phonon density of states corresponding to the same color scheme as in (b).   Experimental data points correspond to the measurements in Ref.~\citenum{Zhao2013}.  The high-symmetry path comes from Ref.~\citenum{Setyawan2010}.  }
\end{figure*}

\subsection{Choice of Pseudopotentials}
We investigate the effects of different pseudopotential approximations on the DFT and DFPT calculation by comparing the OPIUM~\cite{opium_psp}, fhi98PP~\cite{Fuchs1999}, and ONCVPSP~\cite{Hamann2013, Setten2018} generation codes, and find that there are negligible effect on the DFPT calculation for ``well-built'' pseudopotentials~\cite{Grinberg2000,Rappe1990}. The hardest cases are the CDW compounds TiS$_2$~\cite{Dolui2016}, TiSe$_2$~\cite{Guster2018,Chen2015}, NbS$_2$~\cite{Leroux2012,McMullan1985}, and NbSe$_2$~\cite{Johannes2006,Calandra2009} whose specific instabilities are very sensitive to volume and pseudopotential, though their qualitative physics is correct in each case. We do not focus on the CDW aspect or enter into the debates on its nature and details. For completeness and comparison to the others, we thus include both of the Nb and Ti-based CDW TMDs, although we actually treat these materials using their average, translationally-invariant geometry. 

The only unsatisfactory pseudopotentials were those generated with the fhi98pp code~\cite{Perdew1996} for elements W and Ti.  These produced inconsistent results during our relaxations and DFPT calculations. Therefore, unless specifically mentioned, our results use GGA-PBE pseudopotentials, generated with the fhi98pp code (which are slightly softer), for all atoms except W and Ti.  We use a norm-conserving pseudopotential generated with the OPIUM code for W and one generated with the ONCVPSP code for Ti. 

\subsection{Effect of Spin-orbit coupling}

It has been shown that the spin-orbit interaction (SOI) can make a significant correction to the phonon band structure of certain systems~\cite{Verstraete2008, Diaz2007, Diaz2007b}, particularly at the zone edge.  To investigate the effects of the inclusion of SOI on the phonon band structure we calculated the phonon band structure using ONCVPSP pseudopotentials~\cite{Hamann2013} with and without the spin-orbit projectors (see Ref.~\citenum{Grinberg2000} for the SOI method) for both atomic species in several compounds, among them MoTe$_2$ and ReSe$_2$.  The result of this calculation is shown in Fig.~S1 of the SM~\cite{supmat} for MoTe$_2$, the heaviest of the compounds per formula unit investigated here, and ReSe$_2$, the compound with the largest atomic number.  In Fig.~S1 of the SM~\cite{supmat} the phonon band structures are shown without the spin-orbit projectors in black and with the spin-orbit projectors in red.  The inclusion of the SOI made little difference in the lattice parameters, provided a small spin splitting of the valence bands, and had a negligible effect on the phonon band structure of the TMDs, especially the lattice modes when compared to the effect of including the dispersion corrections. Due to the correlation between the changes in the electronic and phonon band structures our calculations show that the SOI modifies the phonon band structure through the electronic band structure and does not have a direct effect on the calculated interatomic force constants.  

\subsection{Effect of the long-range e$^-$-e$^-$ correlation approximation}\label{van_der_waals}

While a precise description of layered semiconductor materials requires a more advanced description of the long-range electron-electron correlation (see Refs.~\citenum{Aryasetiawan1998} and~\citenum{Perdew1985}), DFT in its usual approximations -LDA and GGA- is sufficient to properly describe the vibrational properties of most ionic, metallic, or covalent compounds.  With this in mind, our calculations of the bulk TMDs use three different dispersion schemes given by Grimme~\cite{Grimme2006, Grimme2010, Grimme2011} which are known to have a strong effect~\cite{Troeye2016, Sabatini2016} on electronic and vibrational properties.  These dispersion schemes, known as DFT-D, are based on simple atomic pair-wise terms, with environmental-dependent dispersion coefficients tabulated beforehand using time-dependent DFT, that do not have the computational overhead of other vdW implementations~\cite{Tkatchenko2009, Rydberg2003, Dion2004, Lee2010, Silvestrelli2008, Ambrosetti2012, Furche2001, Fuchs2002, Ren2012}. The result of this calculation for the lattice parameters is shown in the SM~\cite{supmat}.  This calculation and others on the vibrational properties indicate that the DFT-D3 scheme provides the most accurate lattice parameters when compared to measurement. More generally, the performance of the most-popular vdW functional and dispersion schemes have been assessed recently~\cite{Tawfik2018} and DFT-D3 is performing relatively well compared to more elaborate methods. Thus, for all the following calculations, the DFT-D3 scheme is used as part of our DFT and DFPT calculations.

The dispersion corrections influence the phonon frequencies in two different ways: first through geometric changes when compared to calculations with no dispersion corrections, as highlighted in Table~S1 of the SM~\cite{supmat} and second, through their direct contribution to the interatomic force constants as derived in Ref.~\citenum{Troeye2016}. The latter is generally assumed to be negligible due to the small contribution of dispersion corrections to the total energy (~10-100 meV/atom); however, we show here that the dispersion corrections cannot be neglected in the case of TMDs. Still, disentangling the effects of the geometry modification and of the direct contributions to the interatomic force constants needs careful consideration.

To do so, the influence of geometry modification on the phonon frequencies is first estimated using the Gr\"{u}neisen parameter~\cite{Gruneisen1912}, defined as
\begin{equation}\label{grun}
 \gamma_{i}(\boldsymbol{q}) = - \frac{\partial \text{ln} [\omega_{i}(\boldsymbol{q})]}{\partial \text{ln} V},
\end{equation}
where $V$ is the volume of the unit cell and $\omega_i({\bf q})$ is the phonon frequency of the i$^{th}$ mode at momentum ${\bf q}$.  The result of this calculation in the case of MoTe$_2$, shown in Fig.~\ref{Gruneisen}, indicates that the mode and momentum dependent Gr\"{u}neisen parameters are strongly dependent on the type of phonon mode and can vary by an order of magnitude throughout the Brillouin zone, in agreement with experiment~\cite{Sugai1982}.  For clarity, we have separated the mode dependent Gr\"{u}neisen parameters, given by Eq.~\ref{grun}, in Fig.~\ref{Gruneisen} into groups of six modes, based on the corresponding frequencies.  Therefore, the top panel corresponds to the six lowest phonon modes (acoustic and lattice modes), while the middle and top panel corresponds to intralayer modes. From this graph, one can extract that the interlayer modes are extremely sensitive to change of volume (large Gr\"{u}neisen parameters), while the intralayer modes are relatively unaffected by such a change. Therefore, it can be safely assumed that the change of geometry due to the inclusion of dispersion corrections has a relatively negligible role on the phonon frequencies in the TMDs.

Still, the dispersion corrections have a direct contribution to the interatomic force constants and therefore on the phonon frequencies. In order to investigate the role of this contribution, one can simply include or exclude the dispersion corrections during the calculation. The phonon band structures computed with and without dispersion corrections for MoTe$_2$ are represented in Fig.~\ref{MoTe2_lattice_phon}.  In the Raman spectra shown in Fig.~\ref{MoTe2_lattice_phon}(a), the dispersion corrections, and their effects on the electronic and structural properties, are included in both parts of the calculation (DFT and DFPT) since these quantities are related to the derivatives of energy with respect to the electric field, and, as the DFT-D3 correction is independent of the electron density, there are no additional terms in the third order derivatives due to the dispersion correction~\cite{Troeye2016, Troeye2017}. As one can see, the dispersion correction have a drastic influence on the phonon band structure, not only on the intralayer modes but as well for interlayer modes.

\begin{table*}[!tbh]
\caption{Calculated properties of bulk MoS$_2$, ZrS$_2$ and ReS$_2$. Our calculations are given as the first column, experimental measurement as the second column, and the reference to the experimental work in the third column for each material.  We report the in-plane lattice parameters, $a$, and $b$, the out-of-plane  lattice parameter, $c$, the components of the elastic tensor, $c_{ij}$, Young's modulus, $E$, Poisson's ratio, $\nu,$  bulk modulus, $B$, electronic band gap energy, $E_g$, components of the static dielectric tensor, $\epsilon^0_{ij}$, and the optical dielectric tensor, $\epsilon^\infty_{ij}$,  components of the Born effective charge tensor on the transition metal atom, $Z_{ii}$, Debye temperature, $\theta_D$, average sound velocity, v$_{avg}$, and the zero temperature phonon Free-energy, $\Delta F(0)$. The local environment of the tc-TMDs gives rise to two values of each component of the Born effective charge.}
\label{Tab:props}
\centering
\begin{tabular}{|l|d{3.3}  d{3.3} l |d{3.3}  d{3.3} l |d{3.3}  d{3.3} l | }\toprule
       &    \multicolumn{3}{c|}{MoS$_2$}  & \multicolumn{3}{c|}{ZrS$_2$}&  \multicolumn{3}{c|}{ReS$_2$}\\
&\multicolumn{1}{c}{Calc.}&\multicolumn{1}{c}{Exp.}& & \multicolumn{1}{c}{Calc.} & \multicolumn{1}{c}{Exp.}& &\multicolumn{1}{c}{Calc.} &\multicolumn{1}{c}{Exp.} & \\ \hline
a [$\AA$] & 3.162 &3.160 &[\citenum{Dickinson1923}]  &3.687  &3.660 &[\citenum{Greenaway1965}] &6.420  &6.362 & [\citenum{Wildervanck1970}] \\
b [$\AA$] &  &  &  &&&&6.587  &6.465 &[\citenum{Wildervanck1970}] \\
c [$\AA$]  &12.301  &12.290& [\citenum{Dickinson1923}] &5.812 & 5.850 &[\citenum{Greenaway1965}]   &6.999  &6.401 & [\citenum{Wildervanck1970}]\\
$E_g$ [eV]   &0.886   &1.23& [\citenum{Kam1982}]   &1.041 &1.68& [\citenum{Moustafa2009}] & 1.199 & 1.55& [\citenum{Tongay2014}]   \\ \hline
c$_{11}$ [GPa] &210.77 &238 &[\citenum{Feldman1976}]   & 122.00 & &&227.06 & & \\
c$_{22}$ [GPa] &  && &  &   &&249.80  & &\\
c$_{33}$ [GPa] &44.41  &52 &[\citenum{Feldman1976}] &33.58  & &  &27.02  & &\\
c$_{44}$ [GPa] & 17.15 &19 &[\citenum{Feldman1976}] &11.07  &  & &6.39  & &\\
c$_{55}$ [GPa] &  && &  &   &&  6.71& &\\
c$_{66}$ [GPa] & 81.17 &91 &[\citenum{Feldman1976}] &48.92  &&   &  82.81& &\\
E$_x$ [GPa] & 198.65& 240& [\citenum{Aksoy2006}] & 115.68 && &196.41  &&\\
E$_y$ [GPa] &  && &  &&&211.76  &&\\
E$_z$ [GPa] &43.84 & & &32.74  & &&25.54 & &\\
$\nu_{xy}$ &0.22 &0.27 &[\citenum{Aksoy2006}] &0.19  && &0.18 &  &\\
$\nu_{xz}$ & 0.03&& &0.05  && &0.04 &  &\\
$\nu_{yz}$ & & &&  &&&0.05 & & \\
$B$ [GPa]& 66.35 &53.40 &[\citenum{Nayak2014}]  &39.90  & & & 71.14 & & \\    \hline

$\epsilon_{xx}^0$ & 15.53 &15.4 &[\citenum{Wieting1971}] & 37.23&34.50 &[\citenum{Iwasaki1982}] &16.21  & &\\
$\epsilon_{yy}^0$ & & & &  & &&14.36  && \\
$\epsilon_{zz}^0$ &6.87  &6.28 &[\citenum{Wieting1971}] &6.15  &10.20 &[\citenum{Iwasaki1982}]&5.81  && \\
$\epsilon_{xy}^0$ &  & &&  && &-0.19  && \\
$\epsilon_{xz}^0$ &  & & & & &&0.28  & &\\
$\epsilon_{yz}^0$ &  && &  & &&  -0.34& &\\
$\epsilon_{xx}^{\infty}$ & 15.31 &15.2 &[\citenum{Wieting1971}] &9.94  &9.23 &[\citenum{Iwasaki1982}] &15.75  & &\\
$\epsilon_{yy}^{\infty}$ &  & &  & &&& 14.14 & &\\
$\epsilon_{zz}^{\infty}$ & 6.82 &6.25 &[\citenum{Wieting1971}] &5.54  &6.10 &[\citenum{Iwasaki1982}]&7.23  && \\
$\epsilon_{xy}^{\infty}$ &  & &  & &&& -0.05 & &\\
$\epsilon_{xz}^{\infty}$ &  & &  & &&&0.26  & &\\
$\epsilon_{yz}^{\infty}$ &  & &  & &&& -0.31 & &\\
$Z_{xx}^*$ [e] &  -1.09&\lvert1.1\rvert& [\citenum{Wieting1971}] &6.19  &\lvert4.4\rvert&[\citenum{Lucovsky1973}]&-1.53  && \\
 & && & && &-0.56  && \\
$Z_{yy}^*$ [e] & & & &&& &-0.79 && \\
 & & & &&& & 0.47 && \\
$Z_{zz}^*$ [e] &-0.63  &\lvert0.4\rvert &[\citenum{Wieting1971}] & 1.82&& &  -0.38&& \\
 &  && & & && -0.25&& \\\hline

$\theta_D$ [K] & 262.64& 260&[\citenum{Park1996}] & 290.87  & && 147.07 &115& [\citenum{Ho1997}] \\
v$_{avg}$ [km/s]  &4.181 & &&3.996  &&&3.000 & & \\ 
$\Delta F(0)$ &30.45 && &10.96 &&&48.09&&\\
  \botrule      
\end{tabular}
\end{table*}

\begin{figure*}[!t]
\includegraphics[width=1\textwidth]{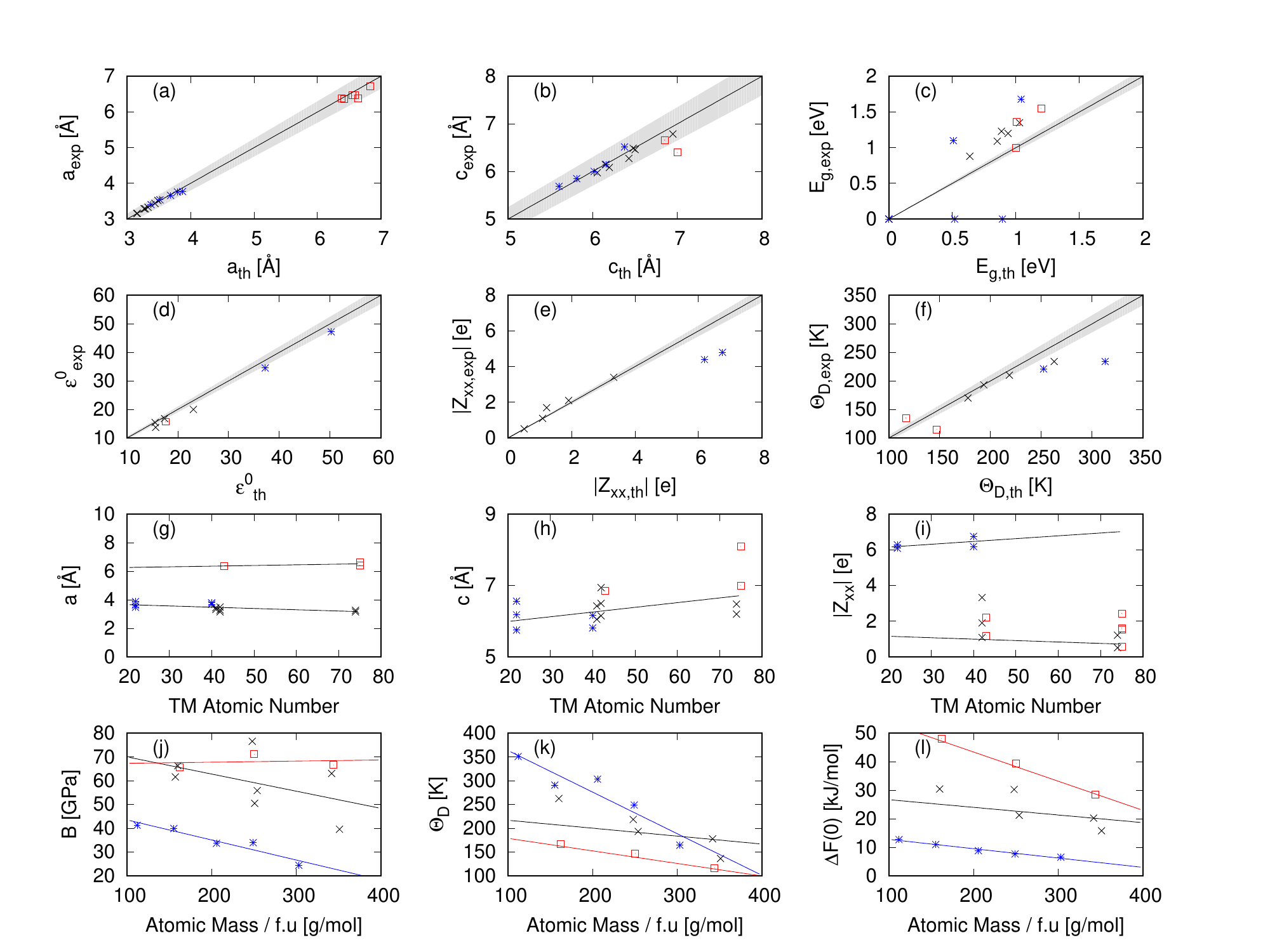} 
\caption{\label{trends}(Color Online) (a) - (f) A comparison between our DFT and experiments for some of the principal vibrational and dielectric properties. Symbols (colors) represent different symmetry classes: h-TMDs (black), t-TMDs (blue), and tc-TMDs (red). (a) in-plane lattice parameter, (b) out-of-plane lattice parameter, (c) indirect energy gap, (d) static dielectric constant, (e) absolute value of the in-plane Born effective charge on the transition metal atom, and (f) Debye temperature. Points correspond to cases where both calculated and experimental data was found.  The light gray wedge represents an error of $\pm$5\%. The overall agreement is very good, but the electronic gap is quite scattered, for reasons described in the text, whereas the dielectric constant is well reproduced. In (g) - (l) we show trends in our calculated data with respect to the atomic number of the metal ion or atomic mass per formula unit (f.u). Lines are linear fits to the available data and the plotted c lattice parameter for the hexagonal compounds in (b) and (h) are half of the unit cell height, for comparison. }
\end{figure*}

Having considered the effects of the normal approximations used in first-principles calculations, and shown which is the most important, we now outline our calculations of the physical properties of the TMDs. For all the calculations found below and in the SM~\cite{supmat}, we use the pseudopotentials mentioned previously, no spin-orbit interaction, and the DFT-D3 dispersion scheme. 

\section{Structural and electro-mechanical properties}\label{properties}

The physical properties of the system, namely the lattice constants, elastic properties and piezoelectric properties, are determined as part of our DFT and DFPT calculation.  In particular, our calculation of the relaxed lattice parameters, shown in Table~\ref{Tab:props} and the SM~\cite{supmat}, show good agreement with experimental measurement~\cite{Dickinson1923, Feldman1976, Aksoy2006, Nayak2014, Brixner1962, Zhao2015, Berkdemir2013, Sourisseau1991, Liu2014, Jellinek1960, Selvi2006, Zhang2016,  Leroux2012,  Greenaway1965,  Scharli1986,  Wildervanck1970,Hart2017}. Indeed, we find that as the atomic number of the transition metal increases for each family of compounds the out-of-plane lattice constants increase in a near linear fashion. A comparison of our calculated lattice parameters with respect to experimental measurement is given in Fig.~\ref{trends}(a) and (b) for the in-plane and out-of-plane lattice parameters with the relationship to the transition metal (TM) atomic number given in Fig.~\ref{trends}(g) and (h). The result of this analysis shows that, for the h-TMDs, our calculated lattice parameters lie within 0.7\% and 2.0\% for the in-plane and out-of-plane lattice parameters respectively. In the case of the t-TMDs, our calculations are within 2.2\% for both sets of lattice parameters. Similarly, for the tc-TMDs, our calculations lie within 2.0\% for the in-plane and out-of-plane lattice parameters (except for the c lattice parameter of ReSe$_2$). The rescaling of the c lattice parameter due to the dispersion corrections, as shown in Table~S1 of the SM~\cite{supmat}, shows clearly why the dispersion corrections are important for accurate calculations of layered materials.  

Our DFPT calculations allow us to determine several elastic properties which are easily compared to experiment.  Since the elastic tensor displays the underlying symmetry of the lattice there are five unique elastic constants for the h and t compounds and 21 unique elastic constants for the tc compounds. Our data tables provide the principal components of the elastic tensor for each of the compounds and the derived elastic properties, such as the Bulk modulus, whose relationship to the elastic tensor is outlined in the SM~\cite{supmat}. These materials are well-known for their use as lubricants, and we find that the elastic coefficient responsible for the sliding between layers, $c_{44}$, is small in these materials, comparable to graphite ($c_{44}$ = 4.25 GPa~\cite{Nihira2003}) and that the magnitude of the bulk modulus, $B$, generally decreases as the mass of the formula unit increases, as shown in Fig.~\ref{trends}(j). Likewise, we find a general overestimation of the bulk modulus when compared to experimental values, contrary to early work by Filippi {\it et. al.}~\cite{Filippi1994} who found a general underestimation.  For the symmetry classes considered here the piezoelectric tensor components are identically zero due to the presence of mirror plane symmetry~\cite{Nye1957, Tromans2011}.
 
\section{Electrical, dielectric, and optical properties}

The electrical, dielectric, and optical properties of the system are central to the most promising applications of the TMDs. Using DFT and DFPT, for the semi-conducting systems, we calculate the band gap energy, dielectric tensors, Born effective charge tensor, nonlinear optical tensor, and Raman susceptibility. For the symmetry classes considered here the nonlinear optical tensor components are identically zero, due to the presence of mirror plane symmetry~\cite{Nye1957, Tromans2011}.  These calculated properties are shown in Table~\ref{Tab:props} and in the SM~\cite{supmat} where we also provide the indirect band gap energy and compare to experimental data~\cite{Kam1982, Wieting1971, Agnihotri1972, Garg1973, Lezama2015, Beal1979, Liang1973, Luttrell2006, Beal1979, Uchida1978, Lucovsky1973, Starnberg1995, Greenaway1965, Iwasaki1982,  Tongay2014, Ho1997, Zhao2015a}.  

\begin{figure}[!t]
\centering
\includegraphics[width=0.45\textwidth]{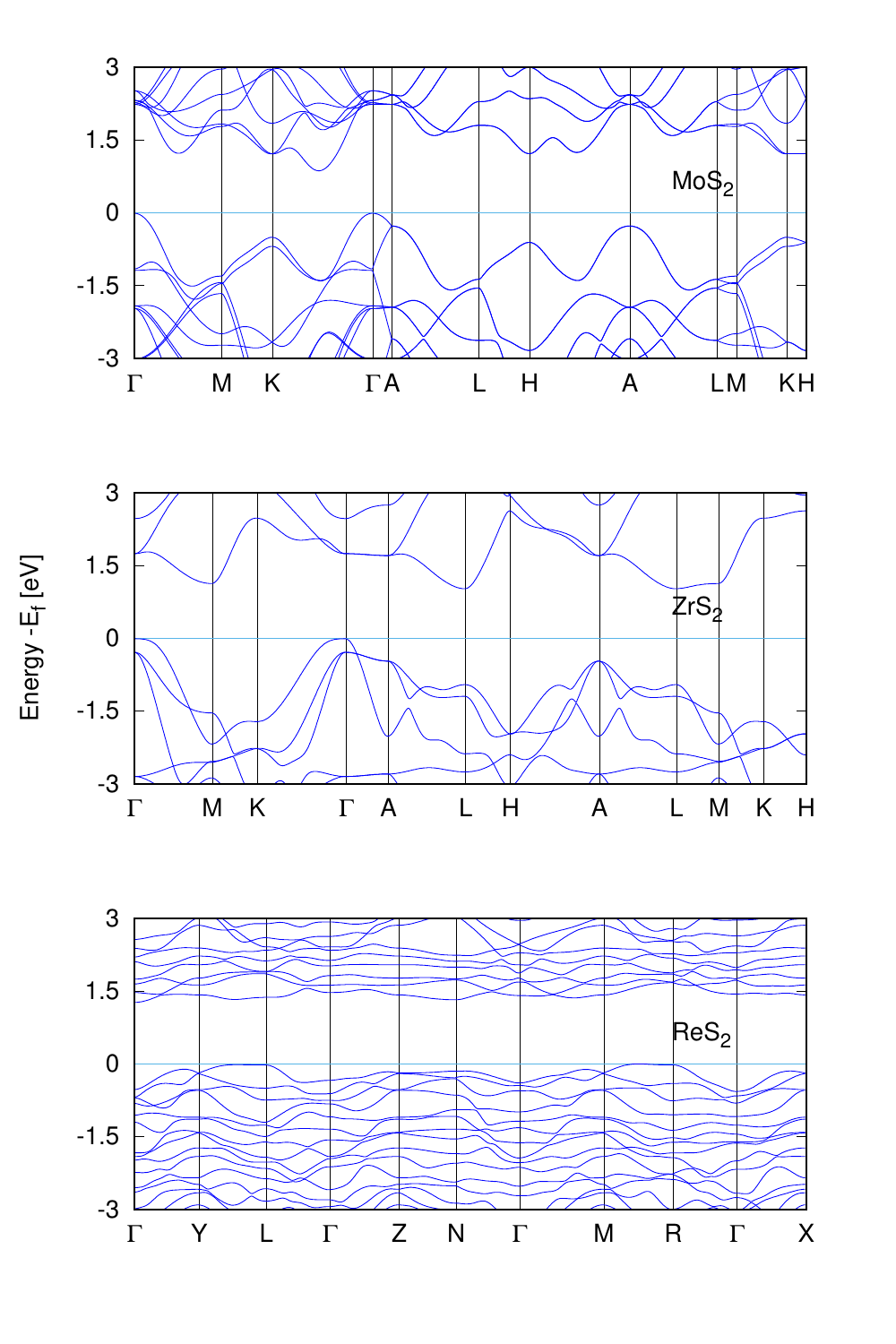} 
\vspace{-1cm}
\caption{\label{electronic_bands}(Color online) Kohn-Sham electronic band structures of the three example compounds plotted along a high symmetry path in k-space~\cite{Setyawan2010}. In each case, the calculated Fermi energy is shifted to the zero of energy as indicated by the light blue line. }
\end{figure}

Our calculation of the indirect energy gap comes from the Kohn-Sham electronic band structure of these materials as shown in Fig.~\ref{electronic_bands} for our three example compounds.  Here, the band structures are plotted along a path in k-space determined by the underlying symmetry of the system. In each case, the zero of energy is set to the Fermi level or to the top of the valence band.  With the exceptions of the metallic compounds~\cite{Leroux2012, Greenaway1965}, these materials all have an indirect band gap originating at $\Gamma$ and ending along the path from M to K. Most of the calculated gap energies are smaller than their experimentally measured counterparts (see Fig.~\ref{trends}(c)). This is due to a fortuitous cancellation of errors between the usual DFT Kohn-Sham noticeable underestimation of the quasiparticle band gap and the large exciton binding energy (the difference between the quasiparticle and optical band gap). Indeed, due to long-range electrostatics in these layered materials, the exciton binding energy, which determines the optical gap, is very large and compensates the DFT underestimation of the real quasiparticle gap~\cite{Saigal2016, Hill2015}. This effect is even stronger in monolayers~\cite{Rigosi2016,Handbicki2015,Kylanpaa2015}. Compared to calculations without any treatment of the long-range e$^-$-e$^-$ correlation, our calculations with GGA and vdW dispersion corrections are more consistent with experiment. Given the limitations of the calculated band structure, our calculations of the zero-frequency dielectric constant compare well to experiment as shown in Fig.~\ref{trends}(d), with much less spread in the data meaning that the static dielectric constant is well represented due, in part, to the fortuitous cancellation of errors.

The Born effective charges, which give the dynamic response of the system to a displacement and electric field perturbation of the system, display signs which are counterintuitive, i.e. the transition-metal takes the negative charge, for the h and tc compounds.  A discussion of the Born effective charge tensor and the origin of the counterintuitive charge is found in Ref.~\citenum{Pike2016}. To compare to the measured value of the Born effective charge requires care, as only the absolute value of the Born effective charge is easily accessible experimentally.  Therefore, experimental values of the effective charge are given in terms of an absolute value and are compared to our calculations in Fig.~\ref{trends}(e) in terms of magnitude only. Fig.~\ref{trends}(i) shows that the magnitude of the Born effective charge depends more on symmetry than on the atomic mass of the transition metal atom, with an increase in the Born effective charge observed for the t-TMDs as a function of atomic number and a general decrease observed for the h- and tc-TMDs. The differences in the local environments of the transition metal atoms in the tc-TMDs give rise to different calculated Born effective charge tensors.  Therefore, two values of the Born effective charge tensor are given for these compounds.

\section{Phonon band structures, the density of states, and Raman spectra}\label{bandstructures}

The thermal and acoustic properties of our compounds come from a calculation of the phonon dispersions. From this calculation, the Helmholtz free energy, $\Delta F$, entropy $S$, and constant-volume specific heat, $C_v$, are determined as a function of temperature using the phonon density of states, $g(\omega)$, as laid out in Ref.~\citenum{Lee1995}. A plot of these quantities is shown in Fig.~S9 of the SM~\cite{supmat}.  The Helmholtz free energy at T= 0K is given in the data tables and shown in Fig.~\ref{trends}(l), which shows a decrease as the atomic mass of the formula unit increases. Our values of the Debye temperature and average sound velocity come from summing up the Debye temperature and average sound velocity at each q point (in reduced coordinates) such that $|q|<0.25$ then dividing by the number of points within that radius, which is determined by the density of our q-point mesh ($30\times30\times30$). This Debye temperature agrees reasonably well with experiment as shown in Fig.~\ref{trends}(f)~\cite{Park1996, Liu2015, Wender1973, Ho1997, Titov2007}.  Similar to the free energy, the Debye temperature decreases with the increasing atomic mass of the formula unit.

\begin{figure*}[!t]
\centering
\includegraphics[width=0.95\textwidth]{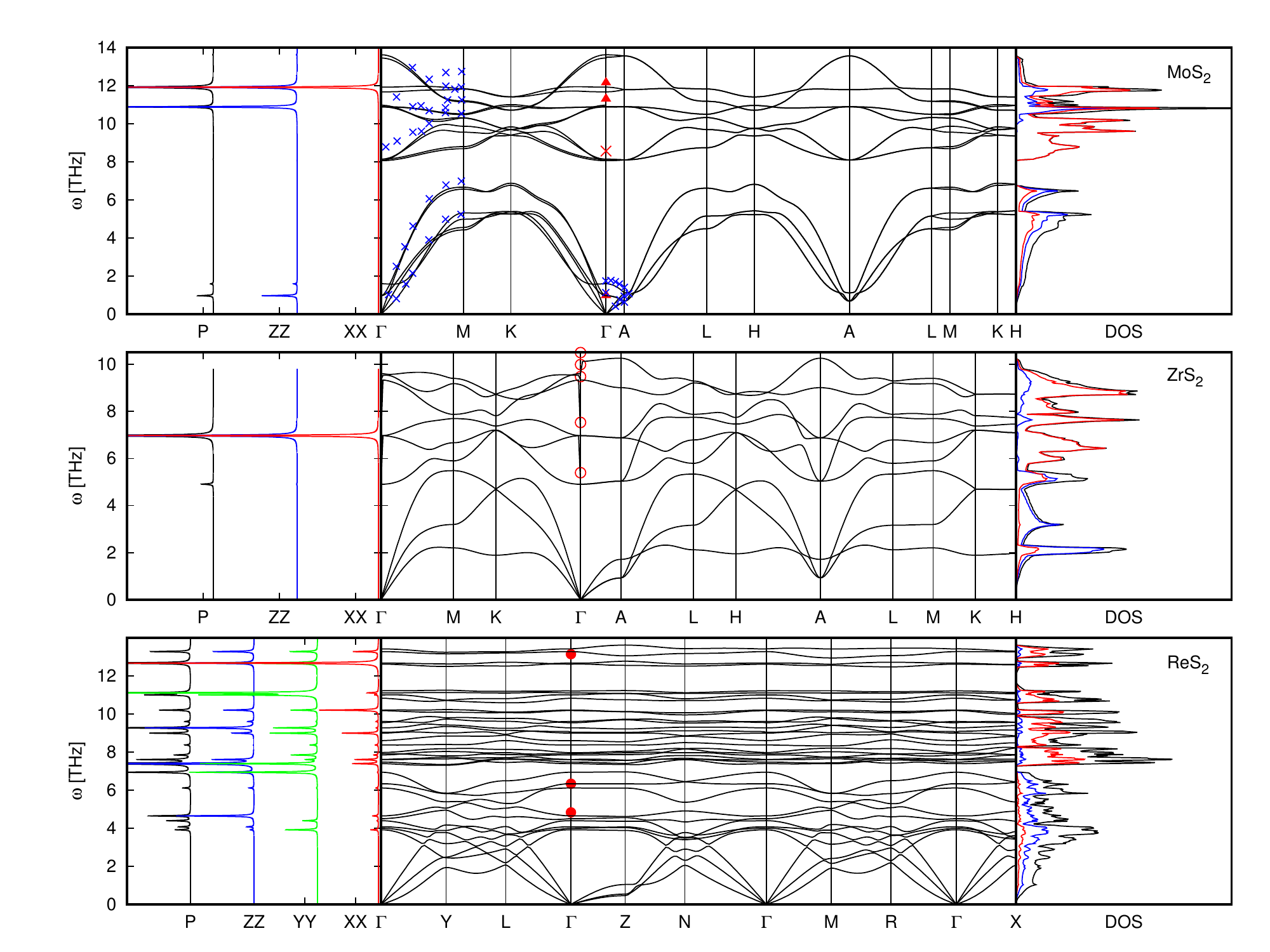} 
\caption{\label{phonon_bands}(Color Online) Raman spectra, phonon band structure, and phonon density of states for our three example compounds plotted along a high symmetry path in q-space. Raman spectra are plotted for non-symmetric polarizations, experimental neutron scattering data from Refs.~\citenum{Wakabayashi1975} and experimental Raman and Infrared data at $\Gamma$ from Refs.~\citenum{Zhao2013,Tonndorf2013,Wolverson2016,Roubi1988}. }
\end{figure*}

In addition to the electronic band structure, our calculations of the phonon band structures, phonon density of states, and Raman spectra allow us to determine the quality of our calculations by comparing to neutron scattering and Raman and Infrared experimental measurements.  In Fig.~\ref{phonon_bands} and the figures in the SM~\cite{supmat}, the calculated phonon band structures, atom-projected phonon density of states, and Raman spectra for the three example compounds are shown and compare to experimental data for the Raman active and Infrared active modes at $\Gamma$ and neutron scattering data at finite \textbf{q}. In particular, neutron scattering data~\cite{Wakabayashi1975, Sourisseau1989, Sourisseau1991, Scharli1986} and Raman and Infrared measurements~\cite{Zhao2013b, Tonndorf2013, Tongay2012, Sekine1980, Zhao2013, Mead1977, Froehlicher2015, Sandoval1992, Sugai1980, Hangyo1983, Roubi1988, Manas2016, Wolverson2016} are used to compare to our calculations.  

In Fig.~\ref{phonon_bands}, and all the Raman spectra in the SM~\cite{supmat} are shown with the atom-projected phonon density of states for the transition metal atoms (in blue) and chalcogen atoms (in red). Our calculations of the phonon band structure, density of states, and Raman spectra for the Ti-based compounds show indications of charge density waves or Kohn anomalies~\cite{Duong2015, Fang1997, Liu2012, Sharma1999, Sugai1980} and, in agreement with recent experimental work, we find no charge density wave in bulk TiTe$_2$~\cite{Chen2017}.  The phonon band structures of the Nb-based compounds, also shown in the SM~\cite{supmat}, reveal several unstable modes due to the metallic nature of the compounds and the limitations of the harmonic approximation when determining the phonon band structure in these materials~\cite{Heil2017}.  

We give the numerical values of our calculated Raman and Infrared active modes in Table S2 of the SM~\cite{supmat} for the h-TMDs to show that our calculations for these modes are within 3.0\% of the experimental measurements. Here the identified phonon modes are in accordance with the Bilbao crystallographic server~\cite{Aroyo2011, Aroyo2006, Aroyo2006a}.  As an additional check of the effect of the dispersion corrections on the phonon frequencies, we show in Table S2 of the SM~\cite{supmat} the Raman and Infrared active modes using both DFT-D3 and DFT-D3(BJ) for the dispersion corrections. These two methods differ in how they treat interactions at short distances, i.e. as $R\rightarrow0$~\cite{Grimme2010, Grimme2011}. From this calculation, we find that using DFT-D3 results in a better agreement with experimental measurements for the low energy phonon branches.

\section{Conclusions}\label{conclusions}
The various approximations used in DFT and DFPT, namely the pseudopotential, spin-orbit, and van der Waals interaction, are often used when investigating the properties of layered materials.  These approximations are important to consider when undertaking calculations of layered materials as they can have a large impact on the structural and vibrational properties of the system.  Here, it was found that the physical properties of the system depend strongly on the van der Waals approximation and to a lesser extent on the spin-orbit approximation.  Therefore, our calculations included the van der Waals interaction in our DFT and DFPT calculations which gives rise to good agreement between our values of the vibrational and dielectric properties when compared to experiment as shown in Fig.~\ref{trends}. Thus our predicted values for the unmeasured quantities are believed to be accurate and useful for future comparison to experimental measurements.

Surprisingly, despite the well-known shortcomings of density-functional theory in the GGA approximation, yielding an inaccurate calculation of the electronic band gap energy, we are still able to accurately calculate the dielectric properties of this material.  Additionally, comparisons with and without the dispersion corrections in the DFPT calculation led to the conclusion that the corresponding additional contributions during the DFPT calculation are critical to the accuracy of the calculated vibrational properties.

In conclusion, our careful testing of the various approximations normally used in DFT and DFPT calculations indicate that care must be taken when calculating the vibrational and dielectric properties of the transition metal dichalcogenides.  Our calculations of the mechanical, electrical, optical, thermal and acoustic properties of a subset of the transition metal dichalcogenides used the \textit{ABINIT} software package, the van der Waals scheme known as DFT-D3, and norm-conserving pseudopotentials.  Due to the sensitivity of the vibrational properties on the lattice parameters and interatomic force constants we use dispersion corrections during our DFT and DFPT calculations for all of the transition metal dichalcogenides and provide experimentally relevant results within our data tables for many unmeasured quantities. It is our hope that this data is used as part of a much larger effort to engineer heterostructures which combine the unique properties of each individual material. 

\section*{Acknowledgments}
The authors gratefully acknowledge funding from the Belgian Fonds National de la Recherche Scientifique FNRS under grant numbers PDR T.1077.15-1/7 (N.A.P and M.J.V) and for an FRIA Grant (B.V.T.). N.A.P would also like to thank the Research Council of Norway through the Frinatek program for funding.  M.J.V and A.D. acknowledge support from ULg and from the Communaut\'{e} Fran\c{c}aise de Belgique (ARC AIMED 15/19-09). Computational resources have been provided by the Universit\'{e} Catholique de Louvain (CISM/UCL); the Consortium des Equipements de Calcul Intensif en F\'{e}d\'{e}ration Wallonie Bruxelles (CECI), funded by FRS-FNRS G.A. 2.5020.11; the Tier-1 supercomputer of the F\'{e}d\'{e}ration Wallonie-Bruxelles, funded by the Walloon Region under G.A. 1117545; and by PRACE-3IP DECI grants, on ARCHER and Salomon (ThermoSpin, ACEID, OPTOGEN, and INTERPHON 3IP G.A. FP7 RI-312763 and 13 G.A. 653838 of H2020).

\bibliography{tmd_source}

\begin{thebibliography}{142}%
\makeatletter
\providecommand \@ifxundefined [1]{%
 \@ifx{#1\undefined}
}%
\providecommand \@ifnum [1]{%
 \ifnum #1\expandafter \@firstoftwo
 \else \expandafter \@secondoftwo
 \fi
}%
\providecommand \@ifx [1]{%
 \ifx #1\expandafter \@firstoftwo
 \else \expandafter \@secondoftwo
 \fi
}%
\providecommand \natexlab [1]{#1}%
\providecommand \enquote  [1]{``#1''}%
\providecommand \bibnamefont  [1]{#1}%
\providecommand \bibfnamefont [1]{#1}%
\providecommand \citenamefont [1]{#1}%
\providecommand \href@noop [0]{\@secondoftwo}%
\providecommand \href [0]{\begingroup \@sanitize@url \@href}%
\providecommand \@href[1]{\@@startlink{#1}\@@href}%
\providecommand \@@href[1]{\endgroup#1\@@endlink}%
\providecommand \@sanitize@url [0]{\catcode `\\12\catcode `\$12\catcode
  `\&12\catcode `\#12\catcode `\^12\catcode `\_12\catcode `\%12\relax}%
\providecommand \@@startlink[1]{}%
\providecommand \@@endlink[0]{}%
\providecommand \url  [0]{\begingroup\@sanitize@url \@url }%
\providecommand \@url [1]{\endgroup\@href {#1}{\urlprefix }}%
\providecommand \urlprefix  [0]{URL }%
\providecommand \Eprint [0]{\href }%
\providecommand \doibase [0]{http://dx.doi.org/}%
\providecommand \selectlanguage [0]{\@gobble}%
\providecommand \bibinfo  [0]{\@secondoftwo}%
\providecommand \bibfield  [0]{\@secondoftwo}%
\providecommand \translation [1]{[#1]}%
\providecommand \BibitemOpen [0]{}%
\providecommand \bibitemStop [0]{}%
\providecommand \bibitemNoStop [0]{.\EOS\space}%
\providecommand \EOS [0]{\spacefactor3000\relax}%
\providecommand \BibitemShut  [1]{\csname bibitem#1\endcsname}%
\let\auto@bib@innerbib\@empty
%</preamble>
\bibitem [{\citenamefont {Chhowalla}\ \emph {et~al.}(2013)\citenamefont
  {Chhowalla}, \citenamefont {Shin}, \citenamefont {Eda}, \citenamefont {Li},
  \citenamefont {Loh},\ and\ \citenamefont {Zhang}}]{Chhowalla2013}%
  \BibitemOpen
  \bibfield  {author} {\bibinfo {author} {\bibfnamefont {M.}~\bibnamefont
  {Chhowalla}}, \bibinfo {author} {\bibfnamefont {H.~S.}\ \bibnamefont {Shin}},
  \bibinfo {author} {\bibfnamefont {G.}~\bibnamefont {Eda}}, \bibinfo {author}
  {\bibfnamefont {L.-J.}\ \bibnamefont {Li}}, \bibinfo {author} {\bibfnamefont
  {K.~P.}\ \bibnamefont {Loh}}, \ and\ \bibinfo {author} {\bibfnamefont
  {H.}~\bibnamefont {Zhang}},\ }\href@noop {} {\bibfield  {journal} {\bibinfo
  {journal} {Nature Chemistry}\ }\textbf {\bibinfo {volume} {5}},\ \bibinfo
  {pages} {263} (\bibinfo {year} {2013})}\BibitemShut {NoStop}%
\bibitem [{\citenamefont {Wang}\ \emph {et~al.}(2012)\citenamefont {Wang},
  \citenamefont {Kalantar-Zadeh}, \citenamefont {Kis}, \citenamefont
  {Coleman},\ and\ \citenamefont {Strano}}]{Wang2012}%
  \BibitemOpen
  \bibfield  {author} {\bibinfo {author} {\bibfnamefont {Q.~H.}\ \bibnamefont
  {Wang}}, \bibinfo {author} {\bibfnamefont {K.}~\bibnamefont
  {Kalantar-Zadeh}}, \bibinfo {author} {\bibfnamefont {A.}~\bibnamefont {Kis}},
  \bibinfo {author} {\bibfnamefont {J.~N.}\ \bibnamefont {Coleman}}, \ and\
  \bibinfo {author} {\bibfnamefont {M.~S.}\ \bibnamefont {Strano}},\
  }\href@noop {} {\bibfield  {journal} {\bibinfo  {journal} {Nature
  Nanotechnology}\ }\textbf {\bibinfo {volume} {7}},\ \bibinfo {pages} {699}
  (\bibinfo {year} {2012})}\BibitemShut {NoStop}%
\bibitem [{\citenamefont {Jariwala}\ \emph {et~al.}(2014)\citenamefont
  {Jariwala}, \citenamefont {Sangwan}, \citenamefont {Lauhon}, \citenamefont
  {Marks},\ and\ \citenamefont {Hersam}}]{Jariwala2014}%
  \BibitemOpen
  \bibfield  {author} {\bibinfo {author} {\bibfnamefont {D.}~\bibnamefont
  {Jariwala}}, \bibinfo {author} {\bibfnamefont {V.~K.}\ \bibnamefont
  {Sangwan}}, \bibinfo {author} {\bibfnamefont {L.~J.}\ \bibnamefont {Lauhon}},
  \bibinfo {author} {\bibfnamefont {T.~J.}\ \bibnamefont {Marks}}, \ and\
  \bibinfo {author} {\bibfnamefont {M.~C.}\ \bibnamefont {Hersam}},\
  }\href@noop {} {\bibfield  {journal} {\bibinfo  {journal} {ACS Nano.}\
  }\textbf {\bibinfo {volume} {8}},\ \bibinfo {pages} {1102} (\bibinfo {year}
  {2014})}\BibitemShut {NoStop}%
\bibitem [{\citenamefont {Rasmussen}\ and\ \citenamefont
  {Thygesen}(2015)}]{Rasmussen2015}%
  \BibitemOpen
  \bibfield  {author} {\bibinfo {author} {\bibfnamefont {F.~A.}\ \bibnamefont
  {Rasmussen}}\ and\ \bibinfo {author} {\bibfnamefont {K.~S.}\ \bibnamefont
  {Thygesen}},\ }\href@noop {} {\bibfield  {journal} {\bibinfo  {journal} {J.
  Phys. Chem. C.}\ }\textbf {\bibinfo {volume} {119}},\ \bibinfo {pages}
  {13169} (\bibinfo {year} {2015})}\BibitemShut {NoStop}%
\bibitem [{\citenamefont {Gupta}\ \emph {et~al.}(2015)\citenamefont {Gupta},
  \citenamefont {Sakthivel},\ and\ \citenamefont {Seal}}]{Gupta2015}%
  \BibitemOpen
  \bibfield  {author} {\bibinfo {author} {\bibfnamefont {A.}~\bibnamefont
  {Gupta}}, \bibinfo {author} {\bibfnamefont {T.}~\bibnamefont {Sakthivel}}, \
  and\ \bibinfo {author} {\bibfnamefont {S.}~\bibnamefont {Seal}},\ }\href@noop
  {} {\bibfield  {journal} {\bibinfo  {journal} {Prog. in Mat. Sci.}\ }\textbf
  {\bibinfo {volume} {73}},\ \bibinfo {pages} {44} (\bibinfo {year}
  {2015})}\BibitemShut {NoStop}%
\bibitem [{\citenamefont {Bhimanapati}\ \emph {et~al.}(2015)\citenamefont
  {Bhimanapati}, \citenamefont {Lin}, \citenamefont {Meunier}, \citenamefont
  {Jung}, \citenamefont {Cha}, \citenamefont {Das}, \citenamefont {Xiao},
  \citenamefont {Son}, \citenamefont {Strano}, \citenamefont {Cooper},
  \citenamefont {Liang}, \citenamefont {Louie}, \citenamefont {Ringe},
  \citenamefont {Zhou}, \citenamefont {Kim}, \citenamefont {Naik},
  \citenamefont {Sumpter}, \citenamefont {Terrones}, \citenamefont {Xia},
  \citenamefont {Wang}, \citenamefont {Zhu}, \citenamefont {Akinwande},
  \citenamefont {Alem}, \citenamefont {Schuller}, \citenamefont {Schaak},
  \citenamefont {Terrones},\ and\ \citenamefont {Robinson}}]{Bhimanapati2015}%
  \BibitemOpen
  \bibfield  {author} {\bibinfo {author} {\bibfnamefont {G.~R.}\ \bibnamefont
  {Bhimanapati}}, \bibinfo {author} {\bibfnamefont {Z.}~\bibnamefont {Lin}},
  \bibinfo {author} {\bibfnamefont {V.}~\bibnamefont {Meunier}}, \bibinfo
  {author} {\bibfnamefont {Y.}~\bibnamefont {Jung}}, \bibinfo {author}
  {\bibfnamefont {J.}~\bibnamefont {Cha}}, \bibinfo {author} {\bibfnamefont
  {S.}~\bibnamefont {Das}}, \bibinfo {author} {\bibfnamefont {D.}~\bibnamefont
  {Xiao}}, \bibinfo {author} {\bibfnamefont {Y.}~\bibnamefont {Son}}, \bibinfo
  {author} {\bibfnamefont {M.~S.}\ \bibnamefont {Strano}}, \bibinfo {author}
  {\bibfnamefont {V.~R.}\ \bibnamefont {Cooper}}, \bibinfo {author}
  {\bibfnamefont {L.}~\bibnamefont {Liang}}, \bibinfo {author} {\bibfnamefont
  {S.~G.}\ \bibnamefont {Louie}}, \bibinfo {author} {\bibfnamefont
  {E.}~\bibnamefont {Ringe}}, \bibinfo {author} {\bibfnamefont
  {W.}~\bibnamefont {Zhou}}, \bibinfo {author} {\bibfnamefont {S.~S.}\
  \bibnamefont {Kim}}, \bibinfo {author} {\bibfnamefont {R.~R.}\ \bibnamefont
  {Naik}}, \bibinfo {author} {\bibfnamefont {B.~G.}\ \bibnamefont {Sumpter}},
  \bibinfo {author} {\bibfnamefont {H.}~\bibnamefont {Terrones}}, \bibinfo
  {author} {\bibfnamefont {F.}~\bibnamefont {Xia}}, \bibinfo {author}
  {\bibfnamefont {Y.}~\bibnamefont {Wang}}, \bibinfo {author} {\bibfnamefont
  {J.}~\bibnamefont {Zhu}}, \bibinfo {author} {\bibfnamefont {D.}~\bibnamefont
  {Akinwande}}, \bibinfo {author} {\bibfnamefont {N.}~\bibnamefont {Alem}},
  \bibinfo {author} {\bibfnamefont {J.~A.}\ \bibnamefont {Schuller}}, \bibinfo
  {author} {\bibfnamefont {R.~E.}\ \bibnamefont {Schaak}}, \bibinfo {author}
  {\bibfnamefont {M.}~\bibnamefont {Terrones}}, \ and\ \bibinfo {author}
  {\bibfnamefont {J.~A.}\ \bibnamefont {Robinson}},\ }\href@noop {} {\bibfield
  {journal} {\bibinfo  {journal} {ACS Nano}\ }\textbf {\bibinfo {volume} {9}},\
  \bibinfo {pages} {11509} (\bibinfo {year} {2015})}\BibitemShut {NoStop}%
\bibitem [{\citenamefont {Yun}\ \emph {et~al.}(2012)\citenamefont {Yun},
  \citenamefont {Han}, \citenamefont {Hong}, \citenamefont {Kim},\ and\
  \citenamefont {Lee}}]{Yun2012}%
  \BibitemOpen
  \bibfield  {author} {\bibinfo {author} {\bibfnamefont {W.~S.}\ \bibnamefont
  {Yun}}, \bibinfo {author} {\bibfnamefont {S.~W.}\ \bibnamefont {Han}},
  \bibinfo {author} {\bibfnamefont {S.~C.}\ \bibnamefont {Hong}}, \bibinfo
  {author} {\bibfnamefont {I.~G.}\ \bibnamefont {Kim}}, \ and\ \bibinfo
  {author} {\bibfnamefont {J.~D.}\ \bibnamefont {Lee}},\ }\href@noop {}
  {\bibfield  {journal} {\bibinfo  {journal} {Phys. Rev. B}\ }\textbf {\bibinfo
  {volume} {85}},\ \bibinfo {pages} {03305} (\bibinfo {year}
  {2012})}\BibitemShut {NoStop}%
\bibitem [{\citenamefont {Ataca}\ \emph {et~al.}(2012)\citenamefont {Ataca},
  \citenamefont {Sahin},\ and\ \citenamefont {Ciraci}}]{Ataca2012}%
  \BibitemOpen
  \bibfield  {author} {\bibinfo {author} {\bibfnamefont {C.}~\bibnamefont
  {Ataca}}, \bibinfo {author} {\bibfnamefont {H.}~\bibnamefont {Sahin}}, \ and\
  \bibinfo {author} {\bibfnamefont {S.}~\bibnamefont {Ciraci}},\ }\href@noop {}
  {\bibfield  {journal} {\bibinfo  {journal} {J. Phys. Chem. C}\ }\textbf
  {\bibinfo {volume} {116}},\ \bibinfo {pages} {8983} (\bibinfo {year}
  {2012})}\BibitemShut {NoStop}%
\bibitem [{\citenamefont {Kuc}(2015)}]{Kuc2015}%
  \BibitemOpen
  \bibfield  {author} {\bibinfo {author} {\bibfnamefont {A.}~\bibnamefont
  {Kuc}},\ }in\ \href@noop {} {\emph {\bibinfo {booktitle} {Chemical Modelling:
  Volume 11}}}\ (\bibinfo  {publisher} {The Royal Society of Chemistry},\
  \bibinfo {year} {2015})\BibitemShut {NoStop}%
\bibitem [{\citenamefont {Radisavjevic}\ \emph {et~al.}(2011)\citenamefont
  {Radisavjevic}, \citenamefont {Radenovic}, \citenamefont {Brivio},
  \citenamefont {Giacometti},\ and\ \citenamefont {Kis}}]{Radisavjevic2011}%
  \BibitemOpen
  \bibfield  {author} {\bibinfo {author} {\bibfnamefont {B.}~\bibnamefont
  {Radisavjevic}}, \bibinfo {author} {\bibfnamefont {A.}~\bibnamefont
  {Radenovic}}, \bibinfo {author} {\bibfnamefont {J.}~\bibnamefont {Brivio}},
  \bibinfo {author} {\bibfnamefont {V.}~\bibnamefont {Giacometti}}, \ and\
  \bibinfo {author} {\bibfnamefont {A.}~\bibnamefont {Kis}},\ }\href@noop {}
  {\bibfield  {journal} {\bibinfo  {journal} {Nature Nanotechnology}\ }\textbf
  {\bibinfo {volume} {6}},\ \bibinfo {pages} {147} (\bibinfo {year}
  {2011})}\BibitemShut {NoStop}%
\bibitem [{\citenamefont {Ganatra}\ and\ \citenamefont
  {Zhang}(2014)}]{Ganatra2014}%
  \BibitemOpen
  \bibfield  {author} {\bibinfo {author} {\bibfnamefont {R.}~\bibnamefont
  {Ganatra}}\ and\ \bibinfo {author} {\bibfnamefont {Q.}~\bibnamefont
  {Zhang}},\ }\href@noop {} {\bibfield  {journal} {\bibinfo  {journal} {ACS
  nano.}\ }\textbf {\bibinfo {volume} {8}},\ \bibinfo {pages} {4074} (\bibinfo
  {year} {2014})}\BibitemShut {NoStop}%
\bibitem [{\citenamefont {Bakel}\ and\ \citenamefont
  {Hosson}(1992)}]{VanBakel1992}%
  \BibitemOpen
  \bibfield  {author} {\bibinfo {author} {\bibfnamefont {G.~P. E. M.~V.}\
  \bibnamefont {Bakel}}\ and\ \bibinfo {author} {\bibfnamefont {J.~T. M.~D.}\
  \bibnamefont {Hosson}},\ }\href@noop {} {\bibfield  {journal} {\bibinfo
  {journal} {Phys. Rev. B}\ }\textbf {\bibinfo {volume} {46}},\ \bibinfo
  {pages} {2001} (\bibinfo {year} {1992})}\BibitemShut {NoStop}%
\bibitem [{\citenamefont {Johannes}\ \emph {et~al.}(2006)\citenamefont
  {Johannes}, \citenamefont {Mazin},\ and\ \citenamefont
  {Howells}}]{Johannes2006}%
  \BibitemOpen
  \bibfield  {author} {\bibinfo {author} {\bibfnamefont {M.~D.}\ \bibnamefont
  {Johannes}}, \bibinfo {author} {\bibfnamefont {I.~I.}\ \bibnamefont {Mazin}},
  \ and\ \bibinfo {author} {\bibfnamefont {C.~A.}\ \bibnamefont {Howells}},\
  }\href@noop {} {\bibfield  {journal} {\bibinfo  {journal} {Phys. Rev. B}\
  }\textbf {\bibinfo {volume} {73}},\ \bibinfo {pages} {205102} (\bibinfo
  {year} {2006})}\BibitemShut {NoStop}%
\bibitem [{\citenamefont {Calandra}\ \emph {et~al.}(2009)\citenamefont
  {Calandra}, \citenamefont {Mazin},\ and\ \citenamefont
  {Mauri}}]{Calandra2009}%
  \BibitemOpen
  \bibfield  {author} {\bibinfo {author} {\bibfnamefont {M.}~\bibnamefont
  {Calandra}}, \bibinfo {author} {\bibfnamefont {I.~I.}\ \bibnamefont {Mazin}},
  \ and\ \bibinfo {author} {\bibfnamefont {F.}~\bibnamefont {Mauri}},\
  }\href@noop {} {\bibfield  {journal} {\bibinfo  {journal} {Phys. Rev. B}\
  }\textbf {\bibinfo {volume} {80}},\ \bibinfo {pages} {241108R} (\bibinfo
  {year} {2009})}\BibitemShut {NoStop}%
\bibitem [{\citenamefont {Zhu}\ \emph {et~al.}(2015)\citenamefont {Zhu},
  \citenamefont {Cao}, \citenamefont {Zhang}, \citenamefont {Plummer},\ and\
  \citenamefont {Guo}}]{Zhu2015}%
  \BibitemOpen
  \bibfield  {author} {\bibinfo {author} {\bibfnamefont {X.}~\bibnamefont
  {Zhu}}, \bibinfo {author} {\bibfnamefont {Y.}~\bibnamefont {Cao}}, \bibinfo
  {author} {\bibfnamefont {J.}~\bibnamefont {Zhang}}, \bibinfo {author}
  {\bibfnamefont {E.~W.}\ \bibnamefont {Plummer}}, \ and\ \bibinfo {author}
  {\bibfnamefont {J.}~\bibnamefont {Guo}},\ }\href@noop {} {\bibfield
  {journal} {\bibinfo  {journal} {PNAS}\ }\textbf {\bibinfo {volume} {112}},\
  \bibinfo {pages} {2367} (\bibinfo {year} {2015})}\BibitemShut {NoStop}%
\bibitem [{\citenamefont {Leroux}\ \emph {et~al.}(2012)\citenamefont {Leroux},
  \citenamefont {Tacon}, \citenamefont {Calandra}, \citenamefont {Measson},
  \citenamefont {Diener}, \citenamefont {Borrissenko}, \citenamefont {Bosak},\
  and\ \citenamefont {Rodiere}}]{Leroux2012}%
  \BibitemOpen
  \bibfield  {author} {\bibinfo {author} {\bibfnamefont {M.}~\bibnamefont
  {Leroux}}, \bibinfo {author} {\bibfnamefont {M.~L.}\ \bibnamefont {Tacon}},
  \bibinfo {author} {\bibfnamefont {M.}~\bibnamefont {Calandra}}, \bibinfo
  {author} {\bibfnamefont {M.-A.}\ \bibnamefont {Measson}}, \bibinfo {author}
  {\bibfnamefont {P.}~\bibnamefont {Diener}}, \bibinfo {author} {\bibfnamefont
  {E.}~\bibnamefont {Borrissenko}}, \bibinfo {author} {\bibfnamefont
  {A.}~\bibnamefont {Bosak}}, \ and\ \bibinfo {author} {\bibfnamefont
  {P.}~\bibnamefont {Rodiere}},\ }\href@noop {} {\bibfield  {journal} {\bibinfo
   {journal} {Phys. Rev. B}\ }\textbf {\bibinfo {volume} {86}},\ \bibinfo
  {pages} {155125} (\bibinfo {year} {2012})}\BibitemShut {NoStop}%
\bibitem [{\citenamefont {McMullan}\ and\ \citenamefont
  {Irwin}(1985)}]{McMullan1985}%
  \BibitemOpen
  \bibfield  {author} {\bibinfo {author} {\bibfnamefont {W.~G.}\ \bibnamefont
  {McMullan}}\ and\ \bibinfo {author} {\bibfnamefont {J.~C.}\ \bibnamefont
  {Irwin}},\ }\href@noop {} {\bibfield  {journal} {\bibinfo  {journal} {Phys.
  Stat. Solidi. B}\ }\textbf {\bibinfo {volume} {129}},\ \bibinfo {pages} {465}
  (\bibinfo {year} {1985})}\BibitemShut {NoStop}%
\bibitem [{\citenamefont {Wang}\ and\ \citenamefont {Chen}(1974)}]{Wang1974}%
  \BibitemOpen
  \bibfield  {author} {\bibinfo {author} {\bibfnamefont {C.~S.}\ \bibnamefont
  {Wang}}\ and\ \bibinfo {author} {\bibfnamefont {J.~M.}\ \bibnamefont
  {Chen}},\ }\href@noop {} {\bibfield  {journal} {\bibinfo  {journal} {Sol.
  Stat. Commun.}\ }\textbf {\bibinfo {volume} {14}},\ \bibinfo {pages} {1145}
  (\bibinfo {year} {1974})}\BibitemShut {NoStop}%
\bibitem [{\citenamefont {Chen}\ \emph {et~al.}(2015)\citenamefont {Chen},
  \citenamefont {Chan}, \citenamefont {Fang}, \citenamefont {Zhang},
  \citenamefont {Chou}, \citenamefont {Mo}, \citenamefont {Hussain},
  \citenamefont {Fedorov},\ and\ \citenamefont {Chaing}}]{Chen2015}%
  \BibitemOpen
  \bibfield  {author} {\bibinfo {author} {\bibfnamefont {P.}~\bibnamefont
  {Chen}}, \bibinfo {author} {\bibfnamefont {Y.-H.}\ \bibnamefont {Chan}},
  \bibinfo {author} {\bibfnamefont {X.-Y.}\ \bibnamefont {Fang}}, \bibinfo
  {author} {\bibfnamefont {Y.}~\bibnamefont {Zhang}}, \bibinfo {author}
  {\bibfnamefont {M.~Y.}\ \bibnamefont {Chou}}, \bibinfo {author}
  {\bibfnamefont {S.-K.}\ \bibnamefont {Mo}}, \bibinfo {author} {\bibfnamefont
  {Z.}~\bibnamefont {Hussain}}, \bibinfo {author} {\bibfnamefont {A.~V.}\
  \bibnamefont {Fedorov}}, \ and\ \bibinfo {author} {\bibfnamefont {T.~C.}\
  \bibnamefont {Chaing}},\ }\href@noop {} {\bibfield  {journal} {\bibinfo
  {journal} {Nat. Commun.}\ }\textbf {\bibinfo {volume} {6}},\ \bibinfo {pages}
  {8943} (\bibinfo {year} {2015})}\BibitemShut {NoStop}%
\bibitem [{\citenamefont {Guster}\ \emph {et~al.}(2018)\citenamefont {Guster},
  \citenamefont {Canadell}, \citenamefont {Pruneda},\ and\ \citenamefont
  {Ordejon}}]{Guster2018}%
  \BibitemOpen
  \bibfield  {author} {\bibinfo {author} {\bibfnamefont {I.~B.}\ \bibnamefont
  {Guster}}, \bibinfo {author} {\bibfnamefont {E.}~\bibnamefont {Canadell}},
  \bibinfo {author} {\bibfnamefont {M.}~\bibnamefont {Pruneda}}, \ and\
  \bibinfo {author} {\bibfnamefont {P.}~\bibnamefont {Ordejon}},\ }\href@noop
  {} {\bibfield  {journal} {\bibinfo  {journal} {2D materials}\ } (\bibinfo
  {year} {2018})}\BibitemShut {NoStop}%
\bibitem [{\citenamefont {Dolui}\ and\ \citenamefont
  {Sanvito}(2016)}]{Dolui2016}%
  \BibitemOpen
  \bibfield  {author} {\bibinfo {author} {\bibfnamefont {K.}~\bibnamefont
  {Dolui}}\ and\ \bibinfo {author} {\bibfnamefont {S.}~\bibnamefont
  {Sanvito}},\ }\href@noop {} {\bibfield  {journal} {\bibinfo  {journal} {E.
  Phys. Lett.}\ }\textbf {\bibinfo {volume} {115}},\ \bibinfo {pages} {47001}
  (\bibinfo {year} {2016})}\BibitemShut {NoStop}%
\bibitem [{\citenamefont {Mounet}\ \emph {et~al.}(2018)\citenamefont {Mounet},
  \citenamefont {Gibertini}, \citenamefont {Schwaller}, \citenamefont {Merkys},
  \citenamefont {Castelli}, \citenamefont {Cepellotti}, \citenamefont {Pizzi},\
  and\ \citenamefont {Marzari}}]{2017_arxiv_mounet_2D}%
  \BibitemOpen
  \bibfield  {author} {\bibinfo {author} {\bibfnamefont {N.}~\bibnamefont
  {Mounet}}, \bibinfo {author} {\bibfnamefont {M.}~\bibnamefont {Gibertini}},
  \bibinfo {author} {\bibfnamefont {P.}~\bibnamefont {Schwaller}}, \bibinfo
  {author} {\bibfnamefont {A.}~\bibnamefont {Merkys}}, \bibinfo {author}
  {\bibfnamefont {I.~E.}\ \bibnamefont {Castelli}}, \bibinfo {author}
  {\bibfnamefont {A.}~\bibnamefont {Cepellotti}}, \bibinfo {author}
  {\bibfnamefont {G.}~\bibnamefont {Pizzi}}, \ and\ \bibinfo {author}
  {\bibfnamefont {N.}~\bibnamefont {Marzari}},\ }\href@noop {} {\bibfield
  {journal} {\bibinfo  {journal} {Nature nanotechnology}\ }\textbf {\bibinfo
  {volume} {13}},\ \bibinfo {pages} {246} (\bibinfo {year} {2018})}\BibitemShut
  {NoStop}%
\bibitem [{\citenamefont {Choudhary}\ \emph {et~al.}(2017)\citenamefont
  {Choudhary}, \citenamefont {Kalish}, \citenamefont {Beams},\ and\
  \citenamefont {Tavazza}}]{2017_choudhary_scirep}%
  \BibitemOpen
  \bibfield  {author} {\bibinfo {author} {\bibfnamefont {K.}~\bibnamefont
  {Choudhary}}, \bibinfo {author} {\bibfnamefont {I.}~\bibnamefont {Kalish}},
  \bibinfo {author} {\bibfnamefont {R.}~\bibnamefont {Beams}}, \ and\ \bibinfo
  {author} {\bibfnamefont {F.}~\bibnamefont {Tavazza}},\ }\href@noop {}
  {\bibfield  {journal} {\bibinfo  {journal} {Sci. Repts.}\ }\textbf {\bibinfo
  {volume} {7}},\ \bibinfo {pages} {5179} (\bibinfo {year} {2017})}\BibitemShut
  {NoStop}%
\bibitem [{\citenamefont {Sofer}\ \emph {et~al.}(2017)\citenamefont {Sofer},
  \citenamefont {Sedmidubsky}, \citenamefont {Luxa}, \citenamefont {Bousa},
  \citenamefont {Huber}, \citenamefont {Lazar}, \citenamefont {Vesely},\ and\
  \citenamefont {Pumera}}]{Sofer2017}%
  \BibitemOpen
  \bibfield  {author} {\bibinfo {author} {\bibfnamefont {Z.}~\bibnamefont
  {Sofer}}, \bibinfo {author} {\bibfnamefont {D.}~\bibnamefont {Sedmidubsky}},
  \bibinfo {author} {\bibfnamefont {J.}~\bibnamefont {Luxa}}, \bibinfo {author}
  {\bibfnamefont {D.}~\bibnamefont {Bousa}}, \bibinfo {author} {\bibfnamefont
  {S.}~\bibnamefont {Huber}}, \bibinfo {author} {\bibfnamefont
  {P.}~\bibnamefont {Lazar}}, \bibinfo {author} {\bibfnamefont
  {M.}~\bibnamefont {Vesely}}, \ and\ \bibinfo {author} {\bibfnamefont
  {M.}~\bibnamefont {Pumera}},\ }\href@noop {} {\bibfield  {journal} {\bibinfo
  {journal} {Chem. Eur. J.}\ }\textbf {\bibinfo {volume} {23}},\ \bibinfo
  {pages} {10177} (\bibinfo {year} {2017})}\BibitemShut {NoStop}%
\bibitem [{\citenamefont {Li}\ \emph {et~al.}(2015)\citenamefont {Li},
  \citenamefont {Chen}, \citenamefont {Shi},\ and\ \citenamefont
  {Li}}]{Li2015}%
  \BibitemOpen
  \bibfield  {author} {\bibinfo {author} {\bibfnamefont {M.}~\bibnamefont
  {Li}}, \bibinfo {author} {\bibfnamefont {C.-H.}\ \bibnamefont {Chen}},
  \bibinfo {author} {\bibfnamefont {Y.}~\bibnamefont {Shi}}, \ and\ \bibinfo
  {author} {\bibfnamefont {L.-J.}\ \bibnamefont {Li}},\ }\href@noop {}
  {\bibfield  {journal} {\bibinfo  {journal} {Mater. Today}\ }\textbf {\bibinfo
  {volume} {19}},\ \bibinfo {pages} {32} (\bibinfo {year} {2015})}\BibitemShut
  {NoStop}%
\bibitem [{\citenamefont {Wei}\ \emph {et~al.}(2015)\citenamefont {Wei},
  \citenamefont {Dai}, \citenamefont {Sun}, \citenamefont {Yin}, \citenamefont
  {Han}, \citenamefont {Huang},\ and\ \citenamefont {Jacob}}]{Wei2015}%
  \BibitemOpen
  \bibfield  {author} {\bibinfo {author} {\bibfnamefont {W.}~\bibnamefont
  {Wei}}, \bibinfo {author} {\bibfnamefont {Y.}~\bibnamefont {Dai}}, \bibinfo
  {author} {\bibfnamefont {Q.}~\bibnamefont {Sun}}, \bibinfo {author}
  {\bibfnamefont {N.}~\bibnamefont {Yin}}, \bibinfo {author} {\bibfnamefont
  {S.}~\bibnamefont {Han}}, \bibinfo {author} {\bibfnamefont {B.}~\bibnamefont
  {Huang}}, \ and\ \bibinfo {author} {\bibfnamefont {T.}~\bibnamefont
  {Jacob}},\ }\href@noop {} {\bibfield  {journal} {\bibinfo  {journal} {Phys.
  Chem. Chem. Phys.}\ }\textbf {\bibinfo {volume} {17}},\ \bibinfo {pages}
  {29380} (\bibinfo {year} {2015})}\BibitemShut {NoStop}%
\bibitem [{\citenamefont {Liu}\ \emph {et~al.}(2016)\citenamefont {Liu},
  \citenamefont {Weiss}, \citenamefont {Duan}, \citenamefont {Cheng},
  \citenamefont {Huang},\ and\ \citenamefont {Duan}}]{Liu2016}%
  \BibitemOpen
  \bibfield  {author} {\bibinfo {author} {\bibfnamefont {Y.}~\bibnamefont
  {Liu}}, \bibinfo {author} {\bibfnamefont {N.~O.}\ \bibnamefont {Weiss}},
  \bibinfo {author} {\bibfnamefont {X.}~\bibnamefont {Duan}}, \bibinfo {author}
  {\bibfnamefont {H.-C.}\ \bibnamefont {Cheng}}, \bibinfo {author}
  {\bibfnamefont {Y.}~\bibnamefont {Huang}}, \ and\ \bibinfo {author}
  {\bibfnamefont {X.}~\bibnamefont {Duan}},\ }\href@noop {} {\bibfield
  {journal} {\bibinfo  {journal} {Nature Review Materials}\ }\textbf {\bibinfo
  {volume} {1}},\ \bibinfo {pages} {16042} (\bibinfo {year}
  {2016})}\BibitemShut {NoStop}%
\bibitem [{\citenamefont {Martin}(2004)}]{Martin2004}%
  \BibitemOpen
  \bibfield  {author} {\bibinfo {author} {\bibfnamefont {R.~M.}\ \bibnamefont
  {Martin}},\ }\href@noop {} {\emph {\bibinfo {title} {Electronic Structure:
  Basic Theory and Practice Methods}}}\ (\bibinfo  {publisher} {Cambridge
  University Press},\ \bibinfo {year} {2004})\BibitemShut {NoStop}%
\bibitem [{\citenamefont {Baroni}\ \emph {et~al.}(2001)\citenamefont {Baroni},
  \citenamefont {de~Gironcoli}, \citenamefont {Corso},\ and\ \citenamefont
  {Giannozzi}}]{Baroni2001}%
  \BibitemOpen
  \bibfield  {author} {\bibinfo {author} {\bibfnamefont {S.}~\bibnamefont
  {Baroni}}, \bibinfo {author} {\bibfnamefont {S.}~\bibnamefont
  {de~Gironcoli}}, \bibinfo {author} {\bibfnamefont {A.~D.}\ \bibnamefont
  {Corso}}, \ and\ \bibinfo {author} {\bibfnamefont {P.}~\bibnamefont
  {Giannozzi}},\ }\href@noop {} {\bibfield  {journal} {\bibinfo  {journal}
  {Rev. Mod. Phys.}\ }\textbf {\bibinfo {volume} {73}},\ \bibinfo {pages} {515}
  (\bibinfo {year} {2001})}\BibitemShut {NoStop}%
\bibitem [{\citenamefont {Grimme}(2006)}]{Grimme2006}%
  \BibitemOpen
  \bibfield  {author} {\bibinfo {author} {\bibfnamefont {S.}~\bibnamefont
  {Grimme}},\ }\href@noop {} {\bibfield  {journal} {\bibinfo  {journal} {J.
  Compt. Chem.}\ }\textbf {\bibinfo {volume} {27}},\ \bibinfo {pages} {1787}
  (\bibinfo {year} {2006})}\BibitemShut {NoStop}%
\bibitem [{\citenamefont {Grimme}\ \emph {et~al.}(2010)\citenamefont {Grimme},
  \citenamefont {Ehrlich},\ and\ \citenamefont {Goerigk}}]{Grimme2010}%
  \BibitemOpen
  \bibfield  {author} {\bibinfo {author} {\bibfnamefont {S.}~\bibnamefont
  {Grimme}}, \bibinfo {author} {\bibfnamefont {S.}~\bibnamefont {Ehrlich}}, \
  and\ \bibinfo {author} {\bibfnamefont {L.}~\bibnamefont {Goerigk}},\
  }\href@noop {} {\bibfield  {journal} {\bibinfo  {journal} {J. Chem. Phys.}\
  }\textbf {\bibinfo {volume} {132}},\ \bibinfo {pages} {154104} (\bibinfo
  {year} {2010})}\BibitemShut {NoStop}%
\bibitem [{\citenamefont {Grimme}\ \emph {et~al.}(2011)\citenamefont {Grimme},
  \citenamefont {Ehrlich},\ and\ \citenamefont {Georigk}}]{Grimme2011}%
  \BibitemOpen
  \bibfield  {author} {\bibinfo {author} {\bibfnamefont {S.}~\bibnamefont
  {Grimme}}, \bibinfo {author} {\bibfnamefont {S.}~\bibnamefont {Ehrlich}}, \
  and\ \bibinfo {author} {\bibfnamefont {L.}~\bibnamefont {Georigk}},\
  }\href@noop {} {\bibfield  {journal} {\bibinfo  {journal} {J. Comp. Chem.}\
  }\textbf {\bibinfo {volume} {32}},\ \bibinfo {pages} {1456} (\bibinfo {year}
  {2011})}\BibitemShut {NoStop}%
\bibitem [{\citenamefont {Setyawan}\ and\ \citenamefont
  {Curtarolo}(2010)}]{Setyawan2010}%
  \BibitemOpen
  \bibfield  {author} {\bibinfo {author} {\bibfnamefont {W.}~\bibnamefont
  {Setyawan}}\ and\ \bibinfo {author} {\bibfnamefont {S.}~\bibnamefont
  {Curtarolo}},\ }\href@noop {} {\bibfield  {journal} {\bibinfo  {journal}
  {Compt. Mat. Sci}\ }\textbf {\bibinfo {volume} {49}},\ \bibinfo {pages} {299}
  (\bibinfo {year} {2010})}\BibitemShut {NoStop}%
\bibitem [{sup()}]{supmat}%
  \BibitemOpen
  \href@noop {} {\emph {\bibinfo {title} {See supplemental material for
  additional details on the calculation methods, derived elastic properties,
  thermal properties, and the numerical results for all of the compounds. The
  supplemental information includes the following
  Refs.~\citenum{Gonze2009,Dickinson1923,Brixner1962,Jellinek1960,Berkdemir2013,
  Wakabayashi1975,Sourisseau1989,Sourisseau1991,Zhao2013b,Tonndorf2013,Tongay2012,Sekine1980,
  Zhao2013,McMullan1985,Wang1974,Sandoval1992,Sugai1980,Hangyo1983,Roubi1988,Manas2016,
  Wolverson2016,Wieting1971,Lucovsky1973,Froehlicher2015,Mead1977,Feldman1976,Aksoy2006,
  Nayak2014,Zhao2015,Selvi2006,Liu2014,Zhang2016,Leroux2012,Greenaway1965,Scharli1986,Wildervanck1970,Kam1982,
  Agnihotri1972,Garg1973,Lezama2015,Beal1979,Liang1973,Luttrell2006,Uchida1978,Iwasaki1982,
  Vaterlaus1985,Starnberg1995,Tongay2014,Ho1997,Zhao2015a,Park1996,Liu2015,
  Wender1973,Titov2007,Nye1957} at [URL].}}}\BibitemShut {Stop}%
\bibitem [{\citenamefont {Gonze}(1997)}]{Gonze1997b}%
  \BibitemOpen
  \bibfield  {author} {\bibinfo {author} {\bibfnamefont {X.}~\bibnamefont
  {Gonze}},\ }\href@noop {} {\bibfield  {journal} {\bibinfo  {journal} {Phys.
  Rev. B}\ }\textbf {\bibinfo {volume} {55}},\ \bibinfo {pages} {10337}
  (\bibinfo {year} {1997})}\BibitemShut {NoStop}%
\bibitem [{\citenamefont {Gonze}\ and\ \citenamefont {Lee}(1997)}]{Gonze1997a}%
  \BibitemOpen
  \bibfield  {author} {\bibinfo {author} {\bibfnamefont {X.}~\bibnamefont
  {Gonze}}\ and\ \bibinfo {author} {\bibfnamefont {C.}~\bibnamefont {Lee}},\
  }\href@noop {} {\bibfield  {journal} {\bibinfo  {journal} {Phys. Rev. B}\
  }\textbf {\bibinfo {volume} {55}},\ \bibinfo {pages} {10355} (\bibinfo {year}
  {1997})}\BibitemShut {NoStop}%
\bibitem [{\citenamefont {Gonze}\ \emph
  {et~al.}(2005{\natexlab{a}})\citenamefont {Gonze}, \citenamefont {Rignanese},
  \citenamefont {Verstraete}, \citenamefont {Beuken}, \citenamefont {Pouillon},
  \citenamefont {Caracas}, \citenamefont {Jollet}, \citenamefont {Torrent},
  \citenamefont {Zerah}, \citenamefont {Mikami}, \citenamefont {Ghosez},
  \citenamefont {Veithen}, \citenamefont {Raty}, \citenamefont {Olevano},
  \citenamefont {Bruneval}, \citenamefont {Reining}, \citenamefont {Godby},
  \citenamefont {Onida}, \citenamefont {Hamann},\ and\ \citenamefont
  {Allan}}]{Gonze2005}%
  \BibitemOpen
  \bibfield  {author} {\bibinfo {author} {\bibfnamefont {X.}~\bibnamefont
  {Gonze}}, \bibinfo {author} {\bibfnamefont {G.-M.}\ \bibnamefont
  {Rignanese}}, \bibinfo {author} {\bibfnamefont {M.}~\bibnamefont
  {Verstraete}}, \bibinfo {author} {\bibfnamefont {J.-M.}\ \bibnamefont
  {Beuken}}, \bibinfo {author} {\bibfnamefont {Y.}~\bibnamefont {Pouillon}},
  \bibinfo {author} {\bibfnamefont {R.}~\bibnamefont {Caracas}}, \bibinfo
  {author} {\bibfnamefont {F.}~\bibnamefont {Jollet}}, \bibinfo {author}
  {\bibfnamefont {M.}~\bibnamefont {Torrent}}, \bibinfo {author} {\bibfnamefont
  {G.}~\bibnamefont {Zerah}}, \bibinfo {author} {\bibfnamefont
  {M.}~\bibnamefont {Mikami}}, \bibinfo {author} {\bibfnamefont
  {P.}~\bibnamefont {Ghosez}}, \bibinfo {author} {\bibfnamefont
  {M.}~\bibnamefont {Veithen}}, \bibinfo {author} {\bibfnamefont {J.-Y.}\
  \bibnamefont {Raty}}, \bibinfo {author} {\bibfnamefont {V.}~\bibnamefont
  {Olevano}}, \bibinfo {author} {\bibfnamefont {F.}~\bibnamefont {Bruneval}},
  \bibinfo {author} {\bibfnamefont {L.}~\bibnamefont {Reining}}, \bibinfo
  {author} {\bibfnamefont {R.}~\bibnamefont {Godby}}, \bibinfo {author}
  {\bibfnamefont {G.}~\bibnamefont {Onida}}, \bibinfo {author} {\bibfnamefont
  {D.}~\bibnamefont {Hamann}}, \ and\ \bibinfo {author} {\bibfnamefont
  {D.}~\bibnamefont {Allan}},\ }\href@noop {} {\bibfield  {journal} {\bibinfo
  {journal} {Z. Kristallogr.}\ }\textbf {\bibinfo {volume} {220}},\ \bibinfo
  {pages} {558} (\bibinfo {year} {2005}{\natexlab{a}})}\BibitemShut {NoStop}%
\bibitem [{\citenamefont {Gonze}\ \emph {et~al.}(2009)\citenamefont {Gonze},
  \citenamefont {Amadon}, \citenamefont {Anglade}, \citenamefont {Beuken},
  \citenamefont {Bottin}, \citenamefont {Boulanger}, \citenamefont {Bruneval},
  \citenamefont {Caliste}, \citenamefont {Caracas}, \citenamefont {Cote},
  \citenamefont {Deutsch}, \citenamefont {Genovese}, \citenamefont {Ghosez},
  \citenamefont {Giantomassi}, \citenamefont {Goedecker}, \citenamefont
  {Hamann}, \citenamefont {Hermet}, \citenamefont {Jollet}, \citenamefont
  {Jomard}, \citenamefont {Leroux}, \citenamefont {Mancini}, \citenamefont
  {Mazevet}, \citenamefont {Oliveira}, \citenamefont {Onida}, \citenamefont
  {Pouillon}, \citenamefont {Rangel}, \citenamefont {Rignanese}, \citenamefont
  {Sangalli}, \citenamefont {Shaltaf}, \citenamefont {Torrent}, \citenamefont
  {Verstraete}, \citenamefont {Zérah},\ and\ \citenamefont
  {Zwanziger}}]{Gonze2009}%
  \BibitemOpen
  \bibfield  {author} {\bibinfo {author} {\bibfnamefont {X.}~\bibnamefont
  {Gonze}}, \bibinfo {author} {\bibfnamefont {B.}~\bibnamefont {Amadon}},
  \bibinfo {author} {\bibfnamefont {P.~M.}\ \bibnamefont {Anglade}}, \bibinfo
  {author} {\bibfnamefont {J.-M.}\ \bibnamefont {Beuken}}, \bibinfo {author}
  {\bibfnamefont {F.}~\bibnamefont {Bottin}}, \bibinfo {author} {\bibfnamefont
  {P.}~\bibnamefont {Boulanger}}, \bibinfo {author} {\bibfnamefont
  {F.}~\bibnamefont {Bruneval}}, \bibinfo {author} {\bibfnamefont
  {D.}~\bibnamefont {Caliste}}, \bibinfo {author} {\bibfnamefont
  {R.}~\bibnamefont {Caracas}}, \bibinfo {author} {\bibfnamefont
  {M.}~\bibnamefont {Cote}}, \bibinfo {author} {\bibfnamefont {T.}~\bibnamefont
  {Deutsch}}, \bibinfo {author} {\bibfnamefont {L.}~\bibnamefont {Genovese}},
  \bibinfo {author} {\bibfnamefont {P.}~\bibnamefont {Ghosez}}, \bibinfo
  {author} {\bibfnamefont {M.}~\bibnamefont {Giantomassi}}, \bibinfo {author}
  {\bibfnamefont {S.}~\bibnamefont {Goedecker}}, \bibinfo {author}
  {\bibfnamefont {D.}~\bibnamefont {Hamann}}, \bibinfo {author} {\bibfnamefont
  {P.}~\bibnamefont {Hermet}}, \bibinfo {author} {\bibfnamefont
  {F.}~\bibnamefont {Jollet}}, \bibinfo {author} {\bibfnamefont
  {G.}~\bibnamefont {Jomard}}, \bibinfo {author} {\bibfnamefont
  {S.}~\bibnamefont {Leroux}}, \bibinfo {author} {\bibfnamefont
  {M.}~\bibnamefont {Mancini}}, \bibinfo {author} {\bibfnamefont
  {S.}~\bibnamefont {Mazevet}}, \bibinfo {author} {\bibfnamefont
  {M.}~\bibnamefont {Oliveira}}, \bibinfo {author} {\bibfnamefont
  {G.}~\bibnamefont {Onida}}, \bibinfo {author} {\bibfnamefont
  {Y.}~\bibnamefont {Pouillon}}, \bibinfo {author} {\bibfnamefont
  {T.}~\bibnamefont {Rangel}}, \bibinfo {author} {\bibfnamefont {G.-M.}\
  \bibnamefont {Rignanese}}, \bibinfo {author} {\bibfnamefont {D.}~\bibnamefont
  {Sangalli}}, \bibinfo {author} {\bibfnamefont {R.}~\bibnamefont {Shaltaf}},
  \bibinfo {author} {\bibfnamefont {M.}~\bibnamefont {Torrent}}, \bibinfo
  {author} {\bibfnamefont {M.}~\bibnamefont {Verstraete}}, \bibinfo {author}
  {\bibfnamefont {G.}~\bibnamefont {Zérah}}, \ and\ \bibinfo {author}
  {\bibfnamefont {J.}~\bibnamefont {Zwanziger}},\ }\href@noop {} {\bibfield
  {journal} {\bibinfo  {journal} {Comp. Phys. Comm.}\ }\textbf {\bibinfo
  {volume} {180}},\ \bibinfo {pages} {2582} (\bibinfo {year}
  {2009})}\BibitemShut {NoStop}%
\bibitem [{\citenamefont {Gonze}\ \emph {et~al.}(2016)\citenamefont {Gonze},
  \citenamefont {Jollet}, \citenamefont {Araujo}, \citenamefont {Adams},
  \citenamefont {Amadon}, \citenamefont {T.Applencourt}, \citenamefont
  {Audouze}, \citenamefont {Beuken}, \citenamefont {Bieder}, \citenamefont
  {Bokhanchuk}, \citenamefont {Bousquet}, \citenamefont {Bruneval},
  \citenamefont {Caliste}, \citenamefont {Cote}, \citenamefont {Dahm},
  \citenamefont {Pieve}, \citenamefont {Delaveau}, \citenamefont {Gennaro},
  \citenamefont {Dorado}, \citenamefont {C.Espejo}, \citenamefont {Geneste},
  \citenamefont {Genovese}, \citenamefont {Gerossier}, \citenamefont
  {Giantomassi}, \citenamefont {Gillet}, \citenamefont {Hamann}, \citenamefont
  {L.He}, \citenamefont {Jomard}, \citenamefont {Janssen}, \citenamefont
  {Roux}, \citenamefont {Levitt}, \citenamefont {Lherbier}, \citenamefont
  {Liu}, \citenamefont {Lukacevic}, \citenamefont {Martin}, \citenamefont
  {Martins}, \citenamefont {Oliveira}, \citenamefont {Ponce}, \citenamefont
  {Pouillon}, \citenamefont {Rangel}, \citenamefont {Rignanese}, \citenamefont
  {Romero}, \citenamefont {Rousseau}, \citenamefont {Rubel}, \citenamefont
  {Shukri}, \citenamefont {Stankovski}, \citenamefont {Torrent}, \citenamefont
  {VanSetten}, \citenamefont {Troeye}, \citenamefont {Verstraete},
  \citenamefont {Waroquiers}, \citenamefont {Wiktor}, \citenamefont {Xu},
  \citenamefont {Zhou},\ and\ \citenamefont {Zwanziger}}]{Gonze2016}%
  \BibitemOpen
  \bibfield  {author} {\bibinfo {author} {\bibfnamefont {X.}~\bibnamefont
  {Gonze}}, \bibinfo {author} {\bibfnamefont {F.}~\bibnamefont {Jollet}},
  \bibinfo {author} {\bibfnamefont {F.~A.}\ \bibnamefont {Araujo}}, \bibinfo
  {author} {\bibfnamefont {D.}~\bibnamefont {Adams}}, \bibinfo {author}
  {\bibfnamefont {B.}~\bibnamefont {Amadon}}, \bibinfo {author} {\bibnamefont
  {T.Applencourt}}, \bibinfo {author} {\bibfnamefont {C.}~\bibnamefont
  {Audouze}}, \bibinfo {author} {\bibfnamefont {J.-M.}\ \bibnamefont {Beuken}},
  \bibinfo {author} {\bibfnamefont {J.}~\bibnamefont {Bieder}}, \bibinfo
  {author} {\bibfnamefont {A.}~\bibnamefont {Bokhanchuk}}, \bibinfo {author}
  {\bibfnamefont {E.}~\bibnamefont {Bousquet}}, \bibinfo {author}
  {\bibfnamefont {F.}~\bibnamefont {Bruneval}}, \bibinfo {author}
  {\bibfnamefont {D.}~\bibnamefont {Caliste}}, \bibinfo {author} {\bibfnamefont
  {M.}~\bibnamefont {Cote}}, \bibinfo {author} {\bibfnamefont {F.}~\bibnamefont
  {Dahm}}, \bibinfo {author} {\bibfnamefont {F.~D.}\ \bibnamefont {Pieve}},
  \bibinfo {author} {\bibfnamefont {M.}~\bibnamefont {Delaveau}}, \bibinfo
  {author} {\bibfnamefont {M.~D.}\ \bibnamefont {Gennaro}}, \bibinfo {author}
  {\bibfnamefont {B.}~\bibnamefont {Dorado}}, \bibinfo {author} {\bibnamefont
  {C.Espejo}}, \bibinfo {author} {\bibfnamefont {G.}~\bibnamefont {Geneste}},
  \bibinfo {author} {\bibfnamefont {L.}~\bibnamefont {Genovese}}, \bibinfo
  {author} {\bibfnamefont {A.}~\bibnamefont {Gerossier}}, \bibinfo {author}
  {\bibfnamefont {M.}~\bibnamefont {Giantomassi}}, \bibinfo {author}
  {\bibfnamefont {Y.}~\bibnamefont {Gillet}}, \bibinfo {author} {\bibfnamefont
  {D.}~\bibnamefont {Hamann}}, \bibinfo {author} {\bibnamefont {L.He}},
  \bibinfo {author} {\bibfnamefont {G.}~\bibnamefont {Jomard}}, \bibinfo
  {author} {\bibfnamefont {J.~L.}\ \bibnamefont {Janssen}}, \bibinfo {author}
  {\bibfnamefont {S.~L.}\ \bibnamefont {Roux}}, \bibinfo {author}
  {\bibfnamefont {A.}~\bibnamefont {Levitt}}, \bibinfo {author} {\bibfnamefont
  {A.}~\bibnamefont {Lherbier}}, \bibinfo {author} {\bibfnamefont
  {F.}~\bibnamefont {Liu}}, \bibinfo {author} {\bibfnamefont {I.}~\bibnamefont
  {Lukacevic}}, \bibinfo {author} {\bibfnamefont {A.}~\bibnamefont {Martin}},
  \bibinfo {author} {\bibfnamefont {C.}~\bibnamefont {Martins}}, \bibinfo
  {author} {\bibfnamefont {M.}~\bibnamefont {Oliveira}}, \bibinfo {author}
  {\bibfnamefont {S.}~\bibnamefont {Ponce}}, \bibinfo {author} {\bibfnamefont
  {Y.}~\bibnamefont {Pouillon}}, \bibinfo {author} {\bibfnamefont
  {T.}~\bibnamefont {Rangel}}, \bibinfo {author} {\bibfnamefont {G.-M.}\
  \bibnamefont {Rignanese}}, \bibinfo {author} {\bibfnamefont {A.}~\bibnamefont
  {Romero}}, \bibinfo {author} {\bibfnamefont {B.}~\bibnamefont {Rousseau}},
  \bibinfo {author} {\bibfnamefont {O.}~\bibnamefont {Rubel}}, \bibinfo
  {author} {\bibfnamefont {A.}~\bibnamefont {Shukri}}, \bibinfo {author}
  {\bibfnamefont {M.}~\bibnamefont {Stankovski}}, \bibinfo {author}
  {\bibfnamefont {M.}~\bibnamefont {Torrent}}, \bibinfo {author} {\bibfnamefont
  {M.}~\bibnamefont {VanSetten}}, \bibinfo {author} {\bibfnamefont {B.~V.}\
  \bibnamefont {Troeye}}, \bibinfo {author} {\bibfnamefont {M.}~\bibnamefont
  {Verstraete}}, \bibinfo {author} {\bibfnamefont {D.}~\bibnamefont
  {Waroquiers}}, \bibinfo {author} {\bibfnamefont {J.}~\bibnamefont {Wiktor}},
  \bibinfo {author} {\bibfnamefont {B.}~\bibnamefont {Xu}}, \bibinfo {author}
  {\bibfnamefont {A.}~\bibnamefont {Zhou}}, \ and\ \bibinfo {author}
  {\bibfnamefont {J.~W.}\ \bibnamefont {Zwanziger}},\ }\href@noop {} {\bibfield
   {journal} {\bibinfo  {journal} {Comp. Phys. Comm.}\ }\textbf {\bibinfo
  {volume} {205}},\ \bibinfo {pages} {106} (\bibinfo {year}
  {2016})}\BibitemShut {NoStop}%
\bibitem [{\citenamefont {Jones}(2015)}]{Jones2015}%
  \BibitemOpen
  \bibfield  {author} {\bibinfo {author} {\bibfnamefont {R.~O.}\ \bibnamefont
  {Jones}},\ }\href@noop {} {\bibfield  {journal} {\bibinfo  {journal} {Rev.
  Mod. Phys.}\ }\textbf {\bibinfo {volume} {87}},\ \bibinfo {pages} {897}
  (\bibinfo {year} {2015})}\BibitemShut {NoStop}%
\bibitem [{\citenamefont {Gonze}\ \emph
  {et~al.}(2005{\natexlab{b}})\citenamefont {Gonze}, \citenamefont
  {Rignanese},\ and\ \citenamefont {Caracas}}]{Gonze2005a}%
  \BibitemOpen
  \bibfield  {author} {\bibinfo {author} {\bibfnamefont {X.}~\bibnamefont
  {Gonze}}, \bibinfo {author} {\bibfnamefont {G.~M.}\ \bibnamefont
  {Rignanese}}, \ and\ \bibinfo {author} {\bibfnamefont {R.}~\bibnamefont
  {Caracas}},\ }\href@noop {} {\bibfield  {journal} {\bibinfo  {journal}
  {Zeitschrift fur Kristallographie}\ }\textbf {\bibinfo {volume} {220}},\
  \bibinfo {pages} {458} (\bibinfo {year} {2005}{\natexlab{b}})}\BibitemShut
  {NoStop}%
\bibitem [{\citenamefont {Verstraete}\ and\ \citenamefont
  {Zanolli}(2014)}]{verstraete_2014_dfpt}%
  \BibitemOpen
  \bibfield  {author} {\bibinfo {author} {\bibfnamefont {M.~J.}\ \bibnamefont
  {Verstraete}}\ and\ \bibinfo {author} {\bibfnamefont {Z.}~\bibnamefont
  {Zanolli}},\ }in\ \href@noop {} {\emph {\bibinfo {booktitle} {Computing
  Solids: Models, Ab-initio Methods and Supercomputing, Lecture Notes of the
  45th Spring School 2014}}}\ (\bibinfo  {publisher} {Schiften des
  Forschungszentrums Julich},\ \bibinfo {year} {2014})\BibitemShut {NoStop}%
\bibitem [{\citenamefont {Monkhorst}\ and\ \citenamefont
  {Pack}(1976)}]{Monkhorst1976}%
  \BibitemOpen
  \bibfield  {author} {\bibinfo {author} {\bibfnamefont {H.~J.}\ \bibnamefont
  {Monkhorst}}\ and\ \bibinfo {author} {\bibfnamefont {J.~D.}\ \bibnamefont
  {Pack}},\ }\href@noop {} {\bibfield  {journal} {\bibinfo  {journal} {Phys.
  Rev. B}\ }\textbf {\bibinfo {volume} {13}},\ \bibinfo {pages} {5188}
  (\bibinfo {year} {1976})}\BibitemShut {NoStop}%
\bibitem [{\citenamefont {Zeng}\ \emph {et~al.}(2013)\citenamefont {Zeng},
  \citenamefont {Liu}, \citenamefont {Dai}, \citenamefont {Yan}, \citenamefont
  {Zhu}, \citenamefont {He}, \citenamefont {Xie}, \citenamefont {Xu},
  \citenamefont {Chen}, \citenamefont {Yao},\ and\ \citenamefont
  {Cui}}]{Zeng2013}%
  \BibitemOpen
  \bibfield  {author} {\bibinfo {author} {\bibfnamefont {H.}~\bibnamefont
  {Zeng}}, \bibinfo {author} {\bibfnamefont {G.-B.}\ \bibnamefont {Liu}},
  \bibinfo {author} {\bibfnamefont {J.}~\bibnamefont {Dai}}, \bibinfo {author}
  {\bibfnamefont {Y.}~\bibnamefont {Yan}}, \bibinfo {author} {\bibfnamefont
  {B.}~\bibnamefont {Zhu}}, \bibinfo {author} {\bibfnamefont {R.}~\bibnamefont
  {He}}, \bibinfo {author} {\bibfnamefont {L.}~\bibnamefont {Xie}}, \bibinfo
  {author} {\bibfnamefont {S.}~\bibnamefont {Xu}}, \bibinfo {author}
  {\bibfnamefont {X.}~\bibnamefont {Chen}}, \bibinfo {author} {\bibfnamefont
  {W.}~\bibnamefont {Yao}}, \ and\ \bibinfo {author} {\bibfnamefont
  {X.}~\bibnamefont {Cui}},\ }\href@noop {} {\bibfield  {journal} {\bibinfo
  {journal} {Sci. Repts}\ }\textbf {\bibinfo {volume} {3}},\ \bibinfo {pages}
  {1608} (\bibinfo {year} {2013})}\BibitemShut {NoStop}%
\bibitem [{\citenamefont {Wolverson}\ and\ \citenamefont
  {Hart}(2016)}]{Wolverson2016}%
  \BibitemOpen
  \bibfield  {author} {\bibinfo {author} {\bibfnamefont {D.}~\bibnamefont
  {Wolverson}}\ and\ \bibinfo {author} {\bibfnamefont {L.~S.}\ \bibnamefont
  {Hart}},\ }\href@noop {} {\bibfield  {journal} {\bibinfo  {journal}
  {Nanoscale Research Letters.}\ }\textbf {\bibinfo {volume} {11}},\ \bibinfo
  {pages} {250} (\bibinfo {year} {2016})}\BibitemShut {NoStop}%
\bibitem [{\citenamefont {Zhao}\ \emph
  {et~al.}(2013{\natexlab{a}})\citenamefont {Zhao}, \citenamefont {Zohreh},
  \citenamefont {Kumar}, \citenamefont {Ren}, \citenamefont {Minglin},
  \citenamefont {Xin}, \citenamefont {Christian}, \citenamefont {Heng},\ and\
  \citenamefont {Goki}}]{Zhao2013}%
  \BibitemOpen
  \bibfield  {author} {\bibinfo {author} {\bibfnamefont {W.}~\bibnamefont
  {Zhao}}, \bibinfo {author} {\bibfnamefont {G.}~\bibnamefont {Zohreh}},
  \bibinfo {author} {\bibfnamefont {A.~K.}\ \bibnamefont {Kumar}}, \bibinfo
  {author} {\bibfnamefont {P.~J.}\ \bibnamefont {Ren}}, \bibinfo {author}
  {\bibfnamefont {T.}~\bibnamefont {Minglin}}, \bibinfo {author} {\bibfnamefont
  {Z.}~\bibnamefont {Xin}}, \bibinfo {author} {\bibfnamefont {K.}~\bibnamefont
  {Christian}}, \bibinfo {author} {\bibfnamefont {T.~P.}\ \bibnamefont {Heng}},
  \ and\ \bibinfo {author} {\bibfnamefont {E.}~\bibnamefont {Goki}},\
  }\href@noop {} {\bibfield  {journal} {\bibinfo  {journal} {Nanoscale}\
  }\textbf {\bibinfo {volume} {5}},\ \bibinfo {pages} {9677} (\bibinfo {year}
  {2013}{\natexlab{a}})}\BibitemShut {NoStop}%
\bibitem [{\citenamefont {Walter}\ and\ \citenamefont {Rappe}()}]{opium_psp}%
  \BibitemOpen
  \bibfield  {author} {\bibinfo {author} {\bibfnamefont {E.~J.}\ \bibnamefont
  {Walter}}\ and\ \bibinfo {author} {\bibfnamefont {A.}~\bibnamefont {Rappe}},\
  }\href {http://opium.sourceforge.net/} {\bibinfo  {journal}
  {http://opium.sourceforge.net/}\ }\BibitemShut {NoStop}%
\bibitem [{\citenamefont {Fuchs}\ and\ \citenamefont
  {Scheffler}(1999)}]{Fuchs1999}%
  \BibitemOpen
\bibfield  {journal} {  }\bibfield  {author} {\bibinfo {author} {\bibfnamefont
  {M.}~\bibnamefont {Fuchs}}\ and\ \bibinfo {author} {\bibfnamefont
  {M.}~\bibnamefont {Scheffler}},\ }\href@noop {} {\bibfield  {journal}
  {\bibinfo  {journal} {Comp. Phys. Comm.}\ }\textbf {\bibinfo {volume}
  {119}},\ \bibinfo {pages} {67} (\bibinfo {year} {1999})}\BibitemShut
  {NoStop}%
\bibitem [{\citenamefont {Hamann}(2013)}]{Hamann2013}%
  \BibitemOpen
  \bibfield  {author} {\bibinfo {author} {\bibfnamefont {D.~R.}\ \bibnamefont
  {Hamann}},\ }\href@noop {} {\bibfield  {journal} {\bibinfo  {journal} {Phys.
  Rev. B}\ }\textbf {\bibinfo {volume} {88}},\ \bibinfo {pages} {085117}
  (\bibinfo {year} {2013})}\BibitemShut {NoStop}%
\bibitem [{\citenamefont {van Setten}\ \emph {et~al.}(2018)\citenamefont {van
  Setten}, \citenamefont {Giantomassi}, \citenamefont {Bousquet}, \citenamefont
  {Verstraete}, \citenamefont {Hamann}, \citenamefont {Gonze},\ and\
  \citenamefont {Rignanese}}]{Setten2018}%
  \BibitemOpen
  \bibfield  {author} {\bibinfo {author} {\bibfnamefont {M.}~\bibnamefont {van
  Setten}}, \bibinfo {author} {\bibfnamefont {M.}~\bibnamefont {Giantomassi}},
  \bibinfo {author} {\bibfnamefont {E.}~\bibnamefont {Bousquet}}, \bibinfo
  {author} {\bibfnamefont {M.}~\bibnamefont {Verstraete}}, \bibinfo {author}
  {\bibfnamefont {D.}~\bibnamefont {Hamann}}, \bibinfo {author} {\bibfnamefont
  {X.}~\bibnamefont {Gonze}}, \ and\ \bibinfo {author} {\bibfnamefont {G.-M.}\
  \bibnamefont {Rignanese}},\ }\href@noop {} {\bibfield  {journal} {\bibinfo
  {journal} {Computer Physics Communications}\ }\textbf {\bibinfo {volume}
  {226}},\ \bibinfo {pages} {39} (\bibinfo {year} {2018})}\BibitemShut
  {NoStop}%
\bibitem [{\citenamefont {Grinberg}\ \emph {et~al.}(2000)\citenamefont
  {Grinberg}, \citenamefont {Ramer},\ and\ \citenamefont
  {Rappe}}]{Grinberg2000}%
  \BibitemOpen
  \bibfield  {author} {\bibinfo {author} {\bibfnamefont {I.}~\bibnamefont
  {Grinberg}}, \bibinfo {author} {\bibfnamefont {J.~N.}\ \bibnamefont {Ramer}},
  \ and\ \bibinfo {author} {\bibfnamefont {A.~M.}\ \bibnamefont {Rappe}},\
  }\href@noop {} {\bibfield  {journal} {\bibinfo  {journal} {Phys. Rev. B}\
  }\textbf {\bibinfo {volume} {62}},\ \bibinfo {pages} {2311} (\bibinfo {year}
  {2000})}\BibitemShut {NoStop}%
\bibitem [{\citenamefont {Rappe}\ \emph {et~al.}(1990)\citenamefont {Rappe},
  \citenamefont {Rabe}, \citenamefont {Kaxiras},\ and\ \citenamefont
  {Joannopoulos}}]{Rappe1990}%
  \BibitemOpen
  \bibfield  {author} {\bibinfo {author} {\bibfnamefont {A.~M.}\ \bibnamefont
  {Rappe}}, \bibinfo {author} {\bibfnamefont {K.~M.}\ \bibnamefont {Rabe}},
  \bibinfo {author} {\bibfnamefont {E.}~\bibnamefont {Kaxiras}}, \ and\
  \bibinfo {author} {\bibfnamefont {J.~D.}\ \bibnamefont {Joannopoulos}},\
  }\href@noop {} {\bibfield  {journal} {\bibinfo  {journal} {Phys. Rev. B}\
  }\textbf {\bibinfo {volume} {41}},\ \bibinfo {pages} {1227} (\bibinfo {year}
  {1990})}\BibitemShut {NoStop}%
\bibitem [{\citenamefont {Perdew}\ \emph {et~al.}(1996)\citenamefont {Perdew},
  \citenamefont {Burke},\ and\ \citenamefont {Ernzerhof}}]{Perdew1996}%
  \BibitemOpen
  \bibfield  {author} {\bibinfo {author} {\bibfnamefont {J.~P.}\ \bibnamefont
  {Perdew}}, \bibinfo {author} {\bibfnamefont {K.}~\bibnamefont {Burke}}, \
  and\ \bibinfo {author} {\bibfnamefont {M.}~\bibnamefont {Ernzerhof}},\
  }\href@noop {} {\bibfield  {journal} {\bibinfo  {journal} {Phys. Rev. Lett.}\
  }\textbf {\bibinfo {volume} {77}},\ \bibinfo {pages} {2865} (\bibinfo {year}
  {1996})}\BibitemShut {NoStop}%
\bibitem [{\citenamefont {Verstraete}\ \emph {et~al.}(2008)\citenamefont
  {Verstraete}, \citenamefont {Torrent}, \citenamefont {Jollet}, \citenamefont
  {Zerah},\ and\ \citenamefont {Gonze}}]{Verstraete2008}%
  \BibitemOpen
  \bibfield  {author} {\bibinfo {author} {\bibfnamefont {M.~J.}\ \bibnamefont
  {Verstraete}}, \bibinfo {author} {\bibfnamefont {M.}~\bibnamefont {Torrent}},
  \bibinfo {author} {\bibfnamefont {F.}~\bibnamefont {Jollet}}, \bibinfo
  {author} {\bibfnamefont {G.}~\bibnamefont {Zerah}}, \ and\ \bibinfo {author}
  {\bibfnamefont {X.}~\bibnamefont {Gonze}},\ }\href@noop {} {\bibfield
  {journal} {\bibinfo  {journal} {Phys. Rev. B}\ }\textbf {\bibinfo {volume}
  {78}},\ \bibinfo {pages} {045119} (\bibinfo {year} {2008})}\BibitemShut
  {NoStop}%
\bibitem [{\citenamefont {Diaz-Sanchez}\ \emph
  {et~al.}(2007{\natexlab{a}})\citenamefont {Diaz-Sanchez}, \citenamefont
  {Romero},\ and\ \citenamefont {Gonze}}]{Diaz2007}%
  \BibitemOpen
  \bibfield  {author} {\bibinfo {author} {\bibfnamefont {L.~E.}\ \bibnamefont
  {Diaz-Sanchez}}, \bibinfo {author} {\bibfnamefont {A.~H.}\ \bibnamefont
  {Romero}}, \ and\ \bibinfo {author} {\bibfnamefont {X.}~\bibnamefont
  {Gonze}},\ }\href@noop {} {\bibfield  {journal} {\bibinfo  {journal} {Phys.
  Rev. B}\ }\textbf {\bibinfo {volume} {76}},\ \bibinfo {pages} {104302}
  (\bibinfo {year} {2007}{\natexlab{a}})}\BibitemShut {NoStop}%
\bibitem [{\citenamefont {Diaz-Sanchez}\ \emph
  {et~al.}(2007{\natexlab{b}})\citenamefont {Diaz-Sanchez}, \citenamefont
  {Romero}, \citenamefont {Cardona}, \citenamefont {Kremer},\ and\
  \citenamefont {Gonze}}]{Diaz2007b}%
  \BibitemOpen
  \bibfield  {author} {\bibinfo {author} {\bibfnamefont {L.~E.}\ \bibnamefont
  {Diaz-Sanchez}}, \bibinfo {author} {\bibfnamefont {A.}~\bibnamefont
  {Romero}}, \bibinfo {author} {\bibfnamefont {M.}~\bibnamefont {Cardona}},
  \bibinfo {author} {\bibfnamefont {R.~K.}\ \bibnamefont {Kremer}}, \ and\
  \bibinfo {author} {\bibfnamefont {X.}~\bibnamefont {Gonze}},\ }\href@noop {}
  {\bibfield  {journal} {\bibinfo  {journal} {Phys. Rev. Lett.}\ }\textbf
  {\bibinfo {volume} {99}},\ \bibinfo {pages} {165504} (\bibinfo {year}
  {2007}{\natexlab{b}})}\BibitemShut {NoStop}%
\bibitem [{\citenamefont {Aryasetiawan}\ and\ \citenamefont
  {Gunnarsson}(1998)}]{Aryasetiawan1998}%
  \BibitemOpen
  \bibfield  {author} {\bibinfo {author} {\bibfnamefont {F.}~\bibnamefont
  {Aryasetiawan}}\ and\ \bibinfo {author} {\bibfnamefont {O.}~\bibnamefont
  {Gunnarsson}},\ }\href@noop {} {\bibfield  {journal} {\bibinfo  {journal}
  {Repts. on Prog. in Phys.}\ }\textbf {\bibinfo {volume} {61}},\ \bibinfo
  {pages} {237} (\bibinfo {year} {1998})}\BibitemShut {NoStop}%
\bibitem [{\citenamefont {Perdew}(1985)}]{Perdew1985}%
  \BibitemOpen
  \bibfield  {author} {\bibinfo {author} {\bibfnamefont {J.~P.}\ \bibnamefont
  {Perdew}},\ }\href@noop {} {\bibfield  {journal} {\bibinfo  {journal} {Inter.
  J. of Quant. Chem.}\ }\textbf {\bibinfo {volume} {28}},\ \bibinfo {pages}
  {497} (\bibinfo {year} {1985})}\BibitemShut {NoStop}%
\bibitem [{\citenamefont {Troeye}\ \emph {et~al.}(2016)\citenamefont {Troeye},
  \citenamefont {Torrent},\ and\ \citenamefont {Gonze}}]{Troeye2016}%
  \BibitemOpen
  \bibfield  {author} {\bibinfo {author} {\bibfnamefont {B.~V.}\ \bibnamefont
  {Troeye}}, \bibinfo {author} {\bibfnamefont {M.}~\bibnamefont {Torrent}}, \
  and\ \bibinfo {author} {\bibfnamefont {X.}~\bibnamefont {Gonze}},\
  }\href@noop {} {\bibfield  {journal} {\bibinfo  {journal} {Phys. Rev. B}\
  }\textbf {\bibinfo {volume} {93}},\ \bibinfo {pages} {144304} (\bibinfo
  {year} {2016})}\BibitemShut {NoStop}%
\bibitem [{\citenamefont {Sabatini}\ \emph {et~al.}(2016)\citenamefont
  {Sabatini}, \citenamefont {Kucukbenli}, \citenamefont {Pham},\ and\
  \citenamefont {de~Gironcoli}}]{Sabatini2016}%
  \BibitemOpen
  \bibfield  {author} {\bibinfo {author} {\bibfnamefont {R.}~\bibnamefont
  {Sabatini}}, \bibinfo {author} {\bibfnamefont {E.}~\bibnamefont
  {Kucukbenli}}, \bibinfo {author} {\bibfnamefont {C.~H.}\ \bibnamefont
  {Pham}}, \ and\ \bibinfo {author} {\bibfnamefont {S.}~\bibnamefont
  {de~Gironcoli}},\ }\href@noop {} {\bibfield  {journal} {\bibinfo  {journal}
  {Phys. Rev. B}\ }\textbf {\bibinfo {volume} {93}},\ \bibinfo {pages} {235120}
  (\bibinfo {year} {2016})}\BibitemShut {NoStop}%
\bibitem [{\citenamefont {Tkatchenko}\ and\ \citenamefont
  {Scheffler}(2009)}]{Tkatchenko2009}%
  \BibitemOpen
  \bibfield  {author} {\bibinfo {author} {\bibfnamefont {A.}~\bibnamefont
  {Tkatchenko}}\ and\ \bibinfo {author} {\bibfnamefont {M.}~\bibnamefont
  {Scheffler}},\ }\href@noop {} {\bibfield  {journal} {\bibinfo  {journal}
  {Phys. Rev. Lett.}\ }\textbf {\bibinfo {volume} {102}},\ \bibinfo {pages}
  {073005} (\bibinfo {year} {2009})}\BibitemShut {NoStop}%
\bibitem [{\citenamefont {Rydberg}\ \emph {et~al.}(2003)\citenamefont
  {Rydberg}, \citenamefont {Dion}, \citenamefont {Jacobson}, \citenamefont
  {Schroder}, \citenamefont {Hyldgaard}, \citenamefont {Simak}, \citenamefont
  {Langerth},\ and\ \citenamefont {Lundqvist}}]{Rydberg2003}%
  \BibitemOpen
  \bibfield  {author} {\bibinfo {author} {\bibfnamefont {H.}~\bibnamefont
  {Rydberg}}, \bibinfo {author} {\bibfnamefont {M.}~\bibnamefont {Dion}},
  \bibinfo {author} {\bibfnamefont {N.}~\bibnamefont {Jacobson}}, \bibinfo
  {author} {\bibfnamefont {E.}~\bibnamefont {Schroder}}, \bibinfo {author}
  {\bibfnamefont {P.}~\bibnamefont {Hyldgaard}}, \bibinfo {author}
  {\bibfnamefont {S.~I.}\ \bibnamefont {Simak}}, \bibinfo {author}
  {\bibfnamefont {D.~C.}\ \bibnamefont {Langerth}}, \ and\ \bibinfo {author}
  {\bibfnamefont {B.~I.}\ \bibnamefont {Lundqvist}},\ }\href@noop {} {\bibfield
   {journal} {\bibinfo  {journal} {Phys. Rev. Lett.}\ }\textbf {\bibinfo
  {volume} {91}},\ \bibinfo {pages} {126402} (\bibinfo {year}
  {2003})}\BibitemShut {NoStop}%
\bibitem [{\citenamefont {Dion}\ \emph {et~al.}(2004)\citenamefont {Dion},
  \citenamefont {Rydberg}, \citenamefont {Schroder}, \citenamefont {Langreth},\
  and\ \citenamefont {Lundqvist}}]{Dion2004}%
  \BibitemOpen
  \bibfield  {author} {\bibinfo {author} {\bibfnamefont {M.}~\bibnamefont
  {Dion}}, \bibinfo {author} {\bibfnamefont {H.}~\bibnamefont {Rydberg}},
  \bibinfo {author} {\bibfnamefont {E.}~\bibnamefont {Schroder}}, \bibinfo
  {author} {\bibfnamefont {D.~C.}\ \bibnamefont {Langreth}}, \ and\ \bibinfo
  {author} {\bibfnamefont {B.~I.}\ \bibnamefont {Lundqvist}},\ }\href@noop {}
  {\bibfield  {journal} {\bibinfo  {journal} {Phys. Rev. Lett.}\ }\textbf
  {\bibinfo {volume} {92}},\ \bibinfo {pages} {246401} (\bibinfo {year}
  {2004})}\BibitemShut {NoStop}%
\bibitem [{\citenamefont {Lee}\ \emph {et~al.}(2010)\citenamefont {Lee},
  \citenamefont {Murray}, \citenamefont {Kong}, \citenamefont {Lundqvist},\
  and\ \citenamefont {Langreth}}]{Lee2010}%
  \BibitemOpen
  \bibfield  {author} {\bibinfo {author} {\bibfnamefont {K.}~\bibnamefont
  {Lee}}, \bibinfo {author} {\bibfnamefont {E.~D.}\ \bibnamefont {Murray}},
  \bibinfo {author} {\bibfnamefont {L.}~\bibnamefont {Kong}}, \bibinfo {author}
  {\bibfnamefont {B.~I.}\ \bibnamefont {Lundqvist}}, \ and\ \bibinfo {author}
  {\bibfnamefont {D.~C.}\ \bibnamefont {Langreth}},\ }\href@noop {} {\bibfield
  {journal} {\bibinfo  {journal} {Phys. Rev. B}\ }\textbf {\bibinfo {volume}
  {82}},\ \bibinfo {pages} {081101} (\bibinfo {year} {2010})}\BibitemShut
  {NoStop}%
\bibitem [{\citenamefont {Silvestrelli}(2008)}]{Silvestrelli2008}%
  \BibitemOpen
  \bibfield  {author} {\bibinfo {author} {\bibfnamefont {P.~L.}\ \bibnamefont
  {Silvestrelli}},\ }\href@noop {} {\bibfield  {journal} {\bibinfo  {journal}
  {Phys. Rev. Lett.}\ }\textbf {\bibinfo {volume} {100}},\ \bibinfo {pages}
  {053002} (\bibinfo {year} {2008})}\BibitemShut {NoStop}%
\bibitem [{\citenamefont {Ambrosetti}\ and\ \citenamefont
  {Silvestrelli}(2012)}]{Ambrosetti2012}%
  \BibitemOpen
  \bibfield  {author} {\bibinfo {author} {\bibfnamefont {A.}~\bibnamefont
  {Ambrosetti}}\ and\ \bibinfo {author} {\bibfnamefont {P.~L.}\ \bibnamefont
  {Silvestrelli}},\ }\href@noop {} {\bibfield  {journal} {\bibinfo  {journal}
  {Phys. Rev. B}\ }\textbf {\bibinfo {volume} {85}},\ \bibinfo {pages} {073101}
  (\bibinfo {year} {2012})}\BibitemShut {NoStop}%
\bibitem [{\citenamefont {Furche}(2001)}]{Furche2001}%
  \BibitemOpen
  \bibfield  {author} {\bibinfo {author} {\bibfnamefont {F.}~\bibnamefont
  {Furche}},\ }\href@noop {} {\bibfield  {journal} {\bibinfo  {journal} {Phys.
  Rev. B}\ }\textbf {\bibinfo {volume} {64}},\ \bibinfo {pages} {095120}
  (\bibinfo {year} {2001})}\BibitemShut {NoStop}%
\bibitem [{\citenamefont {Fuchs}\ and\ \citenamefont
  {Gonze}(2002)}]{Fuchs2002}%
  \BibitemOpen
  \bibfield  {author} {\bibinfo {author} {\bibfnamefont {M.}~\bibnamefont
  {Fuchs}}\ and\ \bibinfo {author} {\bibfnamefont {X.}~\bibnamefont {Gonze}},\
  }\href@noop {} {\bibfield  {journal} {\bibinfo  {journal} {Phys. Rev. B}\
  }\textbf {\bibinfo {volume} {65}},\ \bibinfo {pages} {253109} (\bibinfo
  {year} {2002})}\BibitemShut {NoStop}%
\bibitem [{\citenamefont {Ren}\ \emph {et~al.}(2012)\citenamefont {Ren},
  \citenamefont {Rinke}, \citenamefont {Joas},\ and\ \citenamefont
  {Scheffler}}]{Ren2012}%
  \BibitemOpen
  \bibfield  {author} {\bibinfo {author} {\bibfnamefont {X.}~\bibnamefont
  {Ren}}, \bibinfo {author} {\bibfnamefont {P.}~\bibnamefont {Rinke}}, \bibinfo
  {author} {\bibfnamefont {C.}~\bibnamefont {Joas}}, \ and\ \bibinfo {author}
  {\bibfnamefont {M.}~\bibnamefont {Scheffler}},\ }\href@noop {} {\bibfield
  {journal} {\bibinfo  {journal} {J. Mat. Sci}\ }\textbf {\bibinfo {volume}
  {47}},\ \bibinfo {pages} {7447} (\bibinfo {year} {2012})}\BibitemShut
  {NoStop}%
\bibitem [{\citenamefont {Tawfik}\ \emph {et~al.}(2018)\citenamefont {Tawfik},
  \citenamefont {Gould}, \citenamefont {Stampfl},\ and\ \citenamefont
  {Ford}}]{Tawfik2018}%
  \BibitemOpen
  \bibfield  {author} {\bibinfo {author} {\bibfnamefont {S.~A.}\ \bibnamefont
  {Tawfik}}, \bibinfo {author} {\bibfnamefont {T.}~\bibnamefont {Gould}},
  \bibinfo {author} {\bibfnamefont {C.}~\bibnamefont {Stampfl}}, \ and\
  \bibinfo {author} {\bibfnamefont {M.~J.}\ \bibnamefont {Ford}},\ }\href@noop
  {} {\bibfield  {journal} {\bibinfo  {journal} {Phys. Rev. Materials}\
  }\textbf {\bibinfo {volume} {2}},\ \bibinfo {pages} {034005} (\bibinfo {year}
  {2018})}\BibitemShut {NoStop}%
\bibitem [{\citenamefont {Gruneisen}(1912)}]{Gruneisen1912}%
  \BibitemOpen
  \bibfield  {author} {\bibinfo {author} {\bibfnamefont {E.}~\bibnamefont
  {Gruneisen}},\ }\href@noop {} {\bibfield  {journal} {\bibinfo  {journal}
  {Annalen der Physik}\ }\textbf {\bibinfo {volume} {344}},\ \bibinfo {pages}
  {257} (\bibinfo {year} {1912})}\BibitemShut {NoStop}%
\bibitem [{\citenamefont {Sugai}\ and\ \citenamefont {Ueda}(1982)}]{Sugai1982}%
  \BibitemOpen
  \bibfield  {author} {\bibinfo {author} {\bibfnamefont {S.}~\bibnamefont
  {Sugai}}\ and\ \bibinfo {author} {\bibfnamefont {T.}~\bibnamefont {Ueda}},\
  }\href@noop {} {\bibfield  {journal} {\bibinfo  {journal} {Phys. Rev. B}\
  }\textbf {\bibinfo {volume} {26}},\ \bibinfo {pages} {6554} (\bibinfo {year}
  {1982})}\BibitemShut {NoStop}%
\bibitem [{\citenamefont {Van~Troeye}\ \emph {et~al.}(2017)\citenamefont
  {Van~Troeye}, \citenamefont {van Setten}, \citenamefont {Giantomassi},
  \citenamefont {Torrent}, \citenamefont {Rignanese},\ and\ \citenamefont
  {Gonze}}]{Troeye2017}%
  \BibitemOpen
  \bibfield  {author} {\bibinfo {author} {\bibfnamefont {B.}~\bibnamefont
  {Van~Troeye}}, \bibinfo {author} {\bibfnamefont {M.~J.}\ \bibnamefont {van
  Setten}}, \bibinfo {author} {\bibfnamefont {M.}~\bibnamefont {Giantomassi}},
  \bibinfo {author} {\bibfnamefont {M.}~\bibnamefont {Torrent}}, \bibinfo
  {author} {\bibfnamefont {G.-M.}\ \bibnamefont {Rignanese}}, \ and\ \bibinfo
  {author} {\bibfnamefont {X.}~\bibnamefont {Gonze}},\ }\href@noop {}
  {\bibfield  {journal} {\bibinfo  {journal} {Phys. Rev. B}\ }\textbf {\bibinfo
  {volume} {95}},\ \bibinfo {pages} {024112} (\bibinfo {year}
  {2017})}\BibitemShut {NoStop}%
\bibitem [{\citenamefont {Dickinson}\ and\ \citenamefont
  {Pauling}(1923)}]{Dickinson1923}%
  \BibitemOpen
  \bibfield  {author} {\bibinfo {author} {\bibfnamefont {R.}~\bibnamefont
  {Dickinson}}\ and\ \bibinfo {author} {\bibfnamefont {L.}~\bibnamefont
  {Pauling}},\ }\href@noop {} {\bibfield  {journal} {\bibinfo  {journal} {J.
  Amer. Chem. Soc.}\ }\textbf {\bibinfo {volume} {45}},\ \bibinfo {pages}
  {1466} (\bibinfo {year} {1923})}\BibitemShut {NoStop}%
\bibitem [{\citenamefont {Greenaway}\ and\ \citenamefont
  {Nitzche}(1965)}]{Greenaway1965}%
  \BibitemOpen
  \bibfield  {author} {\bibinfo {author} {\bibfnamefont {D.~L.}\ \bibnamefont
  {Greenaway}}\ and\ \bibinfo {author} {\bibfnamefont {R.}~\bibnamefont
  {Nitzche}},\ }\href@noop {} {\bibfield  {journal} {\bibinfo  {journal} {J.
  Phys. Chem. Sol.}\ }\textbf {\bibinfo {volume} {26}},\ \bibinfo {pages}
  {1445} (\bibinfo {year} {1965})}\BibitemShut {NoStop}%
\bibitem [{\citenamefont {Wildervanck}\ and\ \citenamefont
  {Jellinek}(1971)}]{Wildervanck1970}%
  \BibitemOpen
  \bibfield  {author} {\bibinfo {author} {\bibfnamefont {J.~C.}\ \bibnamefont
  {Wildervanck}}\ and\ \bibinfo {author} {\bibfnamefont {F.}~\bibnamefont
  {Jellinek}},\ }\href@noop {} {\bibfield  {journal} {\bibinfo  {journal} {J.
  of less-common metals}\ }\textbf {\bibinfo {volume} {24}},\ \bibinfo {pages}
  {73} (\bibinfo {year} {1971})}\BibitemShut {NoStop}%
\bibitem [{\citenamefont {Kam}\ and\ \citenamefont
  {Parkinson}(1982)}]{Kam1982}%
  \BibitemOpen
  \bibfield  {author} {\bibinfo {author} {\bibfnamefont {K.~K.}\ \bibnamefont
  {Kam}}\ and\ \bibinfo {author} {\bibfnamefont {B.~A.}\ \bibnamefont
  {Parkinson}},\ }\href@noop {} {\bibfield  {journal} {\bibinfo  {journal} {J.
  Phys. Chem.}\ }\textbf {\bibinfo {volume} {86}},\ \bibinfo {pages} {463}
  (\bibinfo {year} {1982})}\BibitemShut {NoStop}%
\bibitem [{\citenamefont {Moustafa}\ \emph {et~al.}(2009)\citenamefont
  {Moustafa}, \citenamefont {Zandt}, \citenamefont {Janowitz},\ and\
  \citenamefont {Manzke}}]{Moustafa2009}%
  \BibitemOpen
  \bibfield  {author} {\bibinfo {author} {\bibfnamefont {M.}~\bibnamefont
  {Moustafa}}, \bibinfo {author} {\bibfnamefont {T.}~\bibnamefont {Zandt}},
  \bibinfo {author} {\bibfnamefont {C.}~\bibnamefont {Janowitz}}, \ and\
  \bibinfo {author} {\bibfnamefont {R.}~\bibnamefont {Manzke}},\ }\href@noop {}
  {\bibfield  {journal} {\bibinfo  {journal} {Phys. Rev. B}\ }\textbf {\bibinfo
  {volume} {80}},\ \bibinfo {pages} {035206} (\bibinfo {year}
  {2009})}\BibitemShut {NoStop}%
\bibitem [{\citenamefont {Tongay}\ \emph {et~al.}(2014)\citenamefont {Tongay},
  \citenamefont {Sahin}, \citenamefont {Ko}, \citenamefont {Luce},
  \citenamefont {Fan}, \citenamefont {Liu}, \citenamefont {Zhou}, \citenamefont
  {Huang}, \citenamefont {Ho}, \citenamefont {Yan}, \citenamefont {Ogletree},
  \citenamefont {Aloni}, \citenamefont {Ji}, \citenamefont {Li}, \citenamefont
  {Li}, \citenamefont {Peeters},\ and\ \citenamefont {Wu}}]{Tongay2014}%
  \BibitemOpen
  \bibfield  {author} {\bibinfo {author} {\bibfnamefont {S.}~\bibnamefont
  {Tongay}}, \bibinfo {author} {\bibfnamefont {H.}~\bibnamefont {Sahin}},
  \bibinfo {author} {\bibfnamefont {C.}~\bibnamefont {Ko}}, \bibinfo {author}
  {\bibfnamefont {A.}~\bibnamefont {Luce}}, \bibinfo {author} {\bibfnamefont
  {W.}~\bibnamefont {Fan}}, \bibinfo {author} {\bibfnamefont {K.}~\bibnamefont
  {Liu}}, \bibinfo {author} {\bibfnamefont {J.}~\bibnamefont {Zhou}}, \bibinfo
  {author} {\bibfnamefont {Y.-S.}\ \bibnamefont {Huang}}, \bibinfo {author}
  {\bibfnamefont {C.-H.}\ \bibnamefont {Ho}}, \bibinfo {author} {\bibfnamefont
  {J.}~\bibnamefont {Yan}}, \bibinfo {author} {\bibfnamefont {D.~F.}\
  \bibnamefont {Ogletree}}, \bibinfo {author} {\bibfnamefont {S.}~\bibnamefont
  {Aloni}}, \bibinfo {author} {\bibfnamefont {J.}~\bibnamefont {Ji}}, \bibinfo
  {author} {\bibfnamefont {S.}~\bibnamefont {Li}}, \bibinfo {author}
  {\bibfnamefont {J.}~\bibnamefont {Li}}, \bibinfo {author} {\bibfnamefont
  {F.~M.}\ \bibnamefont {Peeters}}, \ and\ \bibinfo {author} {\bibfnamefont
  {J.}~\bibnamefont {Wu}},\ }\href {\doibase
  doi:http://dx.doi.org/10.1038/ncomms4252} {\bibfield  {journal} {\bibinfo
  {journal} {Nature Comm.}\ }\textbf {\bibinfo {volume} {5}},\ \bibinfo {pages}
  {3252} (\bibinfo {year} {2014})}\BibitemShut {NoStop}%
\bibitem [{\citenamefont {Feldman}(1976)}]{Feldman1976}%
  \BibitemOpen
  \bibfield  {author} {\bibinfo {author} {\bibfnamefont {J.}~\bibnamefont
  {Feldman}},\ }\href@noop {} {\bibfield  {journal} {\bibinfo  {journal} {J.
  Phys. Chem. Sol.}\ }\textbf {\bibinfo {volume} {37}},\ \bibinfo {pages}
  {1141} (\bibinfo {year} {1976})}\BibitemShut {NoStop}%
\bibitem [{\citenamefont {Aksoy}\ \emph {et~al.}(2006)\citenamefont {Aksoy},
  \citenamefont {Ma}, \citenamefont {Selvi}, \citenamefont {Chyu},
  \citenamefont {Ertas},\ and\ \citenamefont {White}}]{Aksoy2006}%
  \BibitemOpen
  \bibfield  {author} {\bibinfo {author} {\bibfnamefont {R.}~\bibnamefont
  {Aksoy}}, \bibinfo {author} {\bibfnamefont {Y.}~\bibnamefont {Ma}}, \bibinfo
  {author} {\bibfnamefont {E.}~\bibnamefont {Selvi}}, \bibinfo {author}
  {\bibfnamefont {M.~C.}\ \bibnamefont {Chyu}}, \bibinfo {author}
  {\bibfnamefont {A.}~\bibnamefont {Ertas}}, \ and\ \bibinfo {author}
  {\bibfnamefont {A.}~\bibnamefont {White}},\ }\href@noop {} {\bibfield
  {journal} {\bibinfo  {journal} {J. Phys. Chem. Sol.}\ }\textbf {\bibinfo
  {volume} {67}},\ \bibinfo {pages} {1914} (\bibinfo {year}
  {2006})}\BibitemShut {NoStop}%
\bibitem [{\citenamefont {Nayak}\ \emph {et~al.}(2014)\citenamefont {Nayak},
  \citenamefont {Bhattacharyya}, \citenamefont {Zhu}, \citenamefont {Liu},
  \citenamefont {Wu}, \citenamefont {Pandey}, \citenamefont {Jin},
  \citenamefont {Singh}, \citenamefont {Akinwande},\ and\ \citenamefont
  {Lin}}]{Nayak2014}%
  \BibitemOpen
  \bibfield  {author} {\bibinfo {author} {\bibfnamefont {A.~P.}\ \bibnamefont
  {Nayak}}, \bibinfo {author} {\bibfnamefont {S.}~\bibnamefont
  {Bhattacharyya}}, \bibinfo {author} {\bibfnamefont {J.}~\bibnamefont {Zhu}},
  \bibinfo {author} {\bibfnamefont {J.}~\bibnamefont {Liu}}, \bibinfo {author}
  {\bibfnamefont {X.}~\bibnamefont {Wu}}, \bibinfo {author} {\bibfnamefont
  {T.}~\bibnamefont {Pandey}}, \bibinfo {author} {\bibfnamefont
  {C.}~\bibnamefont {Jin}}, \bibinfo {author} {\bibfnamefont {A.~K.}\
  \bibnamefont {Singh}}, \bibinfo {author} {\bibfnamefont {D.}~\bibnamefont
  {Akinwande}}, \ and\ \bibinfo {author} {\bibfnamefont {J.-F.}\ \bibnamefont
  {Lin}},\ }\href@noop {} {\bibfield  {journal} {\bibinfo  {journal} {Nature
  Comm.}\ }\textbf {\bibinfo {volume} {5}},\ \bibinfo {pages} {3731} (\bibinfo
  {year} {2014})}\BibitemShut {NoStop}%
\bibitem [{\citenamefont {Wieting}\ and\ \citenamefont
  {Verble}(1971)}]{Wieting1971}%
  \BibitemOpen
  \bibfield  {author} {\bibinfo {author} {\bibfnamefont {T.~J.}\ \bibnamefont
  {Wieting}}\ and\ \bibinfo {author} {\bibfnamefont {J.~L.}\ \bibnamefont
  {Verble}},\ }\href@noop {} {\bibfield  {journal} {\bibinfo  {journal} {Phys.
  Rev. B}\ }\textbf {\bibinfo {volume} {3}},\ \bibinfo {pages} {4286} (\bibinfo
  {year} {1971})}\BibitemShut {NoStop}%
\bibitem [{\citenamefont {Iwasaki}\ \emph {et~al.}(1982)\citenamefont
  {Iwasaki}, \citenamefont {Kuroda},\ and\ \citenamefont
  {Nishina}}]{Iwasaki1982}%
  \BibitemOpen
  \bibfield  {author} {\bibinfo {author} {\bibfnamefont {T.}~\bibnamefont
  {Iwasaki}}, \bibinfo {author} {\bibfnamefont {N.}~\bibnamefont {Kuroda}}, \
  and\ \bibinfo {author} {\bibfnamefont {Y.}~\bibnamefont {Nishina}},\
  }\href@noop {} {\bibfield  {journal} {\bibinfo  {journal} {J. Phys. Soc.
  Jpn.}\ }\textbf {\bibinfo {volume} {51}},\ \bibinfo {pages} {2233} (\bibinfo
  {year} {1982})}\BibitemShut {NoStop}%
\bibitem [{\citenamefont {Lucovsky}\ \emph {et~al.}(1973)\citenamefont
  {Lucovsky}, \citenamefont {White}, \citenamefont {Benda},\ and\ \citenamefont
  {Revelli}}]{Lucovsky1973}%
  \BibitemOpen
  \bibfield  {author} {\bibinfo {author} {\bibfnamefont {G.}~\bibnamefont
  {Lucovsky}}, \bibinfo {author} {\bibfnamefont {R.~M.}\ \bibnamefont {White}},
  \bibinfo {author} {\bibfnamefont {J.~A.}\ \bibnamefont {Benda}}, \ and\
  \bibinfo {author} {\bibfnamefont {J.~F.}\ \bibnamefont {Revelli}},\
  }\href@noop {} {\bibfield  {journal} {\bibinfo  {journal} {Phys. Rev. B}\
  }\textbf {\bibinfo {volume} {7}},\ \bibinfo {pages} {3859} (\bibinfo {year}
  {1973})}\BibitemShut {NoStop}%
\bibitem [{\citenamefont {Park}\ \emph {et~al.}(1996)\citenamefont {Park},
  \citenamefont {Richards-Babb}, \citenamefont {Hess}, \citenamefont {Weiss},\
  and\ \citenamefont {Klier}}]{Park1996}%
  \BibitemOpen
  \bibfield  {author} {\bibinfo {author} {\bibfnamefont {K.~T.}\ \bibnamefont
  {Park}}, \bibinfo {author} {\bibfnamefont {M.}~\bibnamefont {Richards-Babb}},
  \bibinfo {author} {\bibfnamefont {J.~S.}\ \bibnamefont {Hess}}, \bibinfo
  {author} {\bibfnamefont {J.}~\bibnamefont {Weiss}}, \ and\ \bibinfo {author}
  {\bibfnamefont {K.}~\bibnamefont {Klier}},\ }\href@noop {} {\bibfield
  {journal} {\bibinfo  {journal} {Phys. Rev. B}\ }\textbf {\bibinfo {volume}
  {54}},\ \bibinfo {pages} {5471} (\bibinfo {year} {1996})}\BibitemShut
  {NoStop}%
\bibitem [{\citenamefont {Ho}\ \emph {et~al.}(1997)\citenamefont {Ho},
  \citenamefont {Liao}, \citenamefont {Huang}, \citenamefont {Yang},\ and\
  \citenamefont {Tong}}]{Ho1997}%
  \BibitemOpen
  \bibfield  {author} {\bibinfo {author} {\bibfnamefont {C.~H.}\ \bibnamefont
  {Ho}}, \bibinfo {author} {\bibfnamefont {P.~C.}\ \bibnamefont {Liao}},
  \bibinfo {author} {\bibfnamefont {Y.~S.}\ \bibnamefont {Huang}}, \bibinfo
  {author} {\bibfnamefont {T.~R.}\ \bibnamefont {Yang}}, \ and\ \bibinfo
  {author} {\bibfnamefont {K.~K.}\ \bibnamefont {Tong}},\ }\href@noop {}
  {\bibfield  {journal} {\bibinfo  {journal} {J. Appl. Phys.}\ }\textbf
  {\bibinfo {volume} {81}},\ \bibinfo {pages} {6380} (\bibinfo {year}
  {1997})}\BibitemShut {NoStop}%
\bibitem [{\citenamefont {Brixner}(1962)}]{Brixner1962}%
  \BibitemOpen
  \bibfield  {author} {\bibinfo {author} {\bibfnamefont {L.~H.}\ \bibnamefont
  {Brixner}},\ }\href@noop {} {\bibfield  {journal} {\bibinfo  {journal} {J.
  Inorg. Nucl. Chem.}\ }\textbf {\bibinfo {volume} {24}},\ \bibinfo {pages}
  {257} (\bibinfo {year} {1962})}\BibitemShut {NoStop}%
\bibitem [{\citenamefont {Zhao}\ \emph
  {et~al.}(2015{\natexlab{a}})\citenamefont {Zhao}, \citenamefont {Zhang},
  \citenamefont {Yuan}, \citenamefont {Wang}, \citenamefont {Lin},
  \citenamefont {Zeng}, \citenamefont {Xu}, \citenamefont {Liu}, \citenamefont
  {Solanki}, \citenamefont {Patel}, \citenamefont {Chi}, \citenamefont
  {Hwang},\ and\ \citenamefont {Mao}}]{Zhao2015}%
  \BibitemOpen
  \bibfield  {author} {\bibinfo {author} {\bibfnamefont {Z.}~\bibnamefont
  {Zhao}}, \bibinfo {author} {\bibfnamefont {H.}~\bibnamefont {Zhang}},
  \bibinfo {author} {\bibfnamefont {H.}~\bibnamefont {Yuan}}, \bibinfo {author}
  {\bibfnamefont {S.}~\bibnamefont {Wang}}, \bibinfo {author} {\bibfnamefont
  {Y.}~\bibnamefont {Lin}}, \bibinfo {author} {\bibfnamefont {Q.}~\bibnamefont
  {Zeng}}, \bibinfo {author} {\bibfnamefont {G.}~\bibnamefont {Xu}}, \bibinfo
  {author} {\bibfnamefont {Z.}~\bibnamefont {Liu}}, \bibinfo {author}
  {\bibfnamefont {G.~K.}\ \bibnamefont {Solanki}}, \bibinfo {author}
  {\bibfnamefont {K.~D.}\ \bibnamefont {Patel}}, \bibinfo {author}
  {\bibfnamefont {Y.}~\bibnamefont {Chi}}, \bibinfo {author} {\bibfnamefont
  {H.~Y.}\ \bibnamefont {Hwang}}, \ and\ \bibinfo {author} {\bibfnamefont
  {W.~L.}\ \bibnamefont {Mao}},\ }\href@noop {} {\bibfield  {journal} {\bibinfo
   {journal} {Nat. Comm.}\ }\textbf {\bibinfo {volume} {6}},\ \bibinfo {pages}
  {7312} (\bibinfo {year} {2015}{\natexlab{a}})}\BibitemShut {NoStop}%
\bibitem [{\citenamefont {Berkdemir}\ \emph {et~al.}(2013)\citenamefont
  {Berkdemir}, \citenamefont {Guiterrez}, \citenamefont {Botello-Mendez},
  \citenamefont {Perea-Lopez}, \citenamefont {Elias}, \citenamefont {Chia},
  \citenamefont {Crespi}, \citenamefont {Lopez-Urias}, \citenamefont
  {Charlier}, \citenamefont {Terrones},\ and\ \citenamefont
  {Terrones}}]{Berkdemir2013}%
  \BibitemOpen
  \bibfield  {author} {\bibinfo {author} {\bibfnamefont {A.}~\bibnamefont
  {Berkdemir}}, \bibinfo {author} {\bibfnamefont {H.~R.}\ \bibnamefont
  {Guiterrez}}, \bibinfo {author} {\bibfnamefont {A.~R.}\ \bibnamefont
  {Botello-Mendez}}, \bibinfo {author} {\bibfnamefont {N.}~\bibnamefont
  {Perea-Lopez}}, \bibinfo {author} {\bibfnamefont {A.~L.}\ \bibnamefont
  {Elias}}, \bibinfo {author} {\bibfnamefont {C.-I.}\ \bibnamefont {Chia}},
  \bibinfo {author} {\bibfnamefont {V.~H.}\ \bibnamefont {Crespi}}, \bibinfo
  {author} {\bibfnamefont {F.}~\bibnamefont {Lopez-Urias}}, \bibinfo {author}
  {\bibfnamefont {J.-C.}\ \bibnamefont {Charlier}}, \bibinfo {author}
  {\bibfnamefont {H.}~\bibnamefont {Terrones}}, \ and\ \bibinfo {author}
  {\bibfnamefont {M.}~\bibnamefont {Terrones}},\ }\href@noop {} {\bibfield
  {journal} {\bibinfo  {journal} {Sci. Repts.}\ }\textbf {\bibinfo {volume}
  {3}},\ \bibinfo {pages} {1755} (\bibinfo {year} {2013})}\BibitemShut
  {NoStop}%
\bibitem [{\citenamefont {Sourisseau}\ \emph {et~al.}(1991)\citenamefont
  {Sourisseau}, \citenamefont {Cruege}, \citenamefont {Foussier},\ and\
  \citenamefont {Alba}}]{Sourisseau1991}%
  \BibitemOpen
  \bibfield  {author} {\bibinfo {author} {\bibfnamefont {C.}~\bibnamefont
  {Sourisseau}}, \bibinfo {author} {\bibfnamefont {F.}~\bibnamefont {Cruege}},
  \bibinfo {author} {\bibfnamefont {M.}~\bibnamefont {Foussier}}, \ and\
  \bibinfo {author} {\bibfnamefont {M.}~\bibnamefont {Alba}},\ }\href@noop {}
  {\bibfield  {journal} {\bibinfo  {journal} {Chem. Phys.}\ }\textbf {\bibinfo
  {volume} {150}},\ \bibinfo {pages} {281} (\bibinfo {year}
  {1991})}\BibitemShut {NoStop}%
\bibitem [{\citenamefont {Liu}\ \emph {et~al.}(2014)\citenamefont {Liu},
  \citenamefont {Yan}, \citenamefont {Chen}, \citenamefont {Fan}, \citenamefont
  {Sun}, \citenamefont {Suh}, \citenamefont {Fu}, \citenamefont {Lee},
  \citenamefont {Zhou}, \citenamefont {Tongay}, \citenamefont {Ji},
  \citenamefont {Neaton},\ and\ \citenamefont {Wu}}]{Liu2014}%
  \BibitemOpen
  \bibfield  {author} {\bibinfo {author} {\bibfnamefont {K.}~\bibnamefont
  {Liu}}, \bibinfo {author} {\bibfnamefont {Q.}~\bibnamefont {Yan}}, \bibinfo
  {author} {\bibfnamefont {M.}~\bibnamefont {Chen}}, \bibinfo {author}
  {\bibfnamefont {W.}~\bibnamefont {Fan}}, \bibinfo {author} {\bibfnamefont
  {Y.}~\bibnamefont {Sun}}, \bibinfo {author} {\bibfnamefont {J.}~\bibnamefont
  {Suh}}, \bibinfo {author} {\bibfnamefont {D.}~\bibnamefont {Fu}}, \bibinfo
  {author} {\bibfnamefont {S.}~\bibnamefont {Lee}}, \bibinfo {author}
  {\bibfnamefont {J.}~\bibnamefont {Zhou}}, \bibinfo {author} {\bibfnamefont
  {S.}~\bibnamefont {Tongay}}, \bibinfo {author} {\bibfnamefont
  {J.}~\bibnamefont {Ji}}, \bibinfo {author} {\bibfnamefont {J.~B.}\
  \bibnamefont {Neaton}}, \ and\ \bibinfo {author} {\bibfnamefont
  {J.}~\bibnamefont {Wu}},\ }\href@noop {} {\bibfield  {journal} {\bibinfo
  {journal} {Nano Lett.}\ }\textbf {\bibinfo {volume} {14}},\ \bibinfo {pages}
  {5097} (\bibinfo {year} {2014})}\BibitemShut {NoStop}%
\bibitem [{\citenamefont {Jellinek}\ \emph {et~al.}(1960)\citenamefont
  {Jellinek}, \citenamefont {Brauer},\ and\ \citenamefont
  {Huller}}]{Jellinek1960}%
  \BibitemOpen
  \bibfield  {author} {\bibinfo {author} {\bibfnamefont {F.}~\bibnamefont
  {Jellinek}}, \bibinfo {author} {\bibfnamefont {G.}~\bibnamefont {Brauer}}, \
  and\ \bibinfo {author} {\bibfnamefont {H.}~\bibnamefont {Huller}},\
  }\href@noop {} {\bibfield  {journal} {\bibinfo  {journal} {Nature}\ }\textbf
  {\bibinfo {volume} {185}},\ \bibinfo {pages} {376} (\bibinfo {year}
  {1960})}\BibitemShut {NoStop}%
\bibitem [{\citenamefont {Selvi}\ \emph {et~al.}(2006)\citenamefont {Selvi},
  \citenamefont {Y.Ma}, \citenamefont {Aksoy}, \citenamefont {Ertas},\ and\
  \citenamefont {White}}]{Selvi2006}%
  \BibitemOpen
  \bibfield  {author} {\bibinfo {author} {\bibfnamefont {E.}~\bibnamefont
  {Selvi}}, \bibinfo {author} {\bibnamefont {Y.Ma}}, \bibinfo {author}
  {\bibfnamefont {R.}~\bibnamefont {Aksoy}}, \bibinfo {author} {\bibfnamefont
  {A.}~\bibnamefont {Ertas}}, \ and\ \bibinfo {author} {\bibfnamefont
  {A.}~\bibnamefont {White}},\ }\href@noop {} {\bibfield  {journal} {\bibinfo
  {journal} {J. of Phys. and Chem of Solids.}\ }\textbf {\bibinfo {volume}
  {67}},\ \bibinfo {pages} {2183} (\bibinfo {year} {2006})}\BibitemShut
  {NoStop}%
\bibitem [{\citenamefont {Zhang}\ \emph {et~al.}(2016)\citenamefont {Zhang},
  \citenamefont {Koutes},\ and\ \citenamefont {Cheung}}]{Zhang2016}%
  \BibitemOpen
  \bibfield  {author} {\bibinfo {author} {\bibfnamefont {R.}~\bibnamefont
  {Zhang}}, \bibinfo {author} {\bibfnamefont {V.}~\bibnamefont {Koutes}}, \
  and\ \bibinfo {author} {\bibfnamefont {R.}~\bibnamefont {Cheung}},\
  }\href@noop {} {\bibfield  {journal} {\bibinfo  {journal} {Appl. Phys.
  Letts.}\ }\textbf {\bibinfo {volume} {108}},\ \bibinfo {pages} {042104}
  (\bibinfo {year} {2016})}\BibitemShut {NoStop}%
\bibitem [{\citenamefont {Scharli}\ and\ \citenamefont
  {Levy}(1986)}]{Scharli1986}%
  \BibitemOpen
  \bibfield  {author} {\bibinfo {author} {\bibfnamefont {M.}~\bibnamefont
  {Scharli}}\ and\ \bibinfo {author} {\bibfnamefont {F.}~\bibnamefont {Levy}},\
  }\href@noop {} {\bibfield  {journal} {\bibinfo  {journal} {Phys. Rev. B}\
  }\textbf {\bibinfo {volume} {33}},\ \bibinfo {pages} {4317} (\bibinfo {year}
  {1986})}\BibitemShut {NoStop}%
\bibitem [{\citenamefont {Hart}\ \emph {et~al.}(2017)\citenamefont {Hart},
  \citenamefont {Webb}, \citenamefont {Dale}, \citenamefont {Bending},
  \citenamefont {Mucha-Kruczynski}, \citenamefont {Wolverson}, \citenamefont
  {Chen}, \citenamefont {Avila},\ and\ \citenamefont {Asensio}}]{Hart2017}%
  \BibitemOpen
  \bibfield  {author} {\bibinfo {author} {\bibfnamefont {L.~S.}\ \bibnamefont
  {Hart}}, \bibinfo {author} {\bibfnamefont {J.~L.}\ \bibnamefont {Webb}},
  \bibinfo {author} {\bibfnamefont {S.}~\bibnamefont {Dale}}, \bibinfo {author}
  {\bibfnamefont {S.~J.}\ \bibnamefont {Bending}}, \bibinfo {author}
  {\bibfnamefont {M.}~\bibnamefont {Mucha-Kruczynski}}, \bibinfo {author}
  {\bibfnamefont {D.}~\bibnamefont {Wolverson}}, \bibinfo {author}
  {\bibfnamefont {C.}~\bibnamefont {Chen}}, \bibinfo {author} {\bibfnamefont
  {J.}~\bibnamefont {Avila}}, \ and\ \bibinfo {author} {\bibfnamefont {M.~C.}\
  \bibnamefont {Asensio}},\ }\href@noop {} {\bibfield  {journal} {\bibinfo
  {journal} {Sci. Rpts.}\ }\textbf {\bibinfo {volume} {7}},\ \bibinfo {pages}
  {5145} (\bibinfo {year} {2017})}\BibitemShut {NoStop}%
\bibitem [{\citenamefont {Nihira}\ and\ \citenamefont
  {Iwata}(2003)}]{Nihira2003}%
  \BibitemOpen
  \bibfield  {author} {\bibinfo {author} {\bibfnamefont {T.}~\bibnamefont
  {Nihira}}\ and\ \bibinfo {author} {\bibfnamefont {T.}~\bibnamefont {Iwata}},\
  }\href@noop {} {\bibfield  {journal} {\bibinfo  {journal} {Phys. Rev. B}\
  }\textbf {\bibinfo {volume} {68}},\ \bibinfo {pages} {134305} (\bibinfo
  {year} {2003})}\BibitemShut {NoStop}%
\bibitem [{\citenamefont {Filippi}\ \emph {et~al.}(1994)\citenamefont
  {Filippi}, \citenamefont {Singh},\ and\ \citenamefont
  {Umrigar}}]{Filippi1994}%
  \BibitemOpen
  \bibfield  {author} {\bibinfo {author} {\bibfnamefont {C.}~\bibnamefont
  {Filippi}}, \bibinfo {author} {\bibfnamefont {D.~J.}\ \bibnamefont {Singh}},
  \ and\ \bibinfo {author} {\bibfnamefont {C.~J.}\ \bibnamefont {Umrigar}},\
  }\href@noop {} {\bibfield  {journal} {\bibinfo  {journal} {Phys. Rev. B}\
  }\textbf {\bibinfo {volume} {50}},\ \bibinfo {pages} {14947} (\bibinfo {year}
  {1994})}\BibitemShut {NoStop}%
\bibitem [{\citenamefont {Nye}(1957)}]{Nye1957}%
  \BibitemOpen
  \bibfield  {author} {\bibinfo {author} {\bibfnamefont {J.~F.}\ \bibnamefont
  {Nye}},\ }\href@noop {} {\emph {\bibinfo {title} {Physical Properties of
  Crystals}}}\ (\bibinfo  {publisher} {Oxford University Press, London},\
  \bibinfo {year} {1957})\BibitemShut {NoStop}%
\bibitem [{\citenamefont {Tromans}(2011)}]{Tromans2011}%
  \BibitemOpen
  \bibfield  {author} {\bibinfo {author} {\bibfnamefont {D.}~\bibnamefont
  {Tromans}},\ }\href@noop {} {\bibfield  {journal} {\bibinfo  {journal}
  {Inter. J. of Res. and Rev. of Appl. Sci.}\ }\textbf {\bibinfo {volume}
  {6}},\ \bibinfo {pages} {462} (\bibinfo {year} {2011})}\BibitemShut {NoStop}%
\bibitem [{\citenamefont {Agnihotri}\ and\ \citenamefont
  {Sehgal}(1972)}]{Agnihotri1972}%
  \BibitemOpen
  \bibfield  {author} {\bibinfo {author} {\bibfnamefont {O.~P.}\ \bibnamefont
  {Agnihotri}}\ and\ \bibinfo {author} {\bibfnamefont {H.~K.}\ \bibnamefont
  {Sehgal}},\ }\href@noop {} {\bibfield  {journal} {\bibinfo  {journal} {Phil.
  Mag.}\ }\textbf {\bibinfo {volume} {26}},\ \bibinfo {pages} {753} (\bibinfo
  {year} {1972})}\BibitemShut {NoStop}%
\bibitem [{\citenamefont {Garg}\ \emph {et~al.}(1973)\citenamefont {Garg},
  \citenamefont {Sehgal},\ and\ \citenamefont {Agnihotri}}]{Garg1973}%
  \BibitemOpen
  \bibfield  {author} {\bibinfo {author} {\bibfnamefont {A.~K.}\ \bibnamefont
  {Garg}}, \bibinfo {author} {\bibfnamefont {H.~K.}\ \bibnamefont {Sehgal}}, \
  and\ \bibinfo {author} {\bibfnamefont {O.~P.}\ \bibnamefont {Agnihotri}},\
  }\href@noop {} {\bibfield  {journal} {\bibinfo  {journal} {Sol. Stat. Comm.}\
  }\textbf {\bibinfo {volume} {12}},\ \bibinfo {pages} {1261} (\bibinfo {year}
  {1973})}\BibitemShut {NoStop}%
\bibitem [{\citenamefont {Lezama}\ \emph {et~al.}(2015)\citenamefont {Lezama},
  \citenamefont {Arora}, \citenamefont {Ubaldini}, \citenamefont {Barreteau},
  \citenamefont {Gianini}, \citenamefont {Potemski},\ and\ \citenamefont
  {Morpurgo}}]{Lezama2015}%
  \BibitemOpen
  \bibfield  {author} {\bibinfo {author} {\bibfnamefont {I.~G.}\ \bibnamefont
  {Lezama}}, \bibinfo {author} {\bibfnamefont {A.}~\bibnamefont {Arora}},
  \bibinfo {author} {\bibfnamefont {A.}~\bibnamefont {Ubaldini}}, \bibinfo
  {author} {\bibfnamefont {C.}~\bibnamefont {Barreteau}}, \bibinfo {author}
  {\bibfnamefont {E.}~\bibnamefont {Gianini}}, \bibinfo {author} {\bibfnamefont
  {M.}~\bibnamefont {Potemski}}, \ and\ \bibinfo {author} {\bibfnamefont
  {A.~F.}\ \bibnamefont {Morpurgo}},\ }\href@noop {} {\bibfield  {journal}
  {\bibinfo  {journal} {Nano Lett.}\ }\textbf {\bibinfo {volume} {15}},\
  \bibinfo {pages} {2336} (\bibinfo {year} {2015})}\BibitemShut {NoStop}%
\bibitem [{\citenamefont {Beal}\ \emph {et~al.}(1976)\citenamefont {Beal},
  \citenamefont {Liang},\ and\ \citenamefont {Hughes}}]{Beal1979}%
  \BibitemOpen
  \bibfield  {author} {\bibinfo {author} {\bibfnamefont {A.~R.}\ \bibnamefont
  {Beal}}, \bibinfo {author} {\bibfnamefont {W.~Y.}\ \bibnamefont {Liang}}, \
  and\ \bibinfo {author} {\bibfnamefont {H.~P.}\ \bibnamefont {Hughes}},\
  }\href@noop {} {\bibfield  {journal} {\bibinfo  {journal} {J. Phys. C: Sol.
  Stat. Phys.}\ }\textbf {\bibinfo {volume} {9}},\ \bibinfo {pages} {2449}
  (\bibinfo {year} {1976})}\BibitemShut {NoStop}%
\bibitem [{\citenamefont {Liang}(1973)}]{Liang1973}%
  \BibitemOpen
  \bibfield  {author} {\bibinfo {author} {\bibfnamefont {W.~Y.}\ \bibnamefont
  {Liang}},\ }\href@noop {} {\bibfield  {journal} {\bibinfo  {journal} {J.
  Phys. C: Sol. Stat. Comm.}\ }\textbf {\bibinfo {volume} {6}},\ \bibinfo
  {pages} {551} (\bibinfo {year} {1973})}\BibitemShut {NoStop}%
\bibitem [{\citenamefont {Luttrell}\ \emph {et~al.}(2006)\citenamefont
  {Luttrell}, \citenamefont {Brown}, \citenamefont {Cao}, \citenamefont {J.~L},
  \citenamefont {Rosentsveig},\ and\ \citenamefont {Tenne}}]{Luttrell2006}%
  \BibitemOpen
  \bibfield  {author} {\bibinfo {author} {\bibfnamefont {R.~D.}\ \bibnamefont
  {Luttrell}}, \bibinfo {author} {\bibfnamefont {S.}~\bibnamefont {Brown}},
  \bibinfo {author} {\bibfnamefont {J.}~\bibnamefont {Cao}}, \bibinfo {author}
  {\bibfnamefont {M.}~\bibnamefont {J.~L}}, \bibinfo {author} {\bibfnamefont
  {R.}~\bibnamefont {Rosentsveig}}, \ and\ \bibinfo {author} {\bibfnamefont
  {R.}~\bibnamefont {Tenne}},\ }\href@noop {} {\bibfield  {journal} {\bibinfo
  {journal} {Phys. Rev. B}\ }\textbf {\bibinfo {volume} {73}},\ \bibinfo
  {pages} {035410} (\bibinfo {year} {2006})}\BibitemShut {NoStop}%
\bibitem [{\citenamefont {Uchida}\ and\ \citenamefont
  {Tanaka}(1978)}]{Uchida1978}%
  \BibitemOpen
  \bibfield  {author} {\bibinfo {author} {\bibfnamefont {S.~I.}\ \bibnamefont
  {Uchida}}\ and\ \bibinfo {author} {\bibfnamefont {S.}~\bibnamefont
  {Tanaka}},\ }\href@noop {} {\bibfield  {journal} {\bibinfo  {journal} {J.
  Phys. Soc. Jpn.}\ }\textbf {\bibinfo {volume} {45}},\ \bibinfo {pages} {153}
  (\bibinfo {year} {1978})}\BibitemShut {NoStop}%
\bibitem [{\citenamefont {Starnberg}\ \emph {et~al.}(1995)\citenamefont
  {Starnberg}, \citenamefont {Brauer},\ and\ \citenamefont
  {Hughes}}]{Starnberg1995}%
  \BibitemOpen
  \bibfield  {author} {\bibinfo {author} {\bibfnamefont {H.~I.}\ \bibnamefont
  {Starnberg}}, \bibinfo {author} {\bibfnamefont {H.~E.}\ \bibnamefont
  {Brauer}}, \ and\ \bibinfo {author} {\bibfnamefont {H.~P.}\ \bibnamefont
  {Hughes}},\ }\href@noop {} {\bibfield  {journal} {\bibinfo  {journal} {J.
  Phys.: Cond Matt.}\ }\textbf {\bibinfo {volume} {8}},\ \bibinfo {pages}
  {1229} (\bibinfo {year} {1995})}\BibitemShut {NoStop}%
\bibitem [{\citenamefont {Zhao}\ \emph
  {et~al.}(2015{\natexlab{b}})\citenamefont {Zhao}, \citenamefont {Wu},
  \citenamefont {Zhong}, \citenamefont {Guo}, \citenamefont {Wang},
  \citenamefont {Xia}, \citenamefont {Yang}, \citenamefont {Tan},\ and\
  \citenamefont {Wang}}]{Zhao2015a}%
  \BibitemOpen
  \bibfield  {author} {\bibinfo {author} {\bibfnamefont {H.}~\bibnamefont
  {Zhao}}, \bibinfo {author} {\bibfnamefont {J.}~\bibnamefont {Wu}}, \bibinfo
  {author} {\bibfnamefont {H.}~\bibnamefont {Zhong}}, \bibinfo {author}
  {\bibfnamefont {Q.}~\bibnamefont {Guo}}, \bibinfo {author} {\bibfnamefont
  {X.}~\bibnamefont {Wang}}, \bibinfo {author} {\bibfnamefont {F.}~\bibnamefont
  {Xia}}, \bibinfo {author} {\bibfnamefont {L.}~\bibnamefont {Yang}}, \bibinfo
  {author} {\bibfnamefont {P.}~\bibnamefont {Tan}}, \ and\ \bibinfo {author}
  {\bibfnamefont {H.}~\bibnamefont {Wang}},\ }\href@noop {} {\bibfield
  {journal} {\bibinfo  {journal} {Nano Res.}\ }\textbf {\bibinfo {volume}
  {8}},\ \bibinfo {pages} {3651} (\bibinfo {year}
  {2015}{\natexlab{b}})}\BibitemShut {NoStop}%
\bibitem [{\citenamefont {Saigal}\ \emph {et~al.}(2016)\citenamefont {Saigal},
  \citenamefont {Sugunakar},\ and\ \citenamefont {Ghosh}}]{Saigal2016}%
  \BibitemOpen
  \bibfield  {author} {\bibinfo {author} {\bibfnamefont {N.}~\bibnamefont
  {Saigal}}, \bibinfo {author} {\bibfnamefont {V.}~\bibnamefont {Sugunakar}}, \
  and\ \bibinfo {author} {\bibfnamefont {S.}~\bibnamefont {Ghosh}},\
  }\href@noop {} {\bibfield  {journal} {\bibinfo  {journal} {Appl. Phys.
  Lett.}\ }\textbf {\bibinfo {volume} {108}},\ \bibinfo {pages} {132105}
  (\bibinfo {year} {2016})}\BibitemShut {NoStop}%
\bibitem [{\citenamefont {Hill}\ \emph {et~al.}(2015)\citenamefont {Hill},
  \citenamefont {Rigosi}, \citenamefont {Roquelet}, \citenamefont {Chernikov},
  \citenamefont {Berkelbach}, \citenamefont {Reichman}, \citenamefont
  {Hybertsen}, \citenamefont {Brus},\ and\ \citenamefont {Heinz}}]{Hill2015}%
  \BibitemOpen
  \bibfield  {author} {\bibinfo {author} {\bibfnamefont {H.~M.}\ \bibnamefont
  {Hill}}, \bibinfo {author} {\bibfnamefont {A.~F.}\ \bibnamefont {Rigosi}},
  \bibinfo {author} {\bibfnamefont {C.}~\bibnamefont {Roquelet}}, \bibinfo
  {author} {\bibfnamefont {A.}~\bibnamefont {Chernikov}}, \bibinfo {author}
  {\bibfnamefont {T.~C.}\ \bibnamefont {Berkelbach}}, \bibinfo {author}
  {\bibfnamefont {D.~R.}\ \bibnamefont {Reichman}}, \bibinfo {author}
  {\bibfnamefont {M.~S.}\ \bibnamefont {Hybertsen}}, \bibinfo {author}
  {\bibfnamefont {L.~E.}\ \bibnamefont {Brus}}, \ and\ \bibinfo {author}
  {\bibfnamefont {T.~F.}\ \bibnamefont {Heinz}},\ }\href@noop {} {\bibfield
  {journal} {\bibinfo  {journal} {Nano Lett.}\ }\textbf {\bibinfo {volume}
  {15}},\ \bibinfo {pages} {2992} (\bibinfo {year} {2015})}\BibitemShut
  {NoStop}%
\bibitem [{\citenamefont {Rigosi}\ \emph {et~al.}(2016)\citenamefont {Rigosi},
  \citenamefont {Hill}, \citenamefont {Rim}, \citenamefont {Flynn},\ and\
  \citenamefont {Heinz}}]{Rigosi2016}%
  \BibitemOpen
  \bibfield  {author} {\bibinfo {author} {\bibfnamefont {A.~F.}\ \bibnamefont
  {Rigosi}}, \bibinfo {author} {\bibfnamefont {H.~M.}\ \bibnamefont {Hill}},
  \bibinfo {author} {\bibfnamefont {K.~T.}\ \bibnamefont {Rim}}, \bibinfo
  {author} {\bibfnamefont {G.~W.}\ \bibnamefont {Flynn}}, \ and\ \bibinfo
  {author} {\bibfnamefont {T.~F.}\ \bibnamefont {Heinz}},\ }\href@noop {}
  {\bibfield  {journal} {\bibinfo  {journal} {Phys. Rev. B}\ }\textbf {\bibinfo
  {volume} {94}},\ \bibinfo {pages} {075440} (\bibinfo {year}
  {2016})}\BibitemShut {NoStop}%
\bibitem [{\citenamefont {Handbicki}\ \emph {et~al.}(2015)\citenamefont
  {Handbicki}, \citenamefont {Currie}, \citenamefont {Kioselglou},
  \citenamefont {Friedman},\ and\ \citenamefont {Jonker}}]{Handbicki2015}%
  \BibitemOpen
  \bibfield  {author} {\bibinfo {author} {\bibfnamefont {A.~T.}\ \bibnamefont
  {Handbicki}}, \bibinfo {author} {\bibfnamefont {M.}~\bibnamefont {Currie}},
  \bibinfo {author} {\bibfnamefont {G.}~\bibnamefont {Kioselglou}}, \bibinfo
  {author} {\bibfnamefont {A.~L.}\ \bibnamefont {Friedman}}, \ and\ \bibinfo
  {author} {\bibfnamefont {B.~T.}\ \bibnamefont {Jonker}},\ }\href@noop {}
  {\bibfield  {journal} {\bibinfo  {journal} {Sol. Stat. Comm.}\ }\textbf
  {\bibinfo {volume} {203}},\ \bibinfo {pages} {16} (\bibinfo {year}
  {2015})}\BibitemShut {NoStop}%
\bibitem [{\citenamefont {Kylanpaa}\ and\ \citenamefont
  {Komsa}(2015)}]{Kylanpaa2015}%
  \BibitemOpen
  \bibfield  {author} {\bibinfo {author} {\bibfnamefont {I.}~\bibnamefont
  {Kylanpaa}}\ and\ \bibinfo {author} {\bibfnamefont {H.-P.}\ \bibnamefont
  {Komsa}},\ }\href@noop {} {\bibfield  {journal} {\bibinfo  {journal} {Phys.
  Rev. B}\ }\textbf {\bibinfo {volume} {92}},\ \bibinfo {pages} {205418}
  (\bibinfo {year} {2015})}\BibitemShut {NoStop}%
\bibitem [{\citenamefont {Pike}\ \emph {et~al.}(2017)\citenamefont {Pike},
  \citenamefont {Troeye}, \citenamefont {Dewandre}, \citenamefont {Gonze},\
  and\ \citenamefont {Verstraete}}]{Pike2016}%
  \BibitemOpen
  \bibfield  {author} {\bibinfo {author} {\bibfnamefont {N.~A.}\ \bibnamefont
  {Pike}}, \bibinfo {author} {\bibfnamefont {B.~V.}\ \bibnamefont {Troeye}},
  \bibinfo {author} {\bibfnamefont {A.}~\bibnamefont {Dewandre}}, \bibinfo
  {author} {\bibfnamefont {X.}~\bibnamefont {Gonze}}, \ and\ \bibinfo {author}
  {\bibfnamefont {M.~J.}\ \bibnamefont {Verstraete}},\ }\href@noop {}
  {\bibfield  {journal} {\bibinfo  {journal} {Phys. Rev. B}\ }\textbf {\bibinfo
  {volume} {95}},\ \bibinfo {pages} {201106R} (\bibinfo {year}
  {2017})}\BibitemShut {NoStop}%
\bibitem [{\citenamefont {Lee}\ and\ \citenamefont {Gonze}(1995)}]{Lee1995}%
  \BibitemOpen
  \bibfield  {author} {\bibinfo {author} {\bibfnamefont {C.}~\bibnamefont
  {Lee}}\ and\ \bibinfo {author} {\bibfnamefont {X.}~\bibnamefont {Gonze}},\
  }\href@noop {} {\bibfield  {journal} {\bibinfo  {journal} {Phys. Rev. B}\
  }\textbf {\bibinfo {volume} {51}},\ \bibinfo {pages} {8610} (\bibinfo {year}
  {1995})}\BibitemShut {NoStop}%
\bibitem [{\citenamefont {Liu}\ \emph {et~al.}(2015)\citenamefont {Liu},
  \citenamefont {Wang}, \citenamefont {Bo}, \citenamefont {Liu}, \citenamefont
  {Yang}, \citenamefont {Huang},\ and\ \citenamefont {Sun}}]{Liu2015}%
  \BibitemOpen
  \bibfield  {author} {\bibinfo {author} {\bibfnamefont {Y.}~\bibnamefont
  {Liu}}, \bibinfo {author} {\bibfnamefont {W.}~\bibnamefont {Wang}}, \bibinfo
  {author} {\bibfnamefont {M.}~\bibnamefont {Bo}}, \bibinfo {author}
  {\bibfnamefont {X.}~\bibnamefont {Liu}}, \bibinfo {author} {\bibfnamefont
  {X.}~\bibnamefont {Yang}}, \bibinfo {author} {\bibfnamefont {Y.}~\bibnamefont
  {Huang}}, \ and\ \bibinfo {author} {\bibfnamefont {C.~Q.}\ \bibnamefont
  {Sun}},\ }\href@noop {} {\bibfield  {journal} {\bibinfo  {journal} {J. Phys.
  Chem. C}\ }\textbf {\bibinfo {volume} {119}},\ \bibinfo {pages} {25071}
  (\bibinfo {year} {2015})}\BibitemShut {NoStop}%
\bibitem [{\citenamefont {Wender}\ and\ \citenamefont
  {Hershkowitz}(1973)}]{Wender1973}%
  \BibitemOpen
  \bibfield  {author} {\bibinfo {author} {\bibfnamefont {S.~A.}\ \bibnamefont
  {Wender}}\ and\ \bibinfo {author} {\bibfnamefont {N.}~\bibnamefont
  {Hershkowitz}},\ }\href@noop {} {\bibfield  {journal} {\bibinfo  {journal}
  {Phys. Rev. B}\ }\textbf {\bibinfo {volume} {8}},\ \bibinfo {pages} {4901}
  (\bibinfo {year} {1973})}\BibitemShut {NoStop}%
\bibitem [{\citenamefont {Titov}\ \emph {et~al.}(2007)\citenamefont {Titov},
  \citenamefont {Skomorokhov}, \citenamefont {Titov}, \citenamefont {Titova},\
  and\ \citenamefont {Semenov}}]{Titov2007}%
  \BibitemOpen
  \bibfield  {author} {\bibinfo {author} {\bibfnamefont {A.~N.}\ \bibnamefont
  {Titov}}, \bibinfo {author} {\bibfnamefont {A.~N.}\ \bibnamefont
  {Skomorokhov}}, \bibinfo {author} {\bibfnamefont {A.~A.}\ \bibnamefont
  {Titov}}, \bibinfo {author} {\bibfnamefont {S.~G.}\ \bibnamefont {Titova}}, \
  and\ \bibinfo {author} {\bibfnamefont {V.~A.}\ \bibnamefont {Semenov}},\
  }\href@noop {} {\bibfield  {journal} {\bibinfo  {journal} {Phys. of the Sol.
  Stat.}\ }\textbf {\bibinfo {volume} {49}},\ \bibinfo {pages} {1532} (\bibinfo
  {year} {2007})}\BibitemShut {NoStop}%
\bibitem [{\citenamefont {Wakabayashi}\ \emph {et~al.}(1975)\citenamefont
  {Wakabayashi}, \citenamefont {Smith},\ and\ \citenamefont
  {Nicklow}}]{Wakabayashi1975}%
  \BibitemOpen
  \bibfield  {author} {\bibinfo {author} {\bibfnamefont {N.}~\bibnamefont
  {Wakabayashi}}, \bibinfo {author} {\bibfnamefont {H.~G.}\ \bibnamefont
  {Smith}}, \ and\ \bibinfo {author} {\bibfnamefont {R.~M.}\ \bibnamefont
  {Nicklow}},\ }\href@noop {} {\bibfield  {journal} {\bibinfo  {journal} {Phys.
  Rev. B}\ }\textbf {\bibinfo {volume} {12}},\ \bibinfo {pages} {656} (\bibinfo
  {year} {1975})}\BibitemShut {NoStop}%
\bibitem [{\citenamefont {Sourisseau}\ \emph {et~al.}(1989)\citenamefont
  {Sourisseau}, \citenamefont {Fouassier}, \citenamefont {Alba}, \citenamefont
  {Ghorayeb},\ and\ \citenamefont {Gorochov}}]{Sourisseau1989}%
  \BibitemOpen
  \bibfield  {author} {\bibinfo {author} {\bibfnamefont {C.}~\bibnamefont
  {Sourisseau}}, \bibinfo {author} {\bibfnamefont {M.}~\bibnamefont
  {Fouassier}}, \bibinfo {author} {\bibfnamefont {M.}~\bibnamefont {Alba}},
  \bibinfo {author} {\bibfnamefont {A.}~\bibnamefont {Ghorayeb}}, \ and\
  \bibinfo {author} {\bibfnamefont {O.}~\bibnamefont {Gorochov}},\ }\href@noop
  {} {\bibfield  {journal} {\bibinfo  {journal} {Mat. Sci. and Eng. B}\
  }\textbf {\bibinfo {volume} {3}},\ \bibinfo {pages} {119} (\bibinfo {year}
  {1989})}\BibitemShut {NoStop}%
\bibitem [{\citenamefont {Zhao}\ \emph
  {et~al.}(2013{\natexlab{b}})\citenamefont {Zhao}, \citenamefont {Luo},
  \citenamefont {Li}, \citenamefont {Zhang}, \citenamefont {Araujo},
  \citenamefont {Gan}, \citenamefont {Wu}, \citenamefont {Zhang}, \citenamefont
  {Quek}, \citenamefont {Dresselhaus},\ and\ \citenamefont
  {Xiong}}]{Zhao2013b}%
  \BibitemOpen
  \bibfield  {author} {\bibinfo {author} {\bibfnamefont {Y.}~\bibnamefont
  {Zhao}}, \bibinfo {author} {\bibfnamefont {X.}~\bibnamefont {Luo}}, \bibinfo
  {author} {\bibfnamefont {H.}~\bibnamefont {Li}}, \bibinfo {author}
  {\bibfnamefont {J.}~\bibnamefont {Zhang}}, \bibinfo {author} {\bibfnamefont
  {P.~T.}\ \bibnamefont {Araujo}}, \bibinfo {author} {\bibfnamefont {C.~K.}\
  \bibnamefont {Gan}}, \bibinfo {author} {\bibfnamefont {J.}~\bibnamefont
  {Wu}}, \bibinfo {author} {\bibfnamefont {H.}~\bibnamefont {Zhang}}, \bibinfo
  {author} {\bibfnamefont {S.~Y.}\ \bibnamefont {Quek}}, \bibinfo {author}
  {\bibfnamefont {M.~S.}\ \bibnamefont {Dresselhaus}}, \ and\ \bibinfo {author}
  {\bibfnamefont {Q.}~\bibnamefont {Xiong}},\ }\href@noop {} {\bibfield
  {journal} {\bibinfo  {journal} {Nano Lett.}\ }\textbf {\bibinfo {volume}
  {13}},\ \bibinfo {pages} {1007} (\bibinfo {year}
  {2013}{\natexlab{b}})}\BibitemShut {NoStop}%
\bibitem [{\citenamefont {Tonndorf}\ \emph {et~al.}(2013)\citenamefont
  {Tonndorf}, \citenamefont {Schmidt}, \citenamefont {Bottger}, \citenamefont
  {Zhang}, \citenamefont {Borner}, \citenamefont {Liebig}, \citenamefont
  {Albrecht}, \citenamefont {Kloc}, \citenamefont {Gordan}, \citenamefont
  {Zahn}, \citenamefont {de~Vasconcellos},\ and\ \citenamefont
  {Bratschitsch}}]{Tonndorf2013}%
  \BibitemOpen
  \bibfield  {author} {\bibinfo {author} {\bibfnamefont {P.}~\bibnamefont
  {Tonndorf}}, \bibinfo {author} {\bibfnamefont {R.}~\bibnamefont {Schmidt}},
  \bibinfo {author} {\bibfnamefont {P.}~\bibnamefont {Bottger}}, \bibinfo
  {author} {\bibfnamefont {X.}~\bibnamefont {Zhang}}, \bibinfo {author}
  {\bibfnamefont {H.}~\bibnamefont {Borner}}, \bibinfo {author} {\bibfnamefont
  {A.}~\bibnamefont {Liebig}}, \bibinfo {author} {\bibfnamefont
  {M.}~\bibnamefont {Albrecht}}, \bibinfo {author} {\bibfnamefont
  {C.}~\bibnamefont {Kloc}}, \bibinfo {author} {\bibfnamefont {O.}~\bibnamefont
  {Gordan}}, \bibinfo {author} {\bibfnamefont {D.~R.~T.}\ \bibnamefont {Zahn}},
  \bibinfo {author} {\bibfnamefont {S.~M.}\ \bibnamefont {de~Vasconcellos}}, \
  and\ \bibinfo {author} {\bibfnamefont {R.}~\bibnamefont {Bratschitsch}},\
  }\href@noop {} {\bibfield  {journal} {\bibinfo  {journal} {Opt. Express}\
  }\textbf {\bibinfo {volume} {21}},\ \bibinfo {pages} {4908} (\bibinfo {year}
  {2013})}\BibitemShut {NoStop}%
\bibitem [{\citenamefont {Tongay}\ \emph {et~al.}(2012)\citenamefont {Tongay},
  \citenamefont {Zhou}, \citenamefont {Ataca}, \citenamefont {Lo},
  \citenamefont {Matthews}, \citenamefont {Li}, \citenamefont {Grossman},\ and\
  \citenamefont {Wu}}]{Tongay2012}%
  \BibitemOpen
  \bibfield  {author} {\bibinfo {author} {\bibfnamefont {S.}~\bibnamefont
  {Tongay}}, \bibinfo {author} {\bibfnamefont {J.}~\bibnamefont {Zhou}},
  \bibinfo {author} {\bibfnamefont {C.}~\bibnamefont {Ataca}}, \bibinfo
  {author} {\bibfnamefont {K.}~\bibnamefont {Lo}}, \bibinfo {author}
  {\bibfnamefont {T.~S.}\ \bibnamefont {Matthews}}, \bibinfo {author}
  {\bibfnamefont {J.}~\bibnamefont {Li}}, \bibinfo {author} {\bibfnamefont
  {J.~C.}\ \bibnamefont {Grossman}}, \ and\ \bibinfo {author} {\bibfnamefont
  {J.}~\bibnamefont {Wu}},\ }\href@noop {} {\bibfield  {journal} {\bibinfo
  {journal} {Nano Lett.}\ }\textbf {\bibinfo {volume} {12}},\ \bibinfo {pages}
  {5576} (\bibinfo {year} {2012})}\BibitemShut {NoStop}%
\bibitem [{\citenamefont {Sekine}\ \emph {et~al.}(1980)\citenamefont {Sekine},
  \citenamefont {Izumi}, \citenamefont {Nakashizu}, \citenamefont
  {Uchinokura},\ and\ \citenamefont {Matsuura}}]{Sekine1980}%
  \BibitemOpen
  \bibfield  {author} {\bibinfo {author} {\bibfnamefont {T.}~\bibnamefont
  {Sekine}}, \bibinfo {author} {\bibfnamefont {M.}~\bibnamefont {Izumi}},
  \bibinfo {author} {\bibfnamefont {T.}~\bibnamefont {Nakashizu}}, \bibinfo
  {author} {\bibfnamefont {K.}~\bibnamefont {Uchinokura}}, \ and\ \bibinfo
  {author} {\bibfnamefont {E.}~\bibnamefont {Matsuura}},\ }\href@noop {}
  {\bibfield  {journal} {\bibinfo  {journal} {J. Phys. Soc. Jpn.}\ }\textbf
  {\bibinfo {volume} {49}},\ \bibinfo {pages} {1069} (\bibinfo {year}
  {1980})}\BibitemShut {NoStop}%
\bibitem [{\citenamefont {Mead}\ and\ \citenamefont {Irwin}(1977)}]{Mead1977}%
  \BibitemOpen
  \bibfield  {author} {\bibinfo {author} {\bibfnamefont {D.~G.}\ \bibnamefont
  {Mead}}\ and\ \bibinfo {author} {\bibfnamefont {J.~C.}\ \bibnamefont
  {Irwin}},\ }\href@noop {} {\bibfield  {journal} {\bibinfo  {journal} {Can. J.
  Phys.}\ }\textbf {\bibinfo {volume} {55}},\ \bibinfo {pages} {379} (\bibinfo
  {year} {1977})}\BibitemShut {NoStop}%
\bibitem [{\citenamefont {Froehlicher}\ \emph {et~al.}(2015)\citenamefont
  {Froehlicher}, \citenamefont {Lorchat}, \citenamefont {Fernique},
  \citenamefont {Joshi}, \citenamefont {Molina-Sanchez}, \citenamefont
  {Wirtz},\ and\ \citenamefont {Berciaud}}]{Froehlicher2015}%
  \BibitemOpen
  \bibfield  {author} {\bibinfo {author} {\bibfnamefont {G.}~\bibnamefont
  {Froehlicher}}, \bibinfo {author} {\bibfnamefont {E.}~\bibnamefont
  {Lorchat}}, \bibinfo {author} {\bibfnamefont {F.}~\bibnamefont {Fernique}},
  \bibinfo {author} {\bibfnamefont {C.}~\bibnamefont {Joshi}}, \bibinfo
  {author} {\bibfnamefont {A.}~\bibnamefont {Molina-Sanchez}}, \bibinfo
  {author} {\bibfnamefont {L.}~\bibnamefont {Wirtz}}, \ and\ \bibinfo {author}
  {\bibfnamefont {S.}~\bibnamefont {Berciaud}},\ }\href@noop {} {\bibfield
  {journal} {\bibinfo  {journal} {Nano Lett.}\ }\textbf {\bibinfo {volume}
  {15}},\ \bibinfo {pages} {6481} (\bibinfo {year} {2015})}\BibitemShut
  {NoStop}%
\bibitem [{\citenamefont {Sandoval}\ \emph {et~al.}(1992)\citenamefont
  {Sandoval}, \citenamefont {Chen},\ and\ \citenamefont
  {Irwin}}]{Sandoval1992}%
  \BibitemOpen
  \bibfield  {author} {\bibinfo {author} {\bibfnamefont {S.~J.}\ \bibnamefont
  {Sandoval}}, \bibinfo {author} {\bibfnamefont {X.~K.}\ \bibnamefont {Chen}},
  \ and\ \bibinfo {author} {\bibfnamefont {J.~C.}\ \bibnamefont {Irwin}},\
  }\href@noop {} {\bibfield  {journal} {\bibinfo  {journal} {Phys. Rev. B}\
  }\textbf {\bibinfo {volume} {45}},\ \bibinfo {pages} {14347} (\bibinfo {year}
  {1992})}\BibitemShut {NoStop}%
\bibitem [{\citenamefont {Sugai}\ \emph {et~al.}(1980)\citenamefont {Sugai},
  \citenamefont {Murase}, \citenamefont {Uchida},\ and\ \citenamefont
  {Tanaka}}]{Sugai1980}%
  \BibitemOpen
  \bibfield  {author} {\bibinfo {author} {\bibfnamefont {S.}~\bibnamefont
  {Sugai}}, \bibinfo {author} {\bibfnamefont {K.}~\bibnamefont {Murase}},
  \bibinfo {author} {\bibfnamefont {S.}~\bibnamefont {Uchida}}, \ and\ \bibinfo
  {author} {\bibfnamefont {S.}~\bibnamefont {Tanaka}},\ }\href@noop {}
  {\bibfield  {journal} {\bibinfo  {journal} {Sol. Stat. Comm.}\ }\textbf
  {\bibinfo {volume} {35}},\ \bibinfo {pages} {433} (\bibinfo {year}
  {1980})}\BibitemShut {NoStop}%
\bibitem [{\citenamefont {Hangyo}\ \emph {et~al.}(1983)\citenamefont {Hangyo},
  \citenamefont {Nakashima},\ and\ \citenamefont {Mitsuishi}}]{Hangyo1983}%
  \BibitemOpen
  \bibfield  {author} {\bibinfo {author} {\bibfnamefont {M.}~\bibnamefont
  {Hangyo}}, \bibinfo {author} {\bibfnamefont {S.-I.}\ \bibnamefont
  {Nakashima}}, \ and\ \bibinfo {author} {\bibfnamefont {A.}~\bibnamefont
  {Mitsuishi}},\ }\href@noop {} {\bibfield  {journal} {\bibinfo  {journal}
  {Ferroelectrics}\ }\textbf {\bibinfo {volume} {52}},\ \bibinfo {pages} {151}
  (\bibinfo {year} {1983})}\BibitemShut {NoStop}%
\bibitem [{\citenamefont {Roubi}\ and\ \citenamefont
  {Carlone}(1988)}]{Roubi1988}%
  \BibitemOpen
  \bibfield  {author} {\bibinfo {author} {\bibfnamefont {L.}~\bibnamefont
  {Roubi}}\ and\ \bibinfo {author} {\bibfnamefont {C.}~\bibnamefont
  {Carlone}},\ }\href@noop {} {\bibfield  {journal} {\bibinfo  {journal} {Phys.
  Rev. B}\ }\textbf {\bibinfo {volume} {37}},\ \bibinfo {pages} {6808}
  (\bibinfo {year} {1988})}\BibitemShut {NoStop}%
\bibitem [{\citenamefont {Manas-Valero}\ \emph {et~al.}(2016)\citenamefont
  {Manas-Valero}, \citenamefont {Garcia-Lopez}, \citenamefont {Cantarero},\
  and\ \citenamefont {Galbiati}}]{Manas2016}%
  \BibitemOpen
  \bibfield  {author} {\bibinfo {author} {\bibfnamefont {S.}~\bibnamefont
  {Manas-Valero}}, \bibinfo {author} {\bibfnamefont {V.}~\bibnamefont
  {Garcia-Lopez}}, \bibinfo {author} {\bibfnamefont {A.}~\bibnamefont
  {Cantarero}}, \ and\ \bibinfo {author} {\bibfnamefont {M.}~\bibnamefont
  {Galbiati}},\ }\href@noop {} {\bibfield  {journal} {\bibinfo  {journal}
  {Appl. Sci.}\ }\textbf {\bibinfo {volume} {6}},\ \bibinfo {pages} {264}
  (\bibinfo {year} {2016})}\BibitemShut {NoStop}%
\bibitem [{\citenamefont {Duong}\ \emph {et~al.}(2015)\citenamefont {Duong},
  \citenamefont {Burghard},\ and\ \citenamefont {Schon}}]{Duong2015}%
  \BibitemOpen
  \bibfield  {author} {\bibinfo {author} {\bibfnamefont {D.~L.}\ \bibnamefont
  {Duong}}, \bibinfo {author} {\bibfnamefont {M.}~\bibnamefont {Burghard}}, \
  and\ \bibinfo {author} {\bibfnamefont {J.~C.}\ \bibnamefont {Schon}},\
  }\href@noop {} {\bibfield  {journal} {\bibinfo  {journal} {Phys. Rev. B}\
  }\textbf {\bibinfo {volume} {92}},\ \bibinfo {pages} {245131} (\bibinfo
  {year} {2015})}\BibitemShut {NoStop}%
\bibitem [{\citenamefont {Fang}\ \emph {et~al.}(1997)\citenamefont {Fang},
  \citenamefont {de~Groot},\ and\ \citenamefont {Haas}}]{Fang1997}%
  \BibitemOpen
  \bibfield  {author} {\bibinfo {author} {\bibfnamefont {C.~M.}\ \bibnamefont
  {Fang}}, \bibinfo {author} {\bibfnamefont {R.~A.}\ \bibnamefont {de~Groot}},
  \ and\ \bibinfo {author} {\bibfnamefont {C.}~\bibnamefont {Haas}},\
  }\href@noop {} {\bibfield  {journal} {\bibinfo  {journal} {Phys. Rev. B}\
  }\textbf {\bibinfo {volume} {56}},\ \bibinfo {pages} {4455} (\bibinfo {year}
  {1997})}\BibitemShut {NoStop}%
\bibitem [{\citenamefont {Liu}\ \emph {et~al.}(2012)\citenamefont {Liu},
  \citenamefont {Porter},\ and\ \citenamefont {Goldberger}}]{Liu2012}%
  \BibitemOpen
  \bibfield  {author} {\bibinfo {author} {\bibfnamefont {Y.~H.}\ \bibnamefont
  {Liu}}, \bibinfo {author} {\bibfnamefont {S.~H.}\ \bibnamefont {Porter}}, \
  and\ \bibinfo {author} {\bibfnamefont {J.~E.}\ \bibnamefont {Goldberger}},\
  }\href@noop {} {\bibfield  {journal} {\bibinfo  {journal} {J. Am. Chem. Soc}\
  }\textbf {\bibinfo {volume} {134}},\ \bibinfo {pages} {5044} (\bibinfo {year}
  {2012})}\BibitemShut {NoStop}%
\bibitem [{\citenamefont {Sharma}\ \emph {et~al.}(1999)\citenamefont {Sharma},
  \citenamefont {Nautiyal}, \citenamefont {Singh}, \citenamefont {Auluck},
  \citenamefont {Blaha},\ and\ \citenamefont {Ambrosh-Draxl}}]{Sharma1999}%
  \BibitemOpen
  \bibfield  {author} {\bibinfo {author} {\bibfnamefont {S.}~\bibnamefont
  {Sharma}}, \bibinfo {author} {\bibfnamefont {T.}~\bibnamefont {Nautiyal}},
  \bibinfo {author} {\bibfnamefont {G.~S.}\ \bibnamefont {Singh}}, \bibinfo
  {author} {\bibfnamefont {S.}~\bibnamefont {Auluck}}, \bibinfo {author}
  {\bibfnamefont {P.}~\bibnamefont {Blaha}}, \ and\ \bibinfo {author}
  {\bibfnamefont {C.}~\bibnamefont {Ambrosh-Draxl}},\ }\href@noop {} {\bibfield
   {journal} {\bibinfo  {journal} {Phys. Rev. B}\ }\textbf {\bibinfo {volume}
  {59}},\ \bibinfo {pages} {14833} (\bibinfo {year} {1999})}\BibitemShut
  {NoStop}%
\bibitem [{\citenamefont {Chen}\ \emph {et~al.}(2017)\citenamefont {Chen},
  \citenamefont {Pai}, \citenamefont {Chan}, \citenamefont {Takayama},
  \citenamefont {Xu}, \citenamefont {Karn}, \citenamefont {Hasegawa},
  \citenamefont {Chou}, \citenamefont {Mo}, \citenamefont {Fedorov},\ and\
  \citenamefont {Chaing}}]{Chen2017}%
  \BibitemOpen
  \bibfield  {author} {\bibinfo {author} {\bibfnamefont {P.}~\bibnamefont
  {Chen}}, \bibinfo {author} {\bibfnamefont {W.~W.}\ \bibnamefont {Pai}},
  \bibinfo {author} {\bibfnamefont {Y.-H.}\ \bibnamefont {Chan}}, \bibinfo
  {author} {\bibfnamefont {A.}~\bibnamefont {Takayama}}, \bibinfo {author}
  {\bibfnamefont {C.-Z.}\ \bibnamefont {Xu}}, \bibinfo {author} {\bibfnamefont
  {A.}~\bibnamefont {Karn}}, \bibinfo {author} {\bibfnamefont {S.}~\bibnamefont
  {Hasegawa}}, \bibinfo {author} {\bibfnamefont {M.~Y.}\ \bibnamefont {Chou}},
  \bibinfo {author} {\bibfnamefont {S.-K.}\ \bibnamefont {Mo}}, \bibinfo
  {author} {\bibfnamefont {A.-V.}\ \bibnamefont {Fedorov}}, \ and\ \bibinfo
  {author} {\bibfnamefont {T.-C.}\ \bibnamefont {Chaing}},\ }\href@noop {}
  {\bibfield  {journal} {\bibinfo  {journal} {Nature. Comm.}\ }\textbf
  {\bibinfo {volume} {8}},\ \bibinfo {pages} {516} (\bibinfo {year}
  {2017})}\BibitemShut {NoStop}%
\bibitem [{\citenamefont {Heil}\ \emph {et~al.}(2017)\citenamefont {Heil},
  \citenamefont {Ponce}, \citenamefont {Lambert}, \citenamefont {Schlipf},
  \citenamefont {Margine},\ and\ \citenamefont {Giustino}}]{Heil2017}%
  \BibitemOpen
  \bibfield  {author} {\bibinfo {author} {\bibfnamefont {C.}~\bibnamefont
  {Heil}}, \bibinfo {author} {\bibfnamefont {S.}~\bibnamefont {Ponce}},
  \bibinfo {author} {\bibfnamefont {H.}~\bibnamefont {Lambert}}, \bibinfo
  {author} {\bibfnamefont {M.}~\bibnamefont {Schlipf}}, \bibinfo {author}
  {\bibfnamefont {E.~R.}\ \bibnamefont {Margine}}, \ and\ \bibinfo {author}
  {\bibfnamefont {F.}~\bibnamefont {Giustino}},\ }\href@noop {} {\bibfield
  {journal} {\bibinfo  {journal} {Phys. Rev. Lett.}\ }\textbf {\bibinfo
  {volume} {119}},\ \bibinfo {pages} {087003} (\bibinfo {year}
  {2017})}\BibitemShut {NoStop}%
\bibitem [{\citenamefont {Aroyo}\ \emph {et~al.}(2011)\citenamefont {Aroyo},
  \citenamefont {Perez-Mato}, \citenamefont {Orobengoa}, \citenamefont {Tasci},
  \citenamefont {de~la Flor},\ and\ \citenamefont {Kirov}}]{Aroyo2011}%
  \BibitemOpen
  \bibfield  {author} {\bibinfo {author} {\bibfnamefont {M.~I.}\ \bibnamefont
  {Aroyo}}, \bibinfo {author} {\bibfnamefont {J.~M.}\ \bibnamefont
  {Perez-Mato}}, \bibinfo {author} {\bibfnamefont {D.}~\bibnamefont
  {Orobengoa}}, \bibinfo {author} {\bibfnamefont {E.}~\bibnamefont {Tasci}},
  \bibinfo {author} {\bibfnamefont {G.}~\bibnamefont {de~la Flor}}, \ and\
  \bibinfo {author} {\bibfnamefont {A.}~\bibnamefont {Kirov}},\ }\href@noop {}
  {\bibfield  {journal} {\bibinfo  {journal} {Bulg. Chem. Commun}\ }\textbf
  {\bibinfo {volume} {43}},\ \bibinfo {pages} {183} (\bibinfo {year}
  {2011})}\BibitemShut {NoStop}%
\bibitem [{\citenamefont {Aroyo}\ \emph
  {et~al.}(2006{\natexlab{a}})\citenamefont {Aroyo}, \citenamefont
  {Perez-Mato}, \citenamefont {Capillas}, \citenamefont {Kroumova},
  \citenamefont {Ivantchev}, \citenamefont {Madariaga}, \citenamefont {Kirov},\
  and\ \citenamefont {Wondratschek}}]{Aroyo2006}%
  \BibitemOpen
  \bibfield  {author} {\bibinfo {author} {\bibfnamefont {M.~I.}\ \bibnamefont
  {Aroyo}}, \bibinfo {author} {\bibfnamefont {J.~M.}\ \bibnamefont
  {Perez-Mato}}, \bibinfo {author} {\bibfnamefont {C.}~\bibnamefont
  {Capillas}}, \bibinfo {author} {\bibfnamefont {E.}~\bibnamefont {Kroumova}},
  \bibinfo {author} {\bibfnamefont {S.}~\bibnamefont {Ivantchev}}, \bibinfo
  {author} {\bibfnamefont {G.}~\bibnamefont {Madariaga}}, \bibinfo {author}
  {\bibfnamefont {A.}~\bibnamefont {Kirov}}, \ and\ \bibinfo {author}
  {\bibfnamefont {H.}~\bibnamefont {Wondratschek}},\ }\href@noop {} {\bibfield
  {journal} {\bibinfo  {journal} {Z. Krist}\ }\textbf {\bibinfo {volume}
  {221}},\ \bibinfo {pages} {15} (\bibinfo {year}
  {2006}{\natexlab{a}})}\BibitemShut {NoStop}%
\bibitem [{\citenamefont {Aroyo}\ \emph
  {et~al.}(2006{\natexlab{b}})\citenamefont {Aroyo}, \citenamefont {Kirov},
  \citenamefont {Capillas}, \citenamefont {Perez-Mato},\ and\ \citenamefont
  {Wondratschek}}]{Aroyo2006a}%
  \BibitemOpen
  \bibfield  {author} {\bibinfo {author} {\bibfnamefont {M.~I.}\ \bibnamefont
  {Aroyo}}, \bibinfo {author} {\bibfnamefont {A.}~\bibnamefont {Kirov}},
  \bibinfo {author} {\bibfnamefont {C.}~\bibnamefont {Capillas}}, \bibinfo
  {author} {\bibfnamefont {J.~M.}\ \bibnamefont {Perez-Mato}}, \ and\ \bibinfo
  {author} {\bibfnamefont {H.}~\bibnamefont {Wondratschek}},\ }\href@noop {}
  {\bibfield  {journal} {\bibinfo  {journal} {Acta Cryst. A.}\ }\textbf
  {\bibinfo {volume} {62}},\ \bibinfo {pages} {115} (\bibinfo {year}
  {2006}{\natexlab{b}})}\BibitemShut {NoStop}%
\end{thebibliography}%

\end{document}